\documentclass[10pt,fleqn,a4paper]{article}

\pdfoutput=1
\usepackage{mathtools}
\usepackage{amsmath}
\usepackage{amssymb}
\usepackage{graphicx}
\usepackage{titlesec}
\usepackage{amsmath}
\usepackage{dsfont}
\usepackage{amsthm}
\usepackage{multicol}
\usepackage[usenames,dvipsnames,svgnames,table]{xcolor}
\usepackage[margin=30pt,font=small,labelfont=bf,labelsep=endash]{caption}
\usepackage{amsthm}
\usepackage{array,multirow}
\usepackage{shuffle}
\usepackage{empheq}
\usepackage[title]{appendix}
\usepackage{tikz}
\usetikzlibrary{calc,matrix}

\newcommand{\cC}{{\cal C}}
\newcommand{\cD}{{\cal D}}
\newcommand{\cE}{{\cal E}}
\newcommand{\cF}{{\cal F}}

\newcommand{\cI}{{\cal I}}
\newcommand{\cJ}{{\cal J}}

\newcommand{\cM}{{\cal M}}
\newcommand{\cN}{{\cal N}}

\newcommand{\cP}{{\cal P}}

\newcommand{\cT}{{\cal T}}

\newcommand{\cZ}{{\cal Z}}

\newcommand{\dC}{\mathds{C}}

\newcommand{\dR}{\mathds{R}}
\newcommand{\dN}{\mathds{N}}

\newcommand{\rd}{\text{d}}
\newcommand{\re}{\text{e}}

\newcommand{\ri}{\text{i}}

\newcommand{\fT}{{\mathfrak T}}

\newcommand{\ie}{{\it i.e.\ }}
\newcommand{\viz}{{\it viz.\ }}

\newcommand{\E}{\mathds{E}}
\let\P\undefined
\newcommand{\P}{\mathds{P}}

\newcommand{\texit}{t_\text{exit}}
\newcommand{\qexit}{q_\text{exit}}
\newcommand{\kmin}{k_\text{min}}
\newcommand{\nmin}{n_\text{min}}
\newcommand{\Femp}{{\cal F}_{\rm\scriptscriptstyle EMP}}
\newcommand{\FQ}{{\cal F}_{\rm\scriptscriptstyle Q}}
\newcommand{\FFC}{{\cal F}_{\rm\scriptscriptstyle FC}}

\newcommand{\pB}{p_{\rm\scriptscriptstyle B}}
\newcommand{\mB}{m_{\rm\scriptscriptstyle B}}
\newcommand{\pQ}{p_{\rm\scriptscriptstyle Q}}
\newcommand{\mQ}{m_{\rm\scriptscriptstyle Q}}
\newcommand{\hatmQ}{{\hat m}_{\rm\scriptscriptstyle Q}}
\newcommand{\mFC}{m_{\rm\scriptscriptstyle FC}}

\numberwithin{equation}{section}
\definecolor{myblue}{rgb}{0.97,0.97,0.97}
\newcommand*\mybluebox[1]{%
\colorbox{myblue}{\hspace{1em}#1\hspace{1em}}}

\begin{document}
\title{On the Power--Law Tails of Vote Distributions \\ in Proportional Elections}
\author{
\\[-0.1cm]
{{{Filippo Palombi$^{a}$\footnote{Corresponding
        author. E--mail: {\tt filippo.palombi@enea.it}\newline}\ \  and Simona Toti$^{b}$}}}\\[1.0ex]
 {{\small{$^a$ENEA -- Italian National Agency for New Technologies, Energy and}}}\\
{{\small{Sustainable Economic Development,}}}
 {\small {{\it Via Enrico Fermi 45, 00044 Frascati -- Italy}}}\\[.1cm]
 {{\small{$^b$ISTAT -- Istituto Nazionale di Statistica,}}}
 {\small {\it Via Cesare Balbo 16, 00184 Rome -- Italy}}\\[.1cm]
}

\date{\today}

\maketitle
\begin{abstract}
  In proportional elections with open lists the excess of preferences received by candidates with respect to the list average is known to follow a universal lognormal distribution. We show that lognormality is broken provided preferences are conditioned to lists with many candidates. In this limit power--law tails emerge. We study the large--list limit in the framework of a quenched approximation of the word--of--mouth model introduced by Fortunato and Castellano (Phys.~Rev.~Lett.~99(13):138701, 2007), where the activism of the agents is mitigated and the noise of the agent--agent interactions is averaged out. Then we argue that our analysis applies {\it mutatis mutandis} to the original model as well.
\end{abstract}

\section{Introduction}

It is well known that vote distributions in proportional elections display scaling and universality features. The discovery is reported in an inspiring paper by Fortunato and Castellano (FC)~\cite{fcscaling} and goes as follows. In a country implementing proportional elections with \emph{open} lists, each party presents lists of candidates in one or more electoral districts. Let $Q$, $v$, $N$ denote respectively the number of candidates in a given list, the number of preferences assigned to a given candidate in the list and the sum of preferences assigned to all of them. The adjective \emph{open} means that $Q$ is not fixed. The variable $x=vQ/N$ measures the excess of votes assigned to a candidate with respect to the average competitor in the same list. FC show that the probability density function (\emph{p.d.f.}) $\Femp(x)$ obtained from the empirical data is independent of $Q$ and $N$ (FC scaling). In addition, they show that $\Femp(x)$ is identical in different countries and years and that it is remarkably well fitted by a lognormal distribution $\ln\cN(\mu,\sigma^2)$ with $\mu = -0.54$ and $\sigma^2 = 1.08$.

In order to understand the origin of this amazing result, the same authors propose a cascade model based on word of mouth, where agents supporting a given candidate strive to persuade their undecided acquaintances to become supporters of the same candidate\footnote{In a more recent paper~\cite{Burghardt} Burghardt, Rand and Girvan consider a contagion--like model, which is equally able to reproduce the universal lognormal distribution.}. The model assumes a network made of $Q$ independent rooted trees, with candidates sitting on the roots and agents on the vertices. Each vertex has a random number $k$ of {\it children}, with $k$ following a power--law distribution $\pi(k)\propto k^{-\alpha}$ (it is understood that~$\alpha>2$ and $k\ge\kmin\ge 1$). At time $t=0$ candidates are the only persuaded agents. Each of them interacts with his/her own neighbours and tries to convince them one by one to vote for him/her. A single interaction may have a positive or negative outcome with probabilities $r$ and $1-r$. At a generic (discrete) time $t>0$, each persuaded agent in the network tries to convince the undecided neighbours according to the same stochastic rule. The process goes on until the overall number of persuaded agents on all trees equals $N$. The final number of preferences assigned to each candidate coincides with the final number of persuaded agents on the corresponding tree. Iterating this process (network generation + word--of--mouth spreading) many times and counting preferences in each iteration yields the conditional distribution $F_{\rm\scriptscriptstyle FC}(x|Q,N)$.

Let $X^{(i)}_n$ be the number of vertices on the $n$th level of the $i$th tree in a given iteration of the algorithm. Since the network is static, $X^{(i)}_n$ is known at time $t=0$ for all $i$ and $n$. Let also $\cP^{(i)}_n(t)$ and ${\cal U}^{(i)}_n(t)$ denote respectively the subsets of persuaded and undecided agents on the $n$th level of the $i$th tree at time~$t$, hence $|\cP^{(i)}_n(t)| + |{\cal U}^{(i)}_n(t)| = X^{(i)}_n$. The dynamics of the number of persuaded agents $Y^{(i)}_n(t) = |\cP^{(i)}_n(t)|$ is quantified by the equations
\begin{empheq}[box=\mybluebox]{align}
  \left\{\begin{array}{l}
  \\[-3.0ex]
 Y^{(i)}_0(t) = 1 \quad \text{ for all } t\ge0 \,,\\[2.2ex]
 Y^{(i)}_n(t) = 0 \quad \text{ if } t<n\,,\\[2.0ex]
 \displaystyle{Y^{(i)}_n(t) = Y^{(i)}_n(t-1)\ + \hskip-0.3cm \sum_{a\in{\cal U}^{(i)}_n(t-1)}\xi_a\ {\bf 1}_{{\cal P}^{(i)}_{n-1}(t-1)}(a')} \quad (a'\text{\small is the parent vertex of } a)\,, \ \text{ if } t\ge n\,,
 \end{array}\right.
\label{eq:FC}
\end{empheq}
where $\{\xi_a\}$ are i.i.d. Bernoulli variables with success probability $r$ and ${\bf 1}_A(x)$ represents the indicator function of $A$, \ie ${\bf 1}_A(x)=1$ if $x\in A$ and ${\bf 1}_A(x)=0$ otherwise. In this framework, the number of preferences assigned to the $i$th candidate at time $t$ is given by
\begin{equation}
  V^{(i)}_{\rm\scriptscriptstyle FC}(t) = \sum_{s=0}^t Y^{(i)}_s(t)\,.
\end{equation}
Eq.~(\ref{eq:FC}) shows at a glance that the FC algorithm is a Markov process. It also shows that the algorithm is not an ordinary branching process, as FC firstly notice in their paper: once an agent gets persuaded, he/she keeps trying to convince his/her undecided acquaintances at all subsequent times. Indeed, the inequality $Y^{(i)}_n(t+1)\ge Y^{(i)}_n(t)$ holds in general for all $n$ and $t$.

The conditional distribution $F_{\rm\scriptscriptstyle FC}(x|Q,N)$ cannot be directly compared with $\Femp(x)$. Since ${Q/N\le x \le Q}$, it follows that $F_{\rm\scriptscriptstyle FC}(x|Q,N) = 0$ for  $x<Q/N$ and $x>Q$. Hence, $F_{\rm\scriptscriptstyle FC}(x|Q,N)$ depends explicitly upon $Q$ and $N$. As such it violates the FC scaling. To get rid of this, FC run their algorithm for all $(Q,N)$ occurring in the empirical datasets and convolve the resulting curves. More precisely, let $p(Q,N)$ denote the empirical probability of $(Q,N)$, namely
\begin{equation}
  p(Q,N) = \frac{\text{no. of pairs } (Q,N) \text{ in the empirical datasets} }{\text{no. of all pairs } (Q',N') \text{ in the empirical datasets}}\,.
\end{equation}
The FC scaling distribution is then given by
\begin{equation}
  \FFC(x) = \sum_{Q,N}\,p(Q,N)\, F_{\rm\scriptscriptstyle FC}(x|Q,N)\,.
  \label{eq:fcconv}
\end{equation}
The agreement of $\FFC(x)$ with $\Femp(x)$ is just perfect provided the model parameters are set to $(\alpha,r,\kmin)=(2.45,0.25,10)$. Convolving $F_{\rm\scriptscriptstyle FC}(x|Q,N)$ with $p(Q,N)$ is truly essential for reproducing the empirical distribution: $F_{\rm\scriptscriptstyle FC}(x|Q,N)$ alone is known to deliver a power--law right tail as $Q\to\infty$; it is only its weighted average over $(Q,N)$ that yields the observed lognormal behaviour. The presence of power--law tails at fixed $(Q,N)$ has been regarded so far as a problem of the model~\cite{ftalk}. 
\begin{figure}[t!]
  \begin{minipage}{0.5\textwidth}
    \centering
    \includegraphics[width=0.9\textwidth]{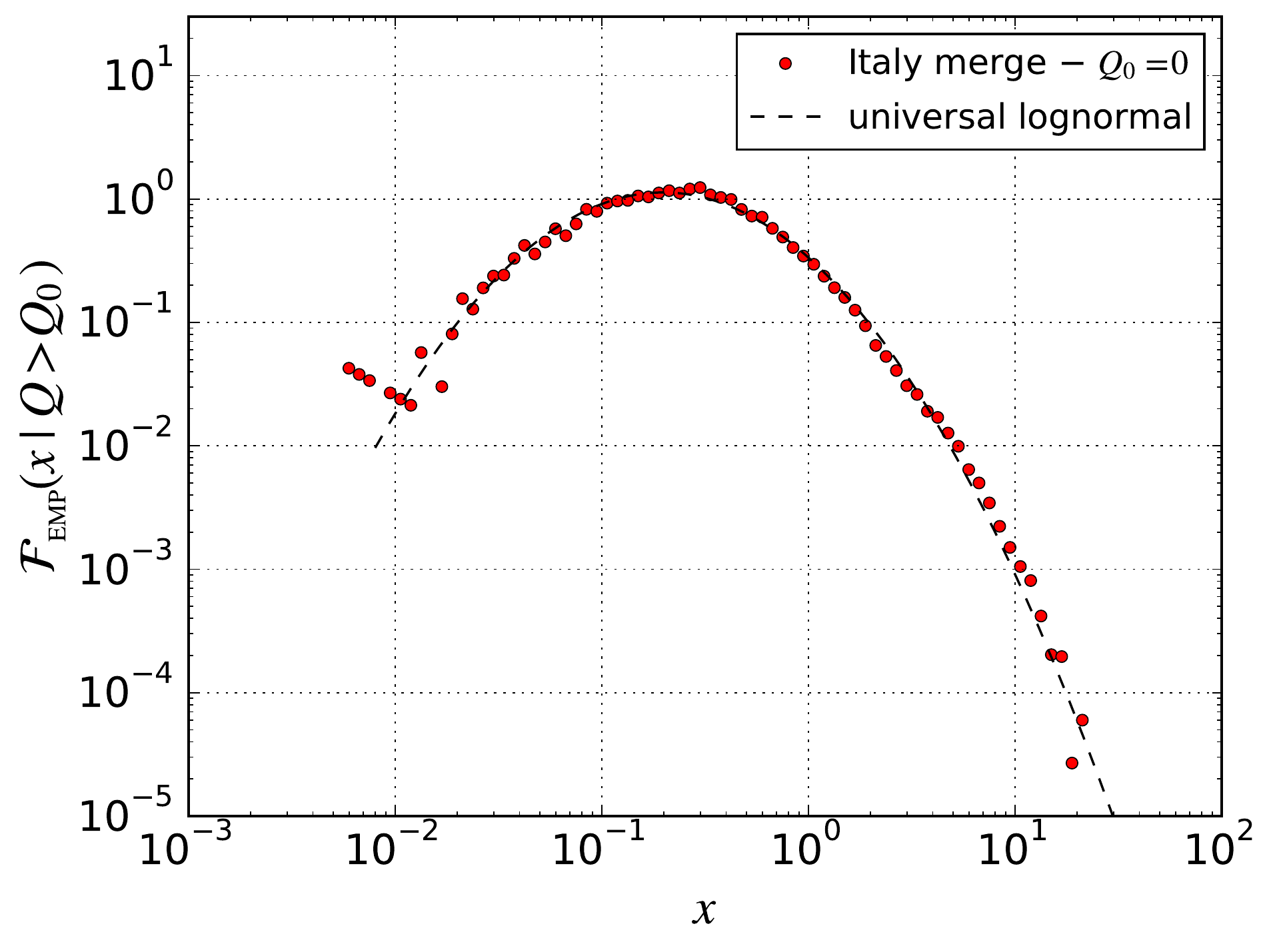}
  \end{minipage}
  \begin{minipage}{0.5\textwidth}
    \centering
    \includegraphics[width=0.9\textwidth]{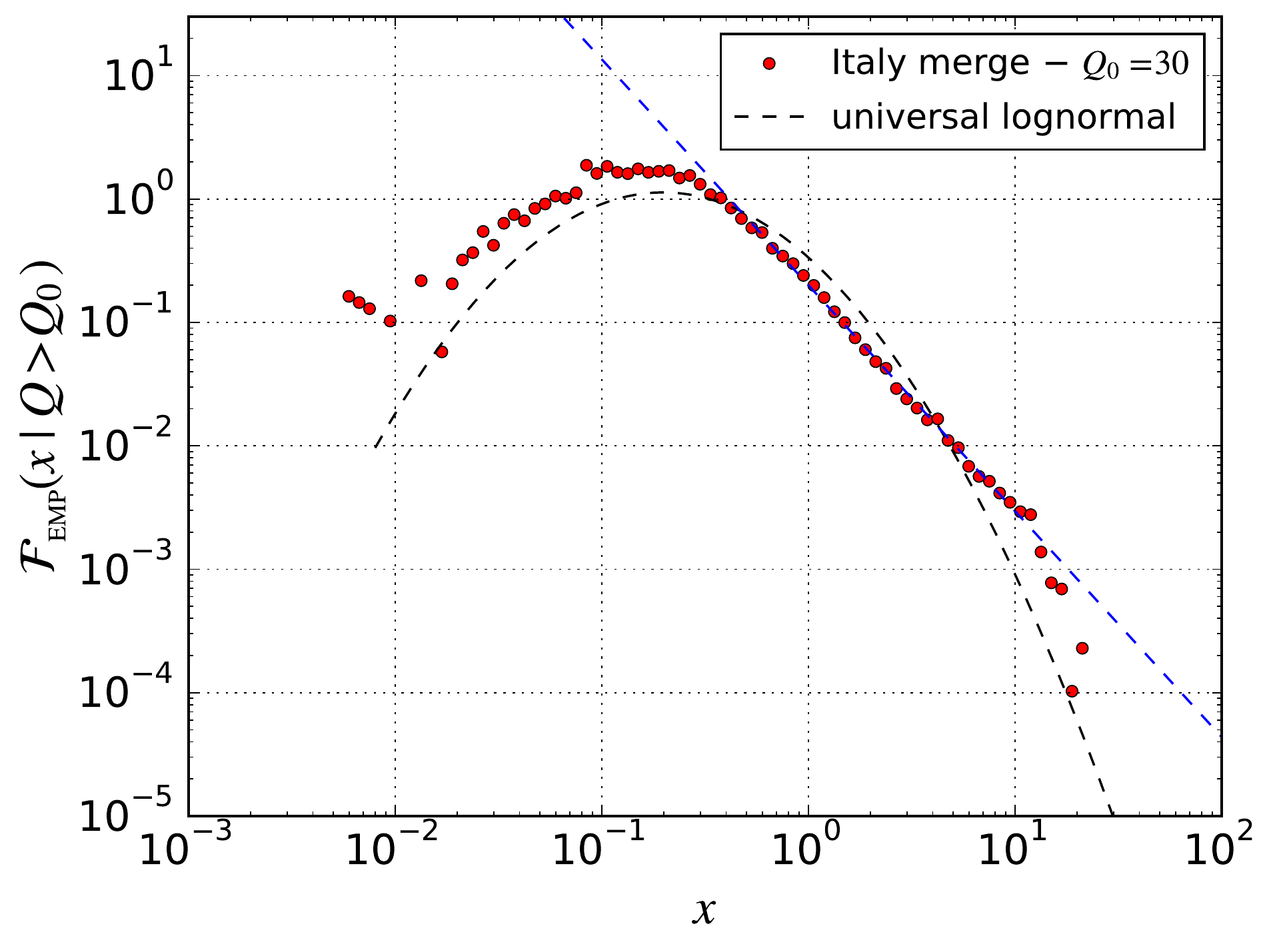}
  \end{minipage}
  \vskip -0.2cm
  \caption{\footnotesize {(Left) complete empirical distribution of the excess--of--votes variable $x$ obtained by merging data of the Italian elections held in 1958, 1972, 1976, 1979, 1987; (Right) conditional empirical distribution of $x$ given $Q>Q_0=30$ obtained by merging the same datasets.}}
  \label{fig:figone}
  \vskip -0.3cm
\end{figure}

On a closer inspection, power--law structures emerge from the empirical data when focusing on a specific subset of them. In place of $\Femp(x)$ we introduce the conditional distribution ${\Femp(x|Q>Q_0)}$, with $Q_0$ properly chosen. This is obtained by selecting data which belong only to lists with $Q>Q_0$. An example is given in Fig.~\ref{fig:figone}, where the empirical distribution of $x$, obtained by merging data from the Italian elections held from 1958 to 1987, is shown for $Q_0=0$ (complete distribution) and $Q_0=30$. Lognormality is manifestly broken in the latter distribution, which is instead characterized by a power--law right tail with final collapse at $x\lesssim Q_0$. With reference to the group U of countries studied in ref.~\cite{fempanal}, it must be added that: {\it i}) a similar behaviour is displayed by data of Polish and Estonian elections; {\it ii}) Danish data have no lists with $Q>Q_0=30$; {\it iii}) Finnish data still display a lognormal conditional distribution given $Q_0=30$, hence they represent a puzzling exception; {\it iv}) Italian data are those with the largest statistics for $Q>Q_0=30$; this is why we plot just them in Fig.~\ref{fig:figone}. 

We propose an interpretation of this effect. We first notice that by construction $\Femp(x|Q>Q_0)$ fulfills the inequality
\begin{equation}
  {\Femp(x)\,<\,\Femp(x|Q>Q_0)}\,, \quad \text{ for }\  x>Q_0\,
  \label{eq:appone}
\end{equation}
(we refer the reader to app.~A for a simple proof of eq.~(\ref{eq:appone})). As a consequence, the opposite inequality ${\Femp(x)\,>\,\Femp(x|Q>Q_0)}$ must hold somewhere in the complementary region $x\le Q_0$, since both $\Femp(x)$ and ${\Femp(x|Q>Q_0)}$ are correctly normalized as \emph{p.d.f.}'s. Secondly, $\Femp(x|Q>Q_0)$ also can be represented as a convolution of conditional distributions $F_{\rm\scriptscriptstyle EMP}(x|Q,N)$ with $Q>Q_0$, \viz
\begin{equation}
  \Femp(x|Q>Q_0) = \sum_{Q>Q_0}\sum_N\, p(Q,N|Q>Q_0)\,F_{\rm\scriptscriptstyle EMP}(x|Q,N)\,,
  \label{eq:FQQ0}
\end{equation}
where the conditional probability $p(Q,N|Q>Q_0)$ is defined for $Q>Q_0$ by
\begin{equation}
  p(Q,N|Q>Q_0) = \frac{\text{no. of pairs } (Q,N) \text{ in the empirical datasets} }{\text{no. of all pairs } (Q',N') \text{ with } Q'>Q_0 \text{ in the empirical datasets}}\,.
  \label{eq:pqncond}
\end{equation}
Although we have insufficient statistics for measuring $F_{\rm\scriptscriptstyle EMP}(x|Q,N)$, we know that it vanishes for $x>Q$ by construction. Hence, we expect it to be contaminated by truncation artefacts for $x\lesssim Q$, amounting to a faster decay just to the left of the truncation point, than we would observe for larger values of $Q$. Independently of how these local distortions are compensated along the domain of the distribution\footnote{They might be absorbed locally in a region close to the boundary $x\lesssim Q$, locally in a different region within the domain of the distribution or globally through a dilution along the whole domain.}, their overall impact is modest provided $Q$ is sufficiently large. Therefore, it is reasonable to assume $F_{\rm\scriptscriptstyle EMP}(x|Q,N)$ to become independent of $Q$ for $x\ll Q$ as $Q\to\infty$. If this is correct, it follows that $\Femp(x|Q>Q_0)$ is almost insensitive to its defining convolution, \ie for $x\ll Q_0$ we must have
\begin{align}
  \Femp(x|Q>Q_0) & = \sum_{Q>Q_0}\sum_N\, p(Q,N|Q>Q_0)\,F_{\rm\scriptscriptstyle EMP}(x|Q,N) \nonumber\\[1.0ex]
  & \hskip -0.0cm \simeq \sum_{N}F_{\rm\scriptscriptstyle EMP}(x|Q_0,N)\sum_{Q>Q_0}p(Q,N|Q>Q_0) \nonumber\\[1.0ex]
  & = \sum_N F_{\rm\scriptscriptstyle EMP}(x|Q_0, N)\,p(N|Q>Q_0) \, \simeq\, F_{\rm\scriptscriptstyle EMP}(x|Q_0,\bar N)\,,
\end{align}
with $\bar N = \E[N|Q>Q_0]\gg 1$ (there is always a positive correlation between $Q$ and $N$ in the empirical data, because parties present lager lists of candidates in larger constituencies). If we believe in the FC model, then $F_{\rm\scriptscriptstyle EMP}(x|Q_0,\bar N)$ is well reproduced by $F_{\rm\scriptscriptstyle FC}(x|Q_0,\bar N)$. Accordingly, $F_{\rm\scriptscriptstyle EMP}(x|Q_0,\bar N)$ has a power--law tail for $1\lesssim x\lesssim Q_0$ and we can conclude that $\Femp(x|Q>Q_0)$ also has a power--law tail in the same region.

The above considerations suggest to regard the power--law tail of $F_{\rm\scriptscriptstyle FC}(x|Q,N)$ not as an unphysical attribute of the model, but -- quite the opposite -- as an essential feature of the phenomenon it describes. Since the complete tail of the distribution emerges only as $Q\to\infty$, we are naturally led to consider the large--list limit
\begin{equation}
\FFC^*(x) = \lim_{Q\to\infty}\lim_{N\to\infty} F_{\rm\scriptscriptstyle FC}(x|Q,N)\,.
\end{equation}
Quite interestingly, this limit represents another realization of the FC scaling. Unlike eq.~(\ref{eq:fcconv}), it does not require any input from the empirical data. The ultimate reason why there exist two different ways to build a scaling function out of $F_{\rm\scriptscriptstyle FC}(x|Q,N)$ is not clear. 

Aim of the present paper is to investigate $\FFC^*(x)$ and its corrections at finite $Q$. Despite the significance of the FC model, the analytic structure of $F_{\rm\scriptscriptstyle FC}(x|Q,N)$ has never been studied in the literature so far, to the best of our knowledge. The analysis we present here is the first attempt to go beyond numerical simulations. Given the complexity of the FC model, in sect.~2 we consider a quenched approximation of it, where the word of mouth generates Galton--Watson trees with Mandelbrot offspring distribution. Albeit simpler, the quenched model is still in good agreement with the empirical data. Based on the exponential scaling of Galton--Watson trees, in sect.~3 we work out the vote distribution in the large--list limit and study how it changes under variations of the stopping rule of the algorithm. In sect.~4 we go back to the original model and show that an analogous solution applies to it \emph{mutatis mutandis}. Specifically, we find that, similar to Galton--Watson trees, also trees resulting from eq.~(\ref{eq:FC}) scale exponentially as $t\to\infty$ and we calculate their growth rate. This is sufficient for studying the FC model in the large--list limit. Finally, in sect.~5 we derive an asymptotic estimate of the power--law tail of the vote distribution in the quenched model. This applies with rather good approximation to the FC model too. We draw our conclusions in sect.~6.

\section{Quenched word--of--mouth model}

\begin{figure}[t!]
  \centering
  \includegraphics[width=0.8\textwidth]{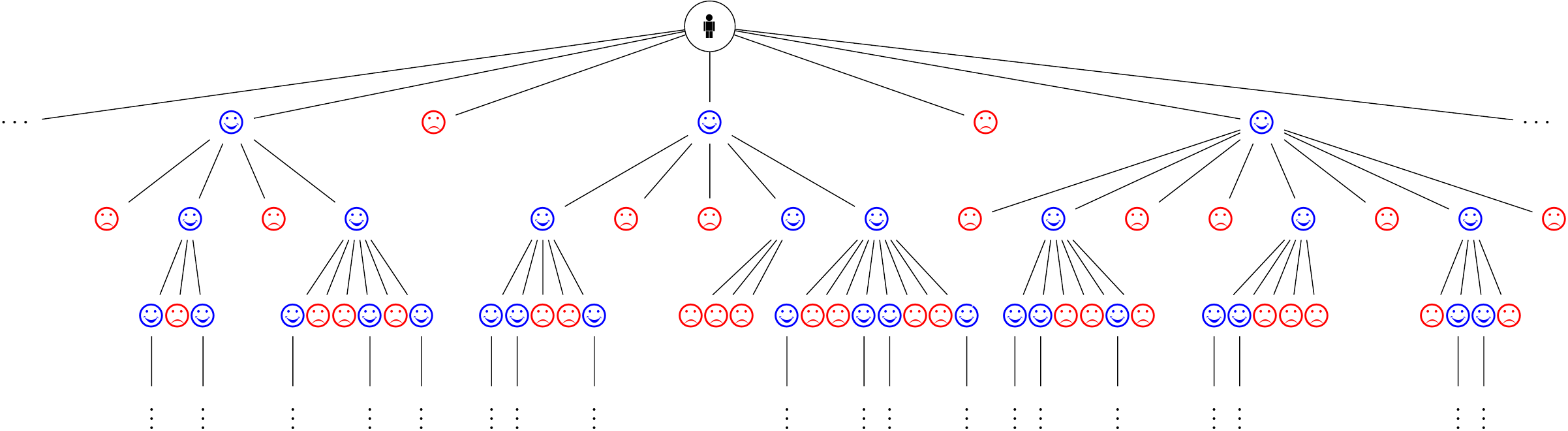}
  \vskip -0.2cm
  \caption{\footnotesize A sample tree with agents freezing after one Bernoulli interaction with their children. Voters (smiling agents) branch according to $\pB(\ell)$. Non--voters (frowning agents) are not regarded as vertices: they are included in the picture only for illustrative purposes.}
  \label{fig:figtwo}
\end{figure}

As mentioned above, the major complication of the FC model is represented by the activism of the agents, who repeatedly endeavor to persuade their undecided acquaintances. A great simplification is achieved if we prevent them from insisting after their first attempt. This turns the model into a standard Galton--Watson process, where the time variable $t$ comes to be identified with the level variable $n$. In this approximation candidates interact with their neighbours at time $t=n=0$, a single interaction being represented by a Bernoulli trial with success probability~$r$.  Hence they freeze, while persuaded agents on level $t=n=1$ interact in turn with their acquaintances lying on the next level, according to the same stochastic prescription. Then they freeze too and so on and so forth. If $Z^{(i)}_t$ denotes the number of persuaded agents on the $t$th level of the $i$th tree\footnote{When not needed, we shall drop the upper index of $Z^{(i)}_t$ and simply write $Z_t$ in place of it.}, the word--of--mouth spreading is described by the simpler equations
\begin{empheq}[box=\mybluebox]{align}
  \left\{\begin{array}{l}
 Z^{(i)}_0 = 1\,,\\[0.2ex]
 \displaystyle{Z^{(i)}_{t+1} = \sum_{j=0}^{Z^{(i)}_t}\xi^{(i)}_{t,j}\,, \quad  t\ge 0}\,,\\[0.0ex]
 \displaystyle{\{\xi^{(i)}_{t,j}\}\ \text { are i.i.d. variables with }\, \xi^{(i)}_{t,j}\,\sim\,\pB(\ell) = C^{-1}\sum_{k=\max\{\ell,\kmin\}}^\infty {k \choose \ell}\frac{r^\ell(1-r)^{k-\ell}}{k^\alpha}}\,,
 \end{array}\right.
\label{eq:BP1}
\end{empheq}
where $\ell\ge 0$ and $C = \sum_{\ell=0}^\infty \sum_{k=\max\{\ell,\kmin\}}^\infty {k \choose \ell}\frac{r^\ell(1-r)^{k-\ell}}{m^\alpha}$ is the normalization constant of the offspring distribution~$\pB(\ell)$. The subindex B (roman font) reminds us that the agent--agent interaction is a Bernoulli trial. A sample tree generated according to the above procedure is depicted in Fig.~\ref{fig:figtwo}. We observe that $\pB(\ell)$ is precisely the probability of contacting $k$ agents with probability~$\pi(k)\propto k^{-\alpha}$ under the condition that $k\ge\kmin$ and of convincing a fraction $\ell\le k$ of them, each with probability $r$, in all possible ways. For later convenience, we let $\mB = \sum_{\ell=0}^\infty\ell\,\pB(\ell)$ denote the average offspring of a vertex. Moreover, in this framework the number of preferences assigned to the $i$th candidate up to level~$t$ is given by
\begin{equation}
  V^{(i)}_t = \sum_{s=0}^t Z^{(i)}_s\,.
  \label{eq:novotes}
\end{equation}

The statistical properties of $Z^{(i)}_t$ (in particular the tail of its distribution) are completely determined by~$\pB(\ell)$. Unfortunately, despite the simplification operated on the original process, the structure of $\pB(\ell)$ is still too complex to allow for analytic calculations. Hence, we proceed to a further reduction of complexity. We notice that if a persuaded agent has many contacts $k$ and interacts with them separately via Bernoulli trials with success probability $r$, on average he/she persuades a fraction $rk$ of them. Indeed, the relation
\begin{equation}
  r \sum_{k=\ell}^\infty{k\choose \ell}\frac{r^\ell(1-r)^{k-\ell}}{k^\alpha}\ \mathop{\approx}^{\ell\to\infty}\ \frac{r^\alpha}{\ell^\alpha}
  \label{eq:probident}
\end{equation}
holds asymptotically. Inspired by this equation, we introduce the integer variable $\nmin = \max\{1,\lfloor r\kmin\rfloor\}$ and in place of eq.~(\ref{eq:BP1}) we consider the Galton--Watson process
\begin{empheq}[box=\mybluebox]{align}
  \left\{\begin{array}{l}
 Z^{(i)}_0\ = 1\,,\\[0.2ex]
 \displaystyle{Z^{(i)}_{t+1}\ = \sum_{j=0}^{Z^{(i)}_t}\xi^{(i)}_{t,j}\,, \quad  t\ge 0}\,,\\[0.0ex]
 \{\xi^{(i)}_{t,j}\}\ \text { are i.i.d. variables with }\, \xi^{(i)}_{t,j}\,\sim\, \displaystyle{\pQ(n) = \frac{1}{\zeta(\alpha,\nmin)}\frac{1}{n^\alpha}}\,,
 \end{array}\right.
\label{eq:BP2}
\end{empheq}
where $n\ge\nmin$ and $\zeta(\alpha,\nmin) = \sum_{n=0}^\infty \frac{1}{(n+\nmin)^\alpha}$ denotes the Hurwitz $\zeta$--function. The offspring distribution $\pQ(n)$ is known in the literature as the \emph{shifted power law} or Mandelbrot distribution~\cite{Mandelbrot}. In the sequel we focus on eq.~(\ref{eq:BP2}) and refer to it as the {\it quenched} model, with the understanding that it has been obtained by quenching both the activism of the agents and the fluctuations of the single interactions characterizing the original FC model. The subindex Q (roman font) in $\pQ(\ell)$ stands for \emph{quenched}. 

\begin{figure}[t!]
  \centering
  \includegraphics[width=0.8\textwidth]{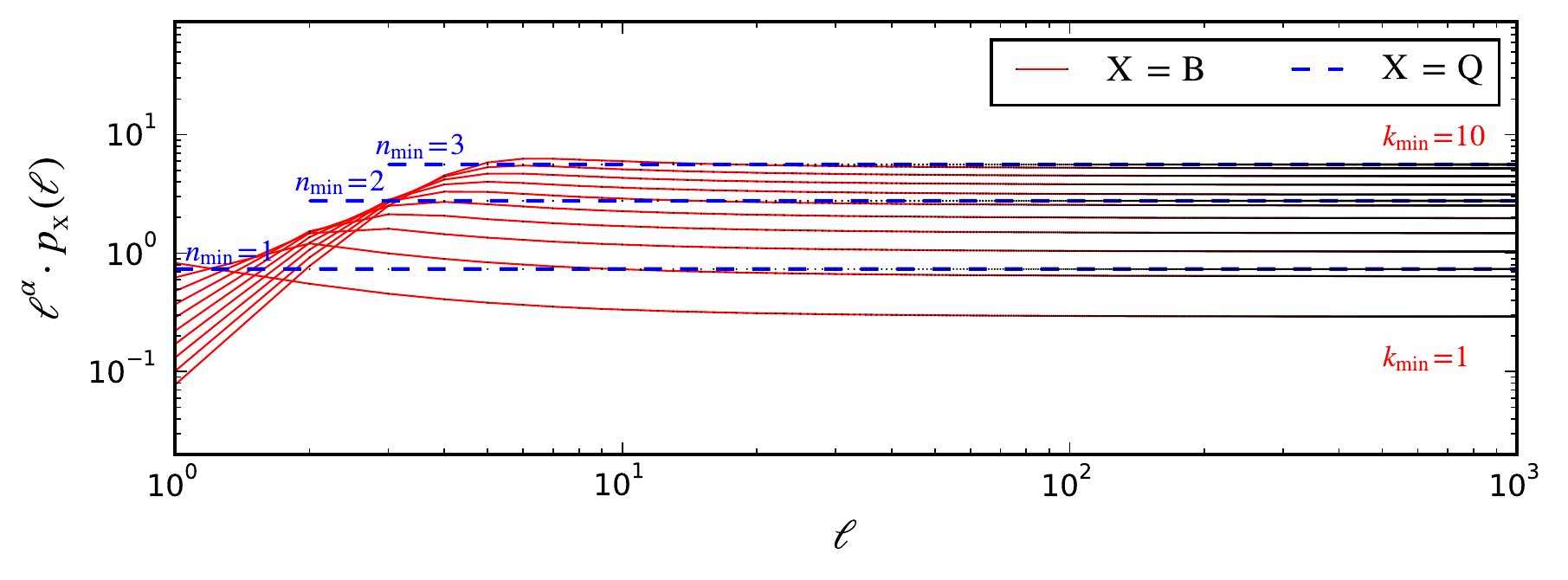}
  \vskip -0.5cm
  \caption{\footnotesize Probabilities $\pB(\ell)$ (continuous lines) and $\pQ(\ell)$ (dashed lines) for $(\alpha,r)=(2.45,0.25)$.}
  \label{fig:figthree}
\end{figure}

In Fig.~\ref{fig:figthree} we compare the offspring distributions $\pB(\ell)$ and $\pQ(n)$ for $\alpha = 2.45$, $r=0.25$ and several choices of $\kmin$ and $\nmin$ just to stress that: {\it i}) $\pQ(n)$ undergoes discrete jumps as $\nmin \to \nmin+1$; {\it ii})~apart from the overall normalization, the behaviour of the two distributions essentially differs only for $\ell\sim {\rm O}(1)$.

We let $\cT$ denote the space of trees (elementary events) and $\cM$ the additive class of subsets of $\cT$ (events). The measure theory in $\cT$ has been first discussed in ref.~\cite{Otter}, to which we refer the reader for details. Eqs.~(\ref{eq:BP1}) and (\ref{eq:BP2})  induce two different probability spaces on $\cT$, namely
\begin{itemize}
\item{eq.~(\ref{eq:BP1}) $\to\ \cP_{\rm\scriptscriptstyle B}(\alpha,r,\kmin) = (\cT,\cM,\P_{\rm\scriptscriptstyle B}\{\,\cdot\,|\alpha,r,\kmin\}\,)$;}\item{eq.~(\ref{eq:BP2}) $\to\ \cP_{\rm\scriptscriptstyle Q}(\alpha,\nmin) = (\cT,\cM,\P_{\rm\scriptscriptstyle Q}\{\,\cdot\,|\alpha,\nmin\}\,)$.}
\end{itemize}
Eq.~(\ref{eq:FC}) induces a family of probability spaces $\cP_{{\rm\scriptscriptstyle FC},t}(\alpha,r,\kmin) = (\cT_t,\cM_t,\P_{{\rm\scriptscriptstyle FC},t}\{\,\cdot\,|\alpha,r,\kmin\}\,)$ parametrized by $t=0,1,2,\ldots$, where $\cT_t$ is the space of finite trees with at most $t$ levels and $\cM_t$ is the additive class of subsets of $\cT_t$.  For simplicity, in the sequel we use the notation $\P\{R\}$ with no subindices and/or parameters to denote the probability of~$R\in\cM$ whenever the context makes it clear to which specific measure we refer. 

\noindent {\bf Remark 2.1.} The network of acquaintances in the FC model is a collection of disjoint and i.i.d. trees, each belonging to $\cP_{\rm\scriptscriptstyle Q}(\alpha,\kmin)$, \ie $X^{(i)}_{n+1} = \sum_{j=0}^{X^{(i)}_n}\xi^{(i)}_{n,j}$ with $\xi^{(i)}_{n,j} \sim \pQ(n)|_{\nmin\,\to\,\kmin}.$ \hfill $\blacksquare$

So far we have discussed the structure of single trees. In the quenched word--of--mouth model we generate trees in parallel level by level: we first draw a realization of $(Z^{(i)}_1)_{i=1}^Q$, hence we draw a realization of $(Z^{(i)}_2)_{i=1}^Q$, \emph{etc}. As $t\to\infty$, the algorithm generates $Q$--tuples of trees, \ie elements of the Cartesian product
\begin{equation}
  \fT = \underbrace{\cT\, \times\, \ldots\, \times\, \cT}_{Q\text{ times }}\,.
\end{equation}
The set $\fT$ is the space of elementary events of the model. The probability measure on $\frak T$ is the product measure obtained in the usual way for independent variables, see ref.~\cite[ch.~IV.6]{Feller}. In the following, we let~${T=(T^{(1)},\ldots,T^{(Q)})}$ denote a generic element of $\fT$. We call $T$ a $Q$--forest. Incidentally, we observe that the stochastic variable $V^{(i)}_t$ is a function of $T^{(i)}$ alone. Therefore, we write $V^{(i)}_t = V_t(T^{(i)})$ whenever appropriate. 

To complete the algorithmic prescription we need to specify a stopping rule to arrest the generation of new levels. In sect.~3 we study, among other things, how the vote distribution depends on this part of the procedure. We consider three possible stopping rules:
\begin{itemize}
\item[SR1:]{the process stops as soon as the overall number of vertices generated on all trees equals $N$. This is the stopping rule adopted by FC in ref.~\cite{fcscaling}. It has the drawback that precisely when the process stops, some trees have developed down to a certain level $\texit$, while the others only down to level $\texit-1$. In most cases there is a tree $T^{(i)}$ for which $Z^{(i)}_{\texit}$ is only partially realized. This complicates the analytic description, as we shall see;}
\item[SR2:]{the process stops as soon as the generation of the earliest level $\texit$, at which the overall number of vertices on all trees reaches $N$, is complete for all trees. Hence, all variables $(Z^{(i)}_{\texit})_{i=1}^Q$ are fully realized. Unfortunately, this rule has the drawback that precisely when the process stops, the overall number $M$ of vertices generated on all trees fluctuates along different iterations of the algorithm, with~$M\ge N$;}
\item[SR3:]{the process stops as soon as the generation of a predefined level $\texit$ is complete for all trees. Similar to the previous case, all variables $(Z^{(i)}_{\texit})_{i=1}^Q$ are fully realized. The overall number $M$ of vertices generated on all trees fluctuates along different iterations, however $M$ is totally unconstrained in this case.}
\end{itemize}
Iterating the above algorithm many times and counting preferences yields the conditional distribution $F_{\rm\scriptscriptstyle Q}(x|Q,N)$ (under SR1 and SR2) or $F_{\rm\scriptscriptstyle Q}(x|Q,\texit)$ (under SR3). We incidentally notice that under SR1 it is no matter whether trees are randomly or sequentially ordered when the level $\texit$ is generated, provided $F_{\rm\scriptscriptstyle Q}(x|Q,N)$ is averaged over all candidates.

\begin{figure}[t!]
  \begin{minipage}{0.5\textwidth}
    \centering
    \includegraphics[width=0.9\textwidth]{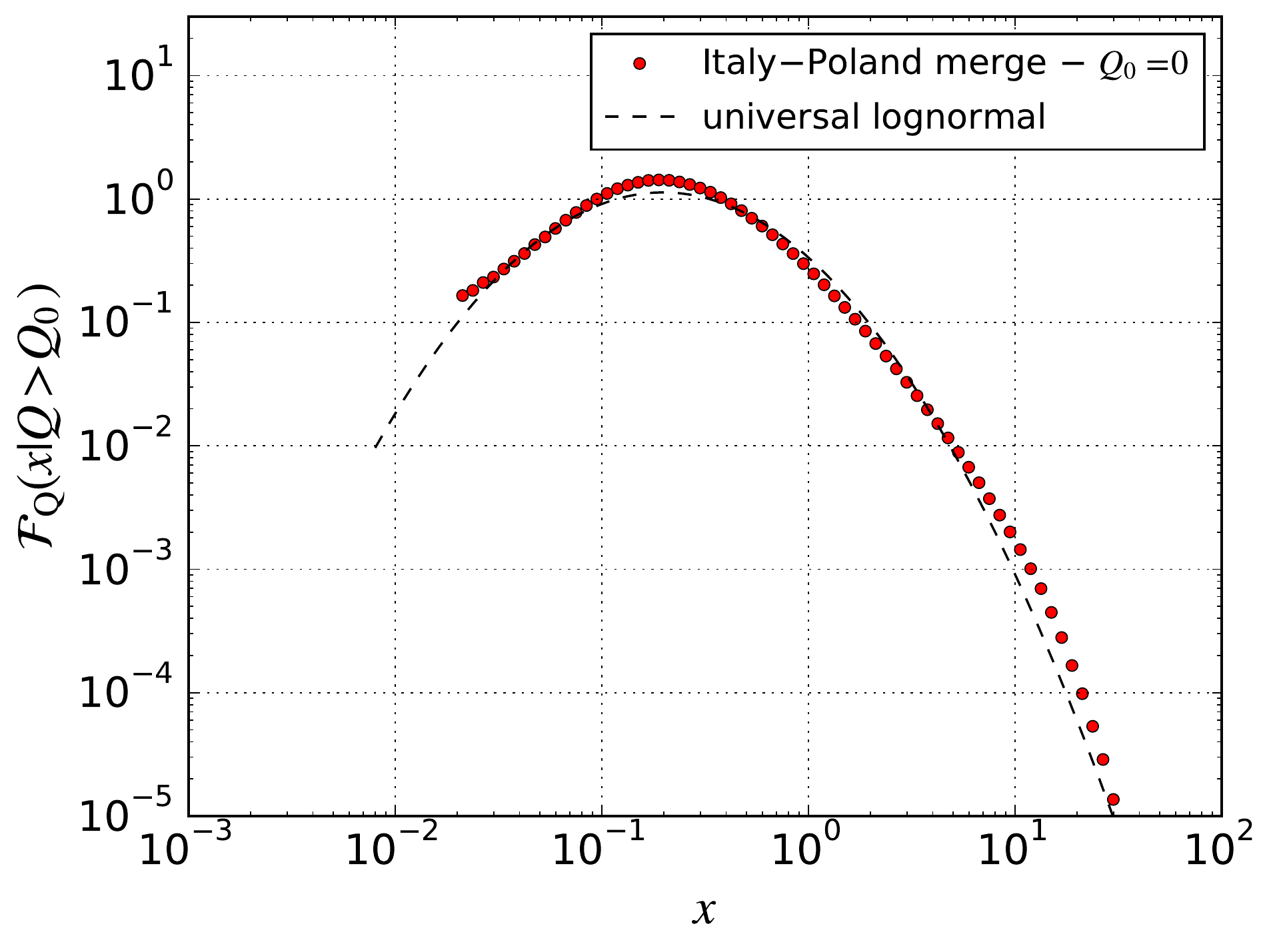}
  \end{minipage}
  \begin{minipage}{0.5\textwidth}
    \centering
    \includegraphics[width=0.9\textwidth]{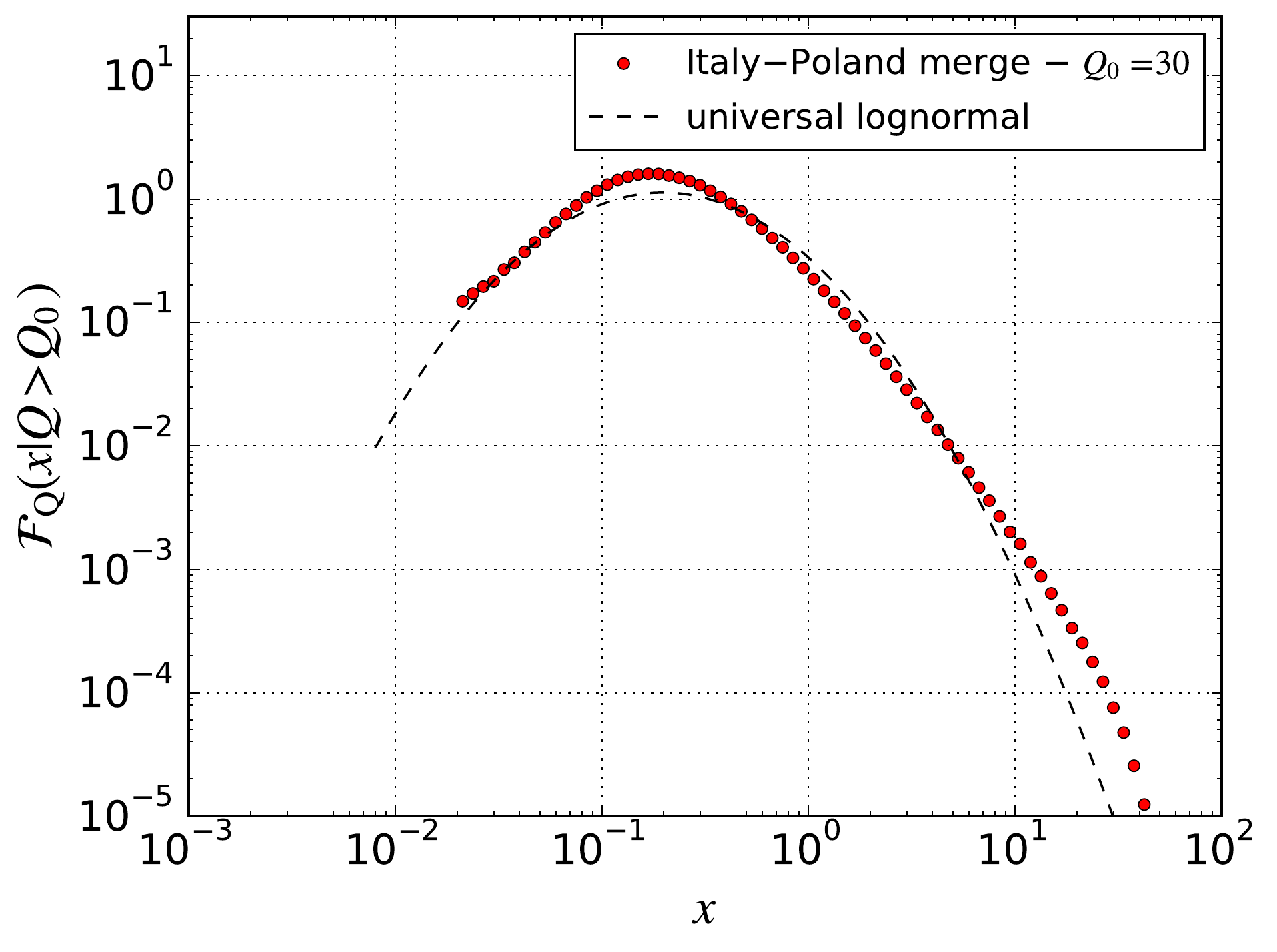}
  \end{minipage}
  \vskip -0.2cm
  \caption{\footnotesize {(Left) complete distribution of excess--of--votes variable $x$ obtained by convolving $F_{\rm\scriptscriptstyle Q}(x|Q,N)$ with the empirical probability~$p(Q,N)$; (Right) conditional distribution of $x$ obtained by convolving $F_{\rm\scriptscriptstyle Q}(x|Q,N)$ with the empirical probability $p(Q,N|Q>Q_0)$ and $Q_0=30$. In both plots, the empirical probability has been obtained by merging datasets of the Italian elections held in 1958, 1972, 1976, 1979, 1987 and the Polish elections held in 2001, 2005, 2007, 2011.}}
  \label{fig:figfour}
\end{figure}

In Fig.~\ref{fig:figfour} we show the distribution $\FQ(x)$ obtained by convolving the conditional distribution $F_{\rm\scriptscriptstyle Q}(x|Q,N)$ (under SR2) with the empirical probability $p(Q,N)$ obtained by merging the datasets of the Italian elections held from 1958 to 1987 jointly with those of the Polish elections held from 2001 to 2011. The curves correspond to a choice of the model parameters $(\alpha,\nmin) = (2.45,3)$. The agreement of the complete distribution (left) with the universal lognormal is not perfect. Yet, it is sufficiently good to let us regard the quenched model as a valid simplification for studying the large--list limit and its corrections at finite $Q$. The emergence of a power--law right tail for $Q_0=30$ (right) is less pronounced than in the empirical data, yet it is still visible.

\noindent {\bf Remark 2.2.} At this point the reader may wonder why the quenched model matches the empirical data so well, in spite of the razor cuts operated on the FC model. To answer this, we notice that from an algorithmic point of view a typical tree~${T^{(i)}\in\cP_{{\rm\scriptscriptstyle FC},\texit}(\alpha,r,\kmin)}$ looks like a hybrid. If $\texit$ is not too small, persuaded voters $Y^{(i)}_n(\texit)$ on the first levels ($n\ll \texit$) are close to their maximum possible value $X^{(i)}_n$, since undecided agents have been contacted several times. This part of the tree looks much like typical elements of $\cP_{\rm\scriptscriptstyle Q}(\alpha,\kmin)$. Of note, the latter are much fatter than typical elements of $\cP_{\rm\scriptscriptstyle Q}(\alpha,\nmin)$. By contrast, undecided agents on the last levels ($n\simeq \texit$) have been contacted just a few times: specifically, those on the $\texit$th level have undergone just one interaction with their parents. Therefore, this part of the tree resembles more typical elements of $\cP_{\rm\scriptscriptstyle B}(\alpha,r,\kmin)$ and these are similar to typical elements of $\cP_{\rm\scriptscriptstyle Q}(\alpha,\nmin)$. Moreover, the last levels are those which contribute most to the overall number of preferences. We conclude that the quenched model works well because trees in $\cP_{\rm\scriptscriptstyle Q}(\alpha,\nmin)$ are similar to trees in $\cP_{{\rm\scriptscriptstyle FC},\texit}(\alpha,r,\kmin)$ precisely where they should be, \ie on the last levels. Notice that the optimal choice of the FC model parameters yields $r\kmin = 2.5$, whereas in the quenched model we have $\nmin=3$. This yields an average compensation of all topological differences. Of course, the above intuitive argument should be substantiated by a quantitative analysis of the growth rate of $Y^{(i)}_n(t)$. We shall discuss this in sect.~4. \hfill $\blacksquare$

\begin{table}[t!]
  \centering
  \begin{tabular}{cc|rrrrr|}
    & & \multicolumn{5}{c|}{$\alpha$} \\[0.2ex]
    \multicolumn{2}{c|}{$\mQ = \frac{\zeta(\alpha-1,\nmin)}{\zeta(\alpha,\nmin)}$} & 2.35 & 2.45 & 2.55 & 2.65 & 2.75\\
    \hline\\[-3.2ex]
    \multirow{3}{*}{\rotatebox{0}{$\nmin$}} & 2 & 6.04419  &  5.06405 &  4.44184 & 4.01235 & 3.69847 \\
    &  3 & 9.80849  &  8.20382 &  7.18374 & 6.47842 & 5.96197 \\[0.0ex]
    &  4 & 13.62090 & 11.38611 &  9.96479 & 8.98148 & 8.26097 \\[0.0ex]
    \hline
  \end{tabular}
  \vskip 0.2cm
  \caption{\footnotesize Values of $\mQ$ for various choices of $(\alpha,\nmin)$.}
  \label{tab:tabone}
  \vskip 0.3cm
\end{table}

\begin{figure}[t!]
  \centering
  \includegraphics[width=0.78\textwidth]{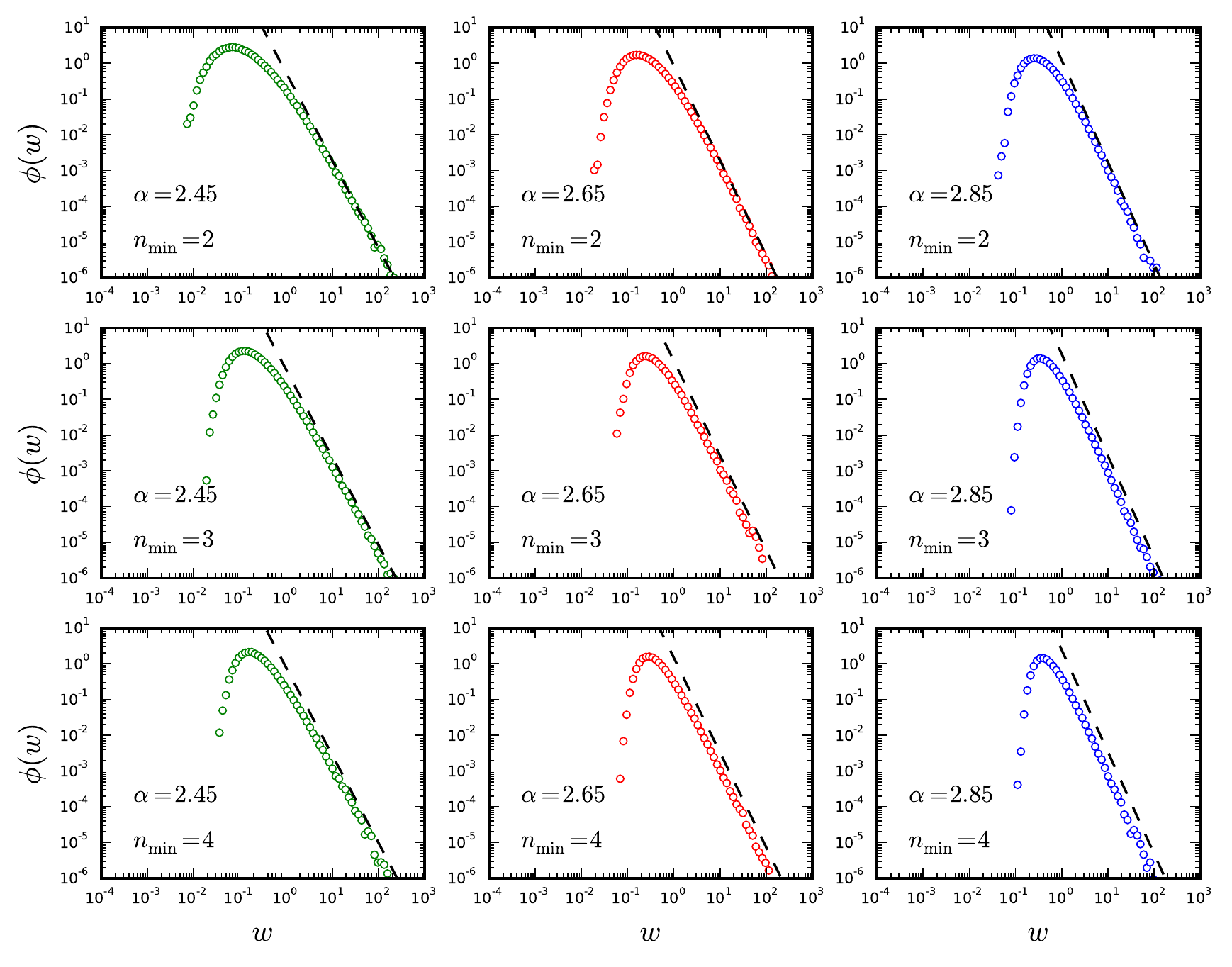}
  \vskip 0.6cm
  \includegraphics[width=0.78\textwidth]{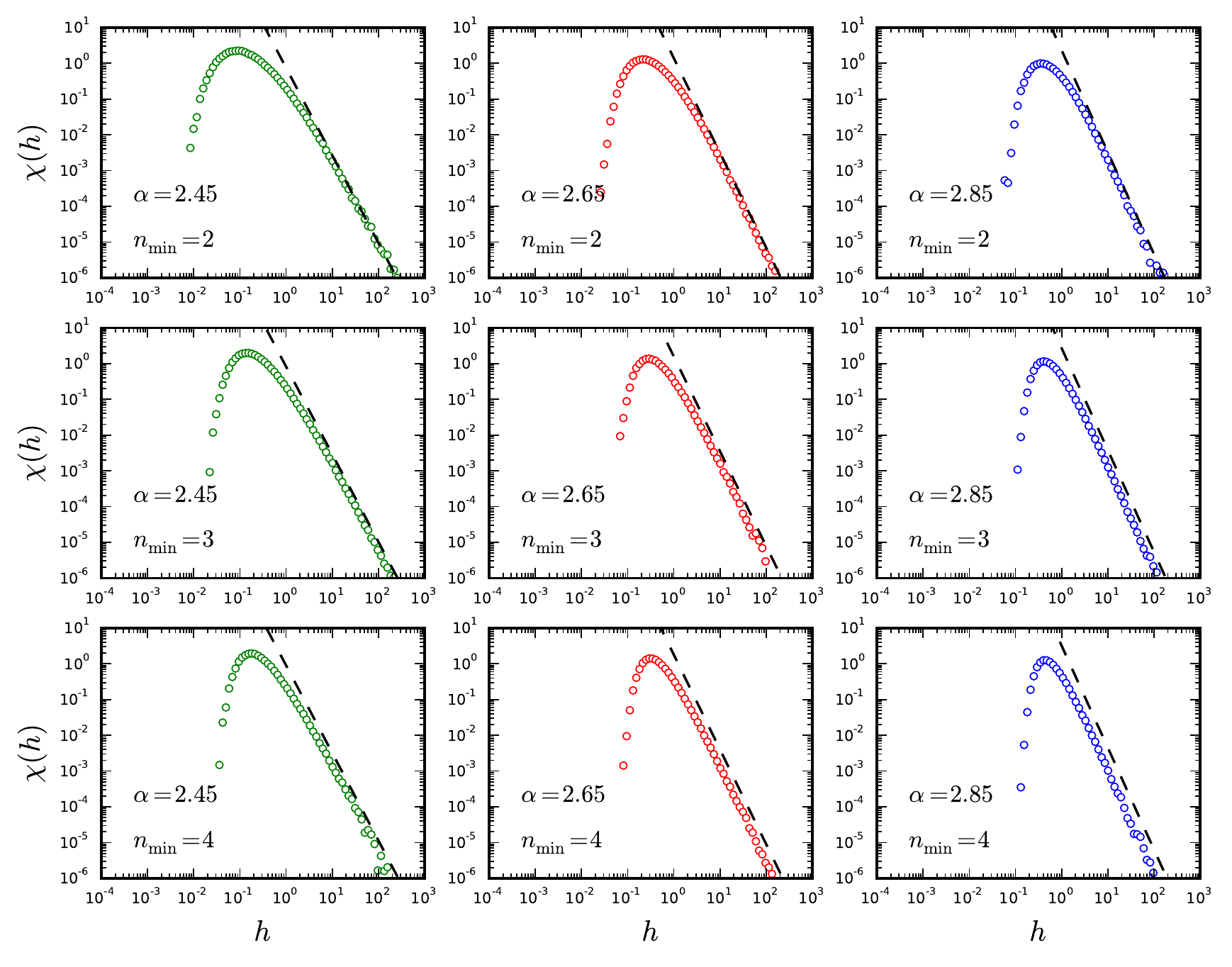}
  \vskip 0.0cm
  \caption{\footnotesize (Top) \emph{p.d.f.} of the limit variable $W$; (Bottom) \emph{p.d.f.} of the limit variable $H$. Dashed lines correspond to the analytic estimates reported resp. in eq.~(\ref{eq:phi}) and eq.~(\ref{eq:chi}).}
  \label{fig:figfive}
\end{figure}

Let us now examine more closely trees in $\cP_{\rm\scriptscriptstyle Q}(\alpha,\nmin)$. The standard toolbox to study them is the theory of branching processes, discussed for instance in the classical monographs by Harris \cite{Harris} and Athreya and Ney~\cite{Athreya}. The average offspring of a vertex is given by
\begin{equation}
  \mQ \equiv \sum_{n=\nmin}^\infty n\,\pQ(n) =  \frac{\zeta(\alpha-1,\nmin)}{\zeta(\alpha,\nmin)}>1\,.
\end{equation}
For the reader's convenience, in Table~\ref{tab:tabone} we collect values of $\mQ$ for a handful of plausible choices of $(\alpha,\nmin)$. Owing to $\mQ>1$, quenched trees are supercritical, \ie on average they fulfill $Z_t\to\infty$ as $t\to\infty$. We recall indeed that 
\begin{equation}
  \E[Z_t] = \E\biggl[\, \E\bigl[ \sum_{j=0}^{Z_{t-1}}\xi_{t,j}\,|\, Z_{t-1}\bigr]\, \biggr] = \E[ \mQ Z_{t-1}] = \mQ\, \E[\,Z_{t-1}]\,.
\label{eq:Ztmt}
\end{equation}
Since $\E[Z_0]=1$, it follows that $\E[Z_t]=\mQ^t$. Hence, we see that $\mQ$ is the exponential growth rate of $Z_t$. We also notice that $\mQ\sim{\rm O}(10)$ if $\alpha$ is not too large and $\nmin$ not too small. As a consequence $Z_t$, $Z_{t+1}$, $Z_{t+2}$, \ldots\, belong to different orders of magnitude  on average. We finally observe that since $\pQ(n) = 0$ for $n<\nmin$, it follows that $\P\{Z_t = 0\} = 0$ for all $t$\footnote{Supercritical Galton--Watson trees with power--law structure are studied in ref.~\cite{LeeKim} under the assumption that the offspring distribution fulfills $p(0)>0$. As a consequence $\lim_{t\to\infty} \P\{Z_t=0\}>0$ in this case, \ie the subspace of finite trees has a positive measure. The authors of ref.~\cite{LeeKim} are interested only in this subspace. Therefore their methodology does not apply here.}. In cases like this, it is appropriate to introduce the variable $W_t = Z_t / \mQ^t$. Obviously, $W_t$ fulfills $\E[W_t]=1$. Moreover, an argument analogous to eq.~(\ref{eq:Ztmt}) proves that $W_t$ is a martingale, that is to say it fulfills
\begin{equation}
  \E[ W_{t}\, |\, W_{t-1} ] = W_{t-1}\,.
\end{equation}
The latter two properties ensure that the sequence $(W_t)_{t\ge 0}$ converges to a finite limit $W$. In Fig.~\ref{fig:figfive} (top) we show the \emph{p.d.f.} $\phi(w)$ of $W$ obtained from numerical simulations of $W_7$ performed at $\alpha=2.45,\,2.65,\,2.85$ and $\nmin=2,3,4$ (simulations suggest that $W_t$ is close to convergence for $t\gtrsim 4$). It is known that the probability generating function of $W$, namely $G_{\rm\scriptscriptstyle W}(z) = \E[z^W]$, obeys the functional equation
\begin{equation}
  G_{\rm\scriptscriptstyle W}(z) = G_{\rm\scriptscriptstyle Q}\left(G_{\rm\scriptscriptstyle W}(z^{1/\mQ})\right)\,,
  \label{eq:gf}
\end{equation}
where $G_{\rm\scriptscriptstyle Q}(z) = \E[z^{Z_1}]=\sum_{n=\nmin}^\infty \pQ(n)z^n$ is the probability generating function of $Z_1$. In principle, eq.~(\ref{eq:gf}) might be used to determine $G_{\rm\scriptscriptstyle W}(z)$. In practice, no off--the--shelf technique is known to solve it, the main difficulty being that it relates $G_{\rm\scriptscriptstyle W}(z)$ at different scales. In sect.~4, we obtain by other means an approximation to $\phi(w)$ which holds in the asymptotic regime $w\to\infty$ and reads
\begin{equation}
  \phi(w) \approx \frac{1}{\zeta(\alpha-1,\nmin)}\frac{1}{w^\alpha}\,.
  \label{eq:phi}
\end{equation}
Dashed lines in Fig.~\ref{fig:figfive} (top) represent eq.~(\ref{eq:phi}). The agreement with simulation data is reasonably good, yet our asymptotic estimate worsens as $\alpha$ and $\nmin$ increase.  

The variable of interest for our aims is the number of preferences $V_t$, introduced in eq.~(\ref{eq:novotes}). From the above discussion, we conclude that $V_t$ diverges as $t\to\infty$ with the same growth rate as $Z_t$. Therefore, it is convenient to introduce the variable
\begin{equation}
  H_t = \frac{V_t}{\mQ^t} = W_t + \frac{1}{\mQ}W_{t-1} + \frac{1}{\mQ^2}W_{t-2} + \ldots\,.
  \label{eq:expH}
\end{equation}
As $t\to\infty$, the sequence $(H_t)_{t\ge0}$ converges to a finite limit $H$. In Fig.~\ref{fig:figfive} (bottom), we show the \emph{p.d.f.} $\chi(h)$ of $H$ obtained from the same numerical simulations described above. The distribution of $H$ looks very similar to that of $W$ and it is easy to understand why. First, we notice that
\begin{equation}
  \E[H_t] = \E[W_t] + \frac{1}{\mQ}\E[W_{t-1}] + \frac{1}{\mQ^2}\E[W_{t-2}] + \ldots = \frac{\mQ - 1/\mQ^{t}}{\mQ-1}\,.
\end{equation}
It follows that $\E[H] = \mQ/(\mQ-1)$. Now, $W_t$, $W_{t-1}$, \ldots\, converge to different copies of $W$ as $t\to\infty$. However, these are not i.i.d. variables since they are correlated by construction ($W_t$ is realized from $W_{t-1}$, which is realized from $W_{t-2}$, \emph{etc}.). Of course, the correlation between $W_t$ and $W_{t-s}$ vanishes as $s$ increases. Since $\mQ\sim{\rm O}(10)$, it follows that terms which are progressively less correlated with $W_t$ are also quickly suppressed in eq.~(\ref{eq:expH}). Hence, it makes sense to approximate $H$ by
\begin{equation}
  H \simeq \frac{\mQ}{\mQ-1}W\,,
\end{equation}
and its \emph{p.d.f.} $\chi(h)$ in the asymptotic regime $h\to\infty$ by
\begin{equation}
  \chi(h) \approx \frac{h_0^{\alpha-1}}{\zeta(\alpha-1,\nmin)}\frac{1}{h^\alpha}\,,
  \label{eq:chi}
\end{equation}
with $h_0 = \mQ/(\mQ-1)$. Dashed lines in Fig.~\ref{fig:figfive} (bottom) represent eq.~(\ref{eq:chi}). We observe that the quality of the approximation is analogous to that obtained for $\phi(w)$. We notice as well that $\chi(h)$ enters the asymptotic regime as $h\gtrsim 1$.

From Fig.~\ref{fig:figfive} (bottom) we also see that a large fraction of the probability mass of $H$ lies in the region $h\lesssim 1$. The partition function $\cZ_H(h) = \P\{H<h\}$ saturates quickly to one as $h>1$, as shown in Table~\ref{tab:tabtwo}, where values of $\cZ_H(1)$ and $\cZ_H(8)$ are reported by way of example for various choices of $(\alpha,\nmin)$.

\begin{center}
\begin{table}[t!]
  \begin{minipage}{0.5\textwidth}
    \hskip 1.0cm\begin{tabular}{cc|rrr|}
      & & \multicolumn{3}{c|}{$\alpha$} \\[0.2ex]
      \multicolumn{2}{c|}{$\P\{H<1\}$} & 2.45 & 2.65 & 2.85 \\
      \hline\\[-3.2ex]
      \multirow{3}{*}{\rotatebox{0}{$\nmin$}} & 2 & 0.788  & 0.698  & 0.616   \\
                                              & 3 & 0.796  & 0.724  & 0.660   \\[0.0ex]
                                              & 4 & 0.798  & 0.734  & 0.683   \\[0.0ex]
      \hline
    \end{tabular}
  \end{minipage}
  \begin{minipage}{0.5\textwidth}
    \hskip 1.0cm \begin{tabular}{cc|rrr|}
      & & \multicolumn{3}{c|}{$\alpha$} \\[0.2ex]
      \multicolumn{2}{c|}{$\P\{H<8\}$} & 2.45 & 2.65 & 2.85 \\
      \hline\\[-3.2ex]
      \multirow{3}{*}{\rotatebox{0}{$\nmin$}} & 2 & 0.981  & 0.982  & 0.984   \\
                                              & 3 & 0.985  & 0.987  & 0.990   \\[0.0ex]
                                              & 4 & 0.986  & 0.989  & 0.992   \\[0.0ex]
      \hline
    \end{tabular}
  \end{minipage}
  \caption{\footnotesize Partition function $\cZ_H(h)=\P\{H<h\}$ for $h=1,8$ and various choices of $(\alpha,\nmin)$.}
  \label{tab:tabtwo}
\end{table}
\end{center}

\vskip -1.1cm
We conclude this section with two short remarks, that will be helpful in sect.~3. In first place, we let $p_t(v) = \P\{V_t = v\}$ denote the probability law of $V_t$ and recall that $V_t$ is an integer variable. Then, for $h = k/\mQ^t$ and $k\in\dN$, we have
\begin{equation}
  p_t(h\mQ^t) = \P\{V_t = h\mQ^t\} = \P\{H_t = h\} \to \chi(h){\rm d}h\,\quad \text{ as } t\to\infty\,,
  \label{eq:scalinglaw}
\end{equation}
since $H$ is a continuous variable. Secondly, the overall number of preferences on all trees is measured by the variable $\sum_{i=1}^QV^{(i)}_t$, which diverges as $t\to\infty$ and also as $Q\to\infty$.  For this reason, we find it convenient to introduce the variable
\vskip -0.7cm
\begin{equation}
  \bar H^{[Q]}_t \equiv \frac{1}{Q}\sum_{i=1}^{Q}H^{(i)}_t\,.
\end{equation}
\vskip -0.3cm
\noindent The sequence $(\bar H^{[Q]}_t)_{t\ge 0}$ converges to a finite limit $\bar H^{[Q]}$ with continuous \emph{p.d.f.} $\bar\chi^{[Q]}(h)$ as $t\to\infty$. We let ${\bar p^{[Q]}_t(v) = \P\bigl\{\sum_{i=1}^QV^{(i)}_t = v\bigr\}}$ for $v\in\dN$. Then, in perfect analogy with eq.~(\ref{eq:scalinglaw}), for $h = k/Q\mQ^t$ and $k\in\dN$ we have ${\bar p^{[Q]}_t(hQ\mQ^t) = \P\bigl\{\sum_{i=1}^Q V^{(i)}_t = hQ\mQ^t\bigr\} = \P\{\bar H^{[Q]}_t = h\} \to \bar\chi^{[Q]}(h)\rd h}$ as $t\to\infty$. We shall say more about the distribution of $\bar H^{[Q]}$ in sect.~3.3.

\section{Large--list limit in the quenched model}

In sect.~2 we mentioned that numerical simulations of $F_{\rm\scriptscriptstyle Q}(x|Q,N)$ under SR1 yield in most cases a tree $T^{(i)}$ for which the variable~$Z^{(i)}_{\texit}$ is only partially realized. This occurs whenever the generation of new vertices stops before those on the $(\texit-1)$th level have fully branched. In such cases the overall number of vertices on $T^{(i)}$ is not correctly counted by the variable~$V^{(i)}_{\texit}$ introduced in eq.~(\ref{eq:novotes}). In order to calculate $F_{\rm\scriptscriptstyle Q}(x|Q,N)$, we either introduce \emph{ad--hoc} variables that count subsets of vertices on the levels of a tree, or we reject \emph{a~priori} all $Q$--forests for which the above issue arises, thus obtaining an estimate of the distribution (the goodness of which we can only assess \emph{a posteriori}). In this work we follow the latter approach. 
There are several possible approximations, lying between the following two extremes:
\begin{itemize}
  \item{the coarsest approximation consists in restricting the set of contributing $Q$--forests  to those for which  \emph{all} of~$(Z^{(i)}_{\texit})_{i=1}^Q$ are fully realized when SR1 is fulfilled. We refer to this set $\cD_{\min}(Q,N)\subset\fT$ as the \emph{minimal ensemble} for SR1.}
  \item{the finest approximation consists in restricting the set of contributing $Q$--forests to those for which \emph{each} of~$(Z^{(i)}_{\texit})_{i=1}^Q$ is either fully realized or not realized at all when SR1 is fulfilled. We refer to this set  $\cD_{\max}(Q,N)\subset\fT$ as the \emph{maximal ensemble} for SR1.}
\end{itemize}
Obviously, we have $\cD_{\min}(Q,N)\subset\cD_{\max}(Q,N)$, therefore we expect the estimate of $F_{\rm\scriptscriptstyle Q}(x|Q,N)$ to be worse in $\cD_{\min}(Q,N)$  than in $\cD_{\max}(Q,N)$. Notice that the above issue does not arise under either SR2 or SR3. In this section, we first work out the large--list limit of $F_{\rm\scriptscriptstyle Q}(x|Q,N)$ in the \emph{minimal ensemble} for SR1. This case study allows us to set up the notation. Then, we calculate the vote distribution in the \emph{maximal ensemble} for SR1 and finally under SR2 and SR3. 

\subsection{$F_{\rm\scriptscriptstyle Q}(x|Q,N)$ in the \emph{minimal ensemble} for SR1}

As previously explained, we assume $Q$ trees and stop the word--of--mouth spreading precisely when the overall number of vertices generated on all trees equals $N$. Accordingly, the space of all possible vote configurations is given by the discrete simplex
\begin{equation}
  \Sigma(Q,N) = \left\{(v_1,\ldots,v_Q)\in \dN^Q:\quad \sum_{k=1}^Q v_k = N\right\}\,.
\end{equation}
Given $v=(v_1,\ldots,v_Q)\in\Sigma(Q,N)$, the probability that at time $t$ the vote variables defined in eq.~(\ref{eq:novotes}) fulfill  $\{V^{(1)}_t = v_1$, \ldots, $V^{(Q)}_t = v_Q\}$ amounts to
\begin{equation}
  \Pi_t(v|Q,N) = \frac{p_t(v_1)\cdots p_t(v_Q)}{\sum_{v\in\Sigma(Q,N)}p_t(v_1)\cdots p_t(v_Q)}\,.
  \label{eq:Pit}
\end{equation}
\noindent It is important to recognize that $\Pi_t(\,\cdot\,|Q,N)$ measures subsets of $\fT$. More precisely, the domain of $\Pi_t(\,\cdot\,|Q,N)$ is the \emph{ensemble} $\cD_t(Q,N)\subset\fT$ defined by
\begin{equation}
  \cD_t(Q,N) = \left\{T\in\fT:\quad \left(V_t(T^{(1)}),\ldots,V_t(T^{(Q)})\right)\in\Sigma(Q,N)\right\}\,.
  \label{eq:DtQN}
\end{equation}
\noindent Sampling $\Pi_t(\,\cdot\,|Q,N)$ is a hard job. In order to generate an acceptable sample $T\in\cD_t(Q,N)$ in a computer simulation, we are supposed to run eq.~(\ref{eq:BP2}) up to the $t$th level (which means drawing \emph{all} variables $(Z^{(i)}_s)_{i=1}^Q$ for $s\le t$) and then check that $\left(V_t(T^{(1)}),\ldots,V_t(T^{(Q)})\right)\in\Sigma(Q,N)$. Needless to say, such samples have an extremely low probability of being generated.  Starting from eq.~(\ref{eq:DtQN}), we define
\begin{equation}
  \cD_{\min}(Q,N) = \bigcup_{t=0}^\infty \cD_t(Q,N)\,.
\end{equation}
This is the \emph{minimal ensemble} for SR1. For all $T\in\cD_{\min}(Q,N)$, SR1 is exactly fulfilled at some level $\texit$ with all of $(Z^{(i)}_{\texit})_{i=1}^Q$ being fully counted. We call it \emph{minimal} since it is the smallest subset of $\frak T$ in which $F_{\rm\scriptscriptstyle Q}(x|Q,N)$ under SR1 can be consistently represented as a functional of the probabilities of $\{(V^{(i)}_t)_{i=1}^Q\}_{t\in\dN}$ alone. Of course, the majority of $T\in\frak T$ contributing to numerical simulations of $F_{\rm\scriptscriptstyle Q}(x|Q,N)$ under SR1 lie outside~$\cD_{\min}(Q,N)$.

Notice that since $\cD_t(Q,N)\cap\cD_{t'}(Q,N) = \emptyset$ for $t'\ne t$, it follows that
\begin{equation}
  \P\{T\in\cD_{\min}(Q,N)\} = \sum_{t=0}^\infty \P\{T\in\cD_t(Q,N)\}\,.
  \label{eq:PDtQN}
\end{equation}
The above probabilities come into play when observing that given $(Q,N)$, the actual level $\texit$ at which the algorithm stops is unknown. For this reason we introduce the stopping time 
\begin{equation}
  \tau_{Q,N} = \min\left\{t\in\dN: \ \  \sum_{i=1}^Q V^{(i)}_t \ge N\right\}\,,
  \label{eq:stoptime}
\end{equation}
representing the earliest level at which the sum of the vote variables defined in eq.~(\ref{eq:novotes}) exceeds $N$. Similar to $V^{(i)}_t$, also $\tau_{Q,N}$ is a stochastic variable, \ie it is a function of $T\in\frak{T}$. As such, it is differently realized in each iteration of the algorithm. Of course, $\tau_{Q,N}$ fulfills
\begin{equation}
  \P\{\tau = t\,|\,\cD_{\min}\} \ \equiv\ \P\{\tau_{Q,N}(T) = t\,|\,T\in\cD_{\min}(Q,N)\} = \frac{\P\{T\in\cD_t(Q,N)\}}{\P\{T\in\cD_{\min}(Q,N)\}}\,.
  \label{eq:PtauDmin}
\end{equation}
Sampling $\P\{\tau=t\,|\,\cD_{\min}\}$ is just as difficult as sampling $\Pi_t(\,\cdot\,|Q,N)$: on the one hand  ${\P\{T\in\cD_{\min}\}\to 0}$ as $Q,N\to\infty$, on the other hand we do not know any algorithmic recipe to \emph{a priori} generate ${Q\text{--forests}}$ ${T\in\cD_{\min}(Q,N)}$ with correct probability. However, we can obtain an estimate of $\P\{\tau=t\,|\,\cD_{\min}\}$ by measuring the frequency of those $T\in{\frak T}$ for which SR1 is fulfilled with either $Z^{(Q)}_{t}$ or $Z^{(1)}_{t+1}$ being partially counted.

Now, given $(Q,N)$ and $x = \ell Q/N$ with $\ell=1,2,\ldots,N$, the \emph{discrete} probability that the excess--of--votes variable yields $x$ in the \emph{minimal ensemble} for SR1 is given by
\begin{equation}
  F^{(i)}_{\rm\scriptscriptstyle Q}(x|Q,N;\cD_{\min}) = \sum_{t=0}^\infty \left[\sum_{v\in\Sigma(Q,N)}\delta_{xN/Q,v_i}\ \Pi_t(v|Q,N)\right]\P\{\tau = t\,|\,\cD_{\min}\}\,,
  \label{eq:Fproblawmin}
\end{equation}
with the $\delta$--symbol within square brackets representing the Kronecker delta. Although we need to select one specific candidate to count preferences, the upper index $(i)$ on the l.h.s. is redundant, since $\Pi_t(v|Q,N)$ is invariant under permutations of the components of the vote vector $v$. Hence, we drop it in the sequel. We wish to work out eq.~(\ref{eq:Fproblawmin}) in the large--list limit, \ie as $Q,N\to\infty$. To this aim, we notice that the simplex constraint can be represented in terms of another $\delta$--symbol: for any function $f(v)$ the identity
\begin{equation}
  \sum_{v\in\Sigma(Q,N)}f(v) = \sum_{v_1,\ldots,v_Q=1}^M\, \delta_{\sum_k v_k,N}\ f(v) = \frac{1}{2\pi}\int_{-\pi}^{\pi}\rd\lambda\, \sum_{v_1,\ldots,v_Q=1}^M\, \re^{\ri\lambda\left(\sum_k v_k - N\right)}\ f(v_1,\ldots,v_Q)\,
  \label{eq:Kharm}
\end{equation}
holds for all integers $M\ge N$. In particular, the rightmost equality follows from representing the Kronecker delta in Fourier harmonics. Upon using eq.~(\ref{eq:Kharm}) and the explicit expression of $\Pi_t(v|Q,N)$, we turn eq.~(\ref{eq:Fproblawmin}) into
\begin{equation}
  F_{\rm\scriptscriptstyle Q}(x|Q,N;\cD_{\min}) = \sum_{t=0}^\infty\, p_t\left(\frac{xN}{Q}\right)\, \left[ \frac{\int_{-\pi}^\pi \rd\lambda\, \re^{-i\lambda \frac{N}{Q} (Q-x)} \prod_{j\ne i}^{1\ldots Q}\sum_{v_j=1}^M \re^{\ri\lambda v_j} p_t(v_j)}{ \int_{-\pi}^\pi \rd\lambda\, \re^{-i\lambda N} \prod_{j=1}^Q \sum_{v_j=1}^M \re^{\ri\lambda v_j} p_t(v_j)}\right]\,\P\{\tau = t\,|\,\cD_{\min}\}\,.
  \label{eq:Foscill}
\end{equation}
Provided $x<Q$, the exponential $\exp\{-\ri\lambda \frac{N}{Q}(Q-x)\}$ at numerator oscillates quickly as $N\to\infty$ and thus lets the integral over $\lambda$ receive contributions only from a region around $\lambda=0$ with size proportional to~${\pi Q/[N(Q-x)]\to 0}$. Analogously, the exponential $\exp\{-\ri\lambda N\}$ at denominator oscillates quickly as ${N\to\infty}$ and thus lets the integral over $\lambda$ receive contributions only from a region around $\lambda=0$ with size proportional to $\pi/N\to 0$\footnote{To make this point clear: we extend $F_{\rm\scriptscriptstyle Q}(x|Q,N)$ to $N\in\dC$ by analytic continuation, then we rotate $N\to-\ri\tilde N$. This turns the oscillating integrals into exponentially damped ones. Everything under the integral sign (except the highly oscillating exponentials) can be then evaluated at $\lambda = 0$ and be taken out of the integrals.  At the end we rotate back $\tilde N\to\ri N$. It is funny to notice that highly oscillating integrals such as eq.~(\ref{eq:Foscill}) arise in applicative contexts which have nothing to do with opinion dynamics. An example is represented by the Bjorken scaling in QCD (see for instance eq.~(9.5) of ref.~\cite{Manohar}), which is analogous in some respect to the FC scaling.}. Therefore, the ratio in square brackets converges quickly to $Q/\left[(Q-x)\sum_{v_i=0}^{M}p_t(v_i)\right]$ as $N\to\infty$. Clearly, in this approximation $F_{\rm\scriptscriptstyle Q}(x|Q,N)$ diverges as $x\to Q$, the reason being that the original integral at numerator stops oscillating in that limit and the above argument fails. However, we can approximate the diverging factor $Q/(Q-x)\to 1$ as $Q\to\infty$. We finally set $M=N$. This yields
\begin{equation}
  F_{\rm\scriptscriptstyle Q}(x|Q,N;\cD_{\min})\, =\, \sum_{t=0}^\infty\, \frac{p_t\left({xN}/{Q}\right)}{\sum_{v=1}^N p_t\left(v\right)}\, \P\{\tau=t\,|\,\cD_{\min}\}\,,\qquad \text{as } Q,N\to\infty\,.
  \label{eq:FQminBefScal}
\end{equation}
To highlight the transition to a continuous distribution, we find it convenient to set $N\to N_s=h_0Q\mQ^s$ (recall that $\E[V_t] = h_0\mQ^t$) and let it diverge along the sequence $s=1,2,3,\ldots,\infty$ (thermodynamic limit). If $N_s$ is not an integer, we define $F_{\rm\scriptscriptstyle Q}(x|Q,N_s;\cD_{\min})$ by linear interpolation between its values at the nearest integers $N_-=\lfloor N_s\rfloor$ and ${N_+=\lceil N_s\rceil}$. This setting is legitimate as well as reasonable: on the one hand we know that preferences distribute symmetrically among the $Q$ trees up to fluctuations, on the other hand the variable $s=\log [N/(h_0Q)]/\log \mQ$ just represents the average level at which the word of mouth stops given $(Q,N)$. In other words, the probability distribution $\P\{\tau=t\,|\,\cD_{\min}\}$ as a function of $t$ peaks precisely at $t\simeq s$ for $N=N_s$. It follows from eq.~(\ref{eq:scalinglaw}) that ${p_t(xN/Q) = p_t(h_0x\mQ^s) = p_t(h_0x\mQ^{s-t}\mQ^t) \to h_0\mQ^{s-t} \chi(h_0x\mQ^{s-t})\rd x}$ as~${s\to\infty}$. In the same limit the probability distribution of $x$ turns into a continuous distribution, namely ${F_{\rm\scriptscriptstyle Q}(x|Q,N_s)\to \FQ(x|Q,N_s)\rd x}$. We conclude that
\begin{empheq}[box=\mybluebox]{align}
  \FQ(x|Q,N_s;\cD_{\min}) \ =\, \hskip -0.1cm \sum_{t=s,s\pm 1, s\pm 2,\ldots}\hskip -0.0cm \dfrac{h_0\mQ^{s-t}\chi(h_0x\mQ^{s-t})}{\int_0^{h_0Q\mQ^{s-t}}\rd y\, \chi(y)}\ \P\{\tau=t\,|\,\cD_{\min}\}\,,\quad \text{ as } Q,s\to\infty\,. 
  \label{eq:fse}
\end{empheq}
The above distribution is correctly normalized, as can be seen upon integrating both sides over $x\in (0,Q)$. Provided $h_0Q\gtrsim \mQ^{t-s+1}$, the integral at denominator can be omitted without significant loss of accuracy (see Table~\ref{tab:tabtwo}).

\noindent {\bf Remark 3.1.} While eq.~(\ref{eq:Fproblawmin}) is exact, eq.~(\ref{eq:fse}) is formally correct provided $x\ll Q$. An important difference between the two distributions arises as we calculate their expectations, which should equal one by symmetry. Indeed, we have
\begin{align}
\E[\,x\,|Q,N;\cD_{\min}] & = \frac{1}{Q}\sum_{i=1}^Q \sum_{x} x\, F^{(i)}_{\rm\scriptscriptstyle Q}(x|Q,N;\cD_{\min}) \nonumber\\[1.0ex]
 & \hskip -1.0cm = \frac{1}{N}\sum_{t=0}^\infty \left[\sum_{v\in\Sigma(Q,N)}\sum_{i=1}^Q v_i\, \Pi_t(v|Q,N)\right]\P\{\tau = t\,|\,\cD_{\min}\} \, = \, \sum_{t=0}^\infty \P\{\tau = t\,|\,\cD_{\min}\}\, =\, 1\,,
\end{align}
whereas
\begin{align}
  \label{eq:expx}
  \cE[\,x\,|Q,N_s;\cD_{\min}] & =  \int_0^Q\rd x\, x\, \FQ(x|Q,N_s;\cD_{\min})\nonumber\\[1.0ex]
  & \hskip -1.8cm = \hskip -0.1cm \sum_{t=s,s\pm 1, s\pm 2,\ldots} \frac{m^{t-s}}{h_0} \dfrac{\int_0^{h_0Q\mQ^{s-t}}\rd y\, y\, \chi(y)}{\int_0^{h_0Q\mQ^{s-t}}\rd y\, \chi(y)}\  \P\{\tau=t\,|\,\cD_{\min}\}\,.
\end{align}
For finite values of $Q$ we have in general $\cE[\,x\,|Q,N_s;\cD_{\min}] \ne 1$. Notice that the ratio of integrals  \linebreak on the r.h.s. of eq.~(\ref{eq:expx}) converges to $h_0$ as $Q\to\infty$. Hence, in this limit we are left with \linebreak ${\cE[\,x\,|Q,N_s;\cD_{\min}] \simeq \sum_t m^{t-s}\,\P\{\tau = t\,|\,\cD_{\min}\}}$. The value of the latter sum depends on how $\tau_{Q,N_s}$ distributes around $s$. In full generality we must regard $\cE[\,x\,|Q,N_s;\cD_{\min}]$ as a measurement of the quality by which eq.~(\ref{eq:fse}) approximates $F_{\rm\scriptscriptstyle Q}(x|Q,N;\cD_{\min})$ in the large--list limit. \hfill $\blacksquare$

In Table~\ref{tab:tabthree} (left) we report an estimate of the probability $\P\{\tau=t\,|\,\cD_{\min}\}$ for  $s=5$, $(\alpha,\nmin) = (2.45,3)$ and ${Q=8,\ldots,256}$. We obtained this estimate from numerical simulations just as explained above, \ie by measuring the frequency of those $T\in\frak T$ for which SR1 is fulfilled with either $Z^{(Q)}_t$ or $Z^{(1)}_{t+1}$ being partially counted. In Table~\ref{tab:tabthree} (right) we report an estimate of the expectation $\cE[\,x\,|Q,N_s;\cD_{\min}]$ corresponding to our estimate of $\P\{\tau=t\,|\,\cD_{\min}\}$ for the same choice of parameters. Analogous simulations at $s=6$ display a rigid shift of the probability distribution of $\tau_{Q,N_6}$ around $t=6$ with no evident change in the structure of the tails. In full generality we conclude that $\tau_{Q,N_s}$ peaks at $t=s$ with increasing probability as $Q\to\infty$, provided $s$ is sufficiently large. By extrapolation we have 
\begin{equation}
  \lim_{Q\to\infty}\lim_{s\to\infty}\P\{\tau=t\,|\,\cD_{\min}\}\ =\ \delta_{t,s}\,.
  \label{eq:tautherm}
\end{equation}
As a consequence, $\cE[\,x\,|Q,N_s;\cD_{\min}]\to 1$ as $Q,s\to\infty$. From Table~\ref{tab:tabthree} we notice that $\cE[\,x\,|Q,N_s;\cD_{\min}]$ deviates from one by less than $20\%$ even at the smallest value of $Q$ among those simulated.

\begin{table}[t!]
  \small
  \centering
  \begin{tabular}{cr|cccc|c|c|c|c|}
    & & \multicolumn{4}{c}{$t$} \\[0.2ex]
    \multicolumn{2}{c|}{$\P\{\tau=t\,|\,\cD_{\min}\}$} & 3 & 4 & 5 & 6 & & $Q$ & $\cE[\,x\,|Q,N_5;\cD_{\min}]$ & $\cC(Q)$ \\
    \cline{1-6}\cline{8-10}\\[-3.2ex]
    \multirow{5}{*}{\rotatebox{0}{$\hskip 0.8cm Q$}}
    &   8 & 0.0018(2)  & 0.0447(8) & 0.9524(8) & 0.0009(1)  & $\phantom{aa}$ &   8 & 1.18(4) & 0.97 \\
    &  16 & 0.0010(2)  & 0.0266(7) & 0.9723(7) & \text{n/a} & $\phantom{aa}$ &  16 & 1.10(3) & 0.97 \\[0.0ex]
    &  32 & 0.0005(1)  & 0.0157(8) & 0.9837(8) & \text{n/a} & $\phantom{aa}$ &  32 & 1.05(3) & 0.98 \\[0.0ex]
    &  64 & 0.0002(1)  & 0.0082(5) & 0.9915(5) & \text{n/a} & $\phantom{aa}$ &  64 & 1.03(3) & 0.99 \\[0.0ex]
    & 128 & \text{n/a} & 0.0048(7) & 0.9951(7) & \text{n/a} & $\phantom{aa}$ & 128 & 1.02(3) & 0.99 \\[0.0ex]
    & 256 & \text{n/a} & 0.0023(6) & 0.9973(5) & \text{n/a} & $\phantom{aa}$ & 256 & 1.00(3) & 0.99 \\[0.0ex]
    \cline{1-6}\cline{8-10}
  \end{tabular}
  \vskip 0.1cm
  \caption{\footnotesize (Left) Approximate probability of $\tau_{Q,N_s}$ in $\cD_{\min}$ for $s=5$, $t=3,\ldots,6$, $(\alpha,\nmin)=(2.45,3)$ and $Q=8,\ldots,256$. Bootstrap uncertainty is quoted in parentheses. N/a entries correspond to measurements for which noise/signal~${\ge 1}$. (Right) Expectation of $x$ in the large--list limit and $\cC(Q)$ corresponding to our estimate of $\P\{\tau=t\,|\,\cD_{\min}\}$.}
  \label{tab:tabthree}
\end{table}

We now go back to eq.~(\ref{eq:fse}). From eq.~(\ref{eq:tautherm}) we infer that
\begin{equation}
  \FQ^*(x) = \lim_{Q\to\infty}\lim_{s\to\infty} \FQ(x|Q,N_s;\cD_{\min}) \,= \, h_0\chi(h_0x)\,.
\end{equation}
Thus we see that $\FQ^*(x)$ has an asymptotic power--law behaviour. The asymptotic regime is reached as soon as $x\gtrsim h_0^{-1}$. Eq.~(\ref{eq:chi}) yields a prediction of the exponent and the scale coefficient of the power law in the large--list limit. If we forget about the structure of $\P\{\tau=t\,|\,\cD_{\min}\}$ for a while and concentrate on the general structure eq.~(\ref{eq:fse}), we see that in the large--list limit $\FQ(x|Q,N_s;\cD_{\min})$ is given by a convolution of $\chi$ at largely separated scales (recall that $\mQ\sim {\rm O}(10)$). Given $x$, we can split the latter into two groups, namely forward ($\mQ x$, $\mQ^2x$, \ldots) and backward ($x/\mQ$, $x/\mQ^2$, \ldots) scales. Explicitly, we have
\begin{align}
  &\hskip -0.5cm \FQ(x|Q,N_s;\cD_{\min})\ = \ \frac{h_0\chi(h_0x)}{\int_0^{h_0Q}\rd y\, \chi(y)}\,\P\left\{\tau=s\,|\,\cD_{\min}\right\} \nonumber\\[2.0ex]
  &\hskip -0.5cm  + \mQ\frac{h_0\chi(h_0\mQ x)}{\int_0^{h_0Q\mQ}\rd y\, \chi(y)}\,\P\left\{\tau=s-1\,|\,\cD_{\min}\right\} + \mQ^2\frac{h_0\chi(h_0\mQ^2x)}{\int_0^{h_0Q\mQ^2}\rd y\, \chi(y)}\,\P\left\{\tau=s-2\,|\,\cD_{\min}\right\}\ + \ldots \hskip 0.6cm {\text{\footnotesize(fw. terms)}} \nonumber\\[2.0ex]
  &\hskip -0.5cm  + \frac{1}{\mQ}\frac{h_0\chi\left(h_0x/\mQ\right)}{\int_0^{h_0Q/\mQ}\rd y\,\chi(y)}\,\P\left\{\tau=s+1\,|\,\cD_{\min}\right\} + \frac{1}{\mQ^2}\frac{h_0\chi\left(h_0x/\mQ^2\right)}{\int_0^{h_0Q/\mQ^2}\rd y\,\chi(y)}\,\P\left\{\tau=s+2\,|\,\cD_{\min}\right\}\ + \ldots \hskip 0.05cm {\text{\footnotesize(bw. terms)}}
  \label{eq:FQexp}
\end{align}
If $\chi(h_0x)$ is in the power--law regime, this is even more the case for $\chi(h_0\mQ^kx)$, $k=1,2,$\,\ldots We thus see~that forward terms are safe: they do not spoil the overall power--law behaviour, do not even change the exponent of the power law and only contribute by modifying its scale coefficient. By contrast, backward terms are potentially dangerous: they shift $x$ towards regions where $\chi$ is not anymore in the asymptotic regime. Backward terms are in principle able to impair the power--law behaviour. Nevertheless, they are suppressed by inverse powers of $\mQ$. This suggests that power--law structures in the scaling vote distribution may emerge even at moderate values of $Q$. Keeping only terms corresponding to $\tau_{Q,N_s}=s,s\pm 1$ and assuming that $x/\mQ \gtrsim 1$ yields the asymptotic limit
\begin{align}
  & \FQ(x|Q,N_s;\cD_{\min}) \simeq\,\frac{\cC(Q)}{\zeta(\alpha-1,\nmin)} \frac{1}{x^\alpha}\,, \quad \text{ as } x,Q,s\to\infty\,,\quad  \text{with}\nonumber\\[2.0ex]
  & \cC(Q) =\, \frac{\P\{\tau=s\,|\,\cD_{\min}\}}{\int_0^{h_0Q}\rd y\,\chi(y)} + \frac{1}{\mQ^{\alpha-1}}\frac{\P\left\{\tau=s-1\,|\,\cD_{\min}\right\}}{\int_0^{h_0Q\mQ}\rd y\,\chi(y)} + \mQ^{\alpha-1}\frac{\P\left\{\tau=s+1\,|\,\cD_{\min}\right\}}{\int_0^{h_0Q/\mQ}\rd y\,\chi(y)}\,.
  \label{eq:FQthreeterms}
\end{align}
The coefficient $\cC(Q)$ represents the only correction to the power--law scaling at finite $Q$. It fulfills $\cC(Q)\simeq 1$ and $\lim_{Q\to\infty}\cC(Q) = 1$. For $Q\gg \mQ$ the integrals at denominator can be omitted, as previously noticed. In the specific case of the \emph{minimal ensemble}, we can set $\cC(Q) = 1$ for all $Q$ without loss of accuracy, as the values reported in Table~\ref{tab:tabthree} (right) suggest. 

\subsection{$F_{\rm\scriptscriptstyle Q}(x|Q,N)$ in the \emph{maximal ensemble} for SR1}

Now that we have shown in a simplified framework how to calculate $F_{\rm\scriptscriptstyle Q}(x|Q,N)$ in the large--list limit under~SR1, we proceed to work out a much better estimate of it. As mentioned at the beginning of the section, the finest approximation in terms of level variables takes into account all $Q$--forests for which \emph{each} of $(Z^{(i)}_{\texit})_{i=1}^Q$ is either fully realized or not realized at all when SR1 is fulfilled. To formalize this, we go through the same steps as we did in last section. Concretely, given $(Q,N)$ we introduce the \emph{ensembles}
\begin{equation}
  \cD_{t,q}(Q,N) = \left\{T\in\fT:\quad \left(V_{t+1}(T^{(1)}),\ldots,V_{t+1}(T^{(q)}),V_t(T^{(q+1)}),\ldots,V_t(T^{(Q)})\right)\in\Sigma(Q,N)\right\}\,,
  \label{eq:DtqQN}
\end{equation}
for $t\ge 0$ and $1\le q\le Q$. Notice that $\cD_{t,Q}(Q,N) = \cD_{t+1}(Q,N)$. Moreover, independently of $t$ or $q$ we expect that $\P\{T\in\cD_{t,q}(Q,N)\}\ll 1$, since $\cD_{t,q}(Q,N)$ is a highly restricted subset of $\fT$. Starting from eq.~(\ref{eq:DtqQN}) we define
\begin{equation}
  \cD_{\max}(Q,N) = \bigcup_{t=0}^\infty\bigcup_{q=1}^Q \cD_{t,q}(Q,N)\,.
\end{equation}
This is the \emph{maximal ensemble} for SR1. For all $T\in\cD_{\max}(Q,N)$, SR1 is exactly fulfilled at some level $\texit$ with each of  $(Z^{(i)}_{\texit})_{i=1}^Q$ being either fully counted or not counted at all. We call it \emph{maximal} since it is the largest subset of $\frak T$ in which $F_{\rm\scriptscriptstyle Q}(x|Q,N)$ under SR1 can be consistently represented as a functional of the probabilities of $\{(V^{(i)}_t)_{i=1}^Q\}_{t\in\dN}$ alone.

Given $v\in\Sigma(Q,N)$, the probability that the vote variables fulfill $\{V^{(1)}_{t+1} = v_1$, \ldots, $V^{(q)}_{t+1} = v_q$, $V^{(q+1)}_t = v_{q+1}$, \ldots, $V^{(Q)}_t = v_Q\}$ amounts to
\begin{equation}
  \Pi_{t,q}(v|Q,N) = \frac{p_{t+1}(v_1)\cdots p_{t+1}(v_q)\,p_t(v_{q+1})\cdots p_t(v_Q)}{\sum_{v\in\Sigma(Q,N)}p_{t+1}(v_1)\cdots p_{t+1}(v_q)\,p_t(v_{q+1})\cdots p_t(v_Q)}\,.
\end{equation}
The domain of $\Pi_{t,q}(v|Q,N)$ is precisely $\cD_{t,q}(Q,N)$. Notice that since $\cD_{t,q} \cap \cD_{t',q'} = \emptyset$ for $(t,q)\ne (t',q')$, it follows that
\begin{equation}
  \P\{T\in\cD_{\max}(Q,N)\} = \sum_{t=0}^\infty\sum_{q=1}^Q \P\{T\in\cD_{t,q}(Q,N)\}\,.
  \label{eq:PDtQN}
\end{equation}
The above probabilities come into play when observing that the actual values $(\texit,\qexit)$ at which the algorithm stops fluctuate randomly along different iterations. For this reason we introduce a fractional stopping time
\begin{equation}
\tau^{\rm\scriptscriptstyle frac}_{Q,N} = \inf\left\{t+\frac{q}{Q}\,:\quad \sum_{i=1}^q V^{(i)}_{t+1} + \sum_{i=q+1}^Q V^{(i)}_{t} \ge N\right\}\,.
\end{equation}
While $\tau^{\rm\scriptscriptstyle frac}_{Q,N}$ is a discrete variable for finite values of $Q$, it becomes continuous as $Q\to\infty$. For finite $Q$ the ``inf'' is actually a ``min'' and since $1\le q\le Q$, there exist no $(t_1,q_1)$, $(t_2,q_2)$ with $(t_1,q_1)\ne (t_2,q_2)$ such that $\tau^{\rm\scriptscriptstyle frac}_{Q,N} = t_1 + q_1/Q = t_2 + q_2/Q$. The probability distribution of $\tau^{\rm\scriptscriptstyle frac}_{Q,N}$ is given by
\begin{equation}
  \P\left\{\tau^{\rm\scriptscriptstyle frac}_{Q,N} = t + \frac{q}{Q}\,\biggr|\, \cD_{\max}\right\} \equiv \P\left\{\tau^{\rm\scriptscriptstyle frac}_{Q,N}(T) = t + \frac{q}{Q}\,\biggr|\, T\in\cD_{\max}(Q,N)\right\} =  \frac{\P\{T\in\cD_{t,q}(Q,N)\}}{\P\{T\in\cD_{\max}(Q,N)\}}\,.
  \label{eq:PtauDmax}
\end{equation}
Although sampling $\P\{\tau^{\rm\scriptscriptstyle frac}_{Q,N} = t + q/Q\,|\,\cD_{\max}\}$ is very difficult (for the same reasons as previously), we can obtain an estimate of it by measuring the frequency of those $T\in{\frak T}$ for which SR1 is fulfilled with $Z^{(q+1)}_{t+1}$ being partially counted.

Now, given $(Q,N)$ and $x = \ell Q/N$ with $\ell = 1,2,\ldots,N$, the discrete probability that the excess--of--votes variable yields $x$ in the \emph{maximal ensemble} for SR1 is given by
\begin{equation}
  F_{\rm\scriptscriptstyle Q}(x|Q,N;\cD_{\max}) = \frac{1}{Q}\sum_{i=1}^Q\sum_{t=0}^\infty \sum_{q=1}^Q\left[\sum_{v\in\Sigma(Q,N)}\delta_{xN/Q,v_i}\Pi_{t,q}(v|Q,N)\right]\P\left\{\tau^{\rm\scriptscriptstyle frac}_{Q,N} = t + \frac{q}{Q}\,\biggr|\, \cD_{\max}\right\}\,.
\end{equation}
By the same arguments used in sect.~3.1 we can show that the above expression boils down to
\begin{equation}
  F_{\rm\scriptscriptstyle Q}(x|Q,N;\cD_{\max}) = \sum_{t=0}^\infty \sum_{q=1}^Q \left[\frac{q}{Q}\,\frac{p_{t+1}\left(xN/Q\right)}{\sum_{v=1}^Np_{t+1}(v)} + \frac{Q-q}{Q}\,\frac{p_t\left(xN/Q\right)}{\sum_{v=1}^Np_t(v)}\right]\P\left\{\tau^{\rm\scriptscriptstyle frac}_{Q,N} = t + \frac{q}{Q}\,\biggr|\, \cD_{\max}\right\}\,,
\end{equation}
as $Q,N\to\infty$. Each term on the r.h.s. combines linearly the probabilities of the vote variables $V_t$ and $V_{t+1}$ for some $t$. We can rearrange the sum so as to avoid such mixing. This yields
\begin{equation}
  F_{\rm\scriptscriptstyle Q}(x|Q,N;\cD_{\max}) = \sum_{t=0}^\infty \, \frac{p_t\left(xN/Q\right)}{\sum_{v=1}^Np_t(v)}\ \P_{\rm int}\{\tau = t\,|\,\cD_{\max}\}\,,
  \label{eq:FQmaxBefScal}
\end{equation}
with the integrated probability $\P_{\rm int}\{\tau = t\,|\,\cD_{\max}\}$ being defined by
\begin{equation}
  \P_{\rm int}\{\tau = t\,|\,\cD_{\max}\} \equiv \frac{1}{Q}\sum_{q=1}^Q \left[q\,\P\left\{\tau^{\rm\scriptscriptstyle frac}_{Q,N} = t-1+\frac{q}{Q}\biggr|\cD_{\max}\right\} + (Q-q)\,\P\left\{\tau^{\rm\scriptscriptstyle frac}_{Q,N} = t+\frac{q}{Q}\biggr|\cD_{\max}\right\}\right]\,.
  \label{eq:Pint}
\end{equation}
\begin{table}[t!]
  \small
  \centering
  \begin{tabular}{cr|cccc|c|c|c|c|}
    & & \multicolumn{4}{c}{$t$} \\[0.2ex]
    \multicolumn{2}{c|}{$\P_{\rm int}\{\tau=t\,|\,\cD_{\max}\}$} & 3 & 4 & 5 & 6 & & $Q$ & $\cE[\,x\,|Q,N_5;\cD_{\max}]$ & $\cC(Q)$ \\
    \cline{1-6}\cline{8-10}\\[-3.2ex]
    \multirow{5}{*}{\rotatebox{0}{$\hskip 0.8cm Q$}}
    &   8 & 0.0031(1) & 0.1194(5) & 0.7787(6) & 0.0987(3) & $\phantom{aa}$ &   8 & 0.81(4) & 2.79 \\
    &  16 & 0.0023(1) & 0.1093(5) & 0.8126(4) & 0.0756(2) & $\phantom{aa}$ &  16 & 0.91(3) & 1.82 \\[0.0ex]
    &  32 & 0.0015(1) & 0.1012(4) & 0.8418(4) & 0.0554(1) & $\phantom{aa}$ &  32 & 0.97(3) & 1.84 \\[0.0ex]
    &  64 & 0.0012(1) & 0.0940(5) & 0.8650(4) & 0.0397(1) & $\phantom{aa}$ &  64 & 1.00(3) & 0.87 \\[0.0ex]
    & 128 & 0.0010(1) & 0.0863(4) & 0.8850(4) & 0.0277(1) & $\phantom{aa}$ & 128 & 1.00(3) & 0.88 \\[0.0ex]
    & 256 & 0.0007(1) & 0.0778(5) & 0.9029(4) & 0.0187(1) & $\phantom{aa}$ & 256 & 0.97(3) & 0.90 \\[0.0ex]
    \cline{1-6}\cline{8-10}
  \end{tabular}
  \vskip 0.1cm
  \caption{\footnotesize (Left) Approximate integrated probability of ${\tau^{\rm\scriptscriptstyle frac}_{Q,N_s}}$ in $\cD_{\max}$ for $s=5$, $t=3,\ldots,6$, $(\alpha,\nmin)=(2.45,3)$ and $Q=8,\ldots,256$. Bootstrap uncertainty is quoted in parentheses; (Right) Expectation of $x$ in the large--list limit and $\cC(Q)$ corresponding to our estimate of ${\P_{\text{int}}\{\tau=t\,|\,{\cal D}_{\max}\}}$.}
  \label{tab:tabfour}
\end{table}
\hskip -0.11cm We see that eq.~(\ref{eq:FQmaxBefScal}) differs from eq.~(\ref{eq:FQminBefScal}) only in that $\P\{\tau = t\,|\,\cD_{\min}\}$ is replaced by $\P_{\rm int}\{\tau = t\,|\,\cD_{\max}\}$. We also observe that $\P_{\rm int}\{\tau = t\,|\,\cD_{\max}\}$ is a convex average of the probabilities of $\tau^{\rm\scriptscriptstyle frac}_{Q,N}$ around $t$. The first term in square brackets on the r.h.s. of eq.~(\ref{eq:Pint}) has maximal weight for $q=Q$ and for this value of $q$ it yields $Q\P\{\tau^{\rm\scriptscriptstyle frac}_{Q,N} = t\,|\,\cD_{\max}\}$. The second term in square bracket has maximal weight for $q=1$ and for this value of $q$ it yields $(Q-1)\P\{\tau^{\rm\scriptscriptstyle frac}_{Q,N} = t + 1/Q\,|\,\cD_{\max}\} \simeq Q \P\{\tau^{\rm\scriptscriptstyle frac}_{Q,N} = t\,|\,\cD_{\max}\}$ as $Q\to\infty$. The other terms of eq.~(\ref{eq:Pint}) are weighted by increasingly lower coefficients and yield the probability of the stopping time at fractional points which are increasingly far from $t$ within the interval $(t-1,t+1)$. It can be easily checked that $\sum_{t=0}^\infty \P_{\rm int}\{\tau=t\,|\,\cD_{\rm max}\} \, = \, 1$. 

\begin{figure}[t!]
 \centering
 \includegraphics[width=0.83\textwidth]{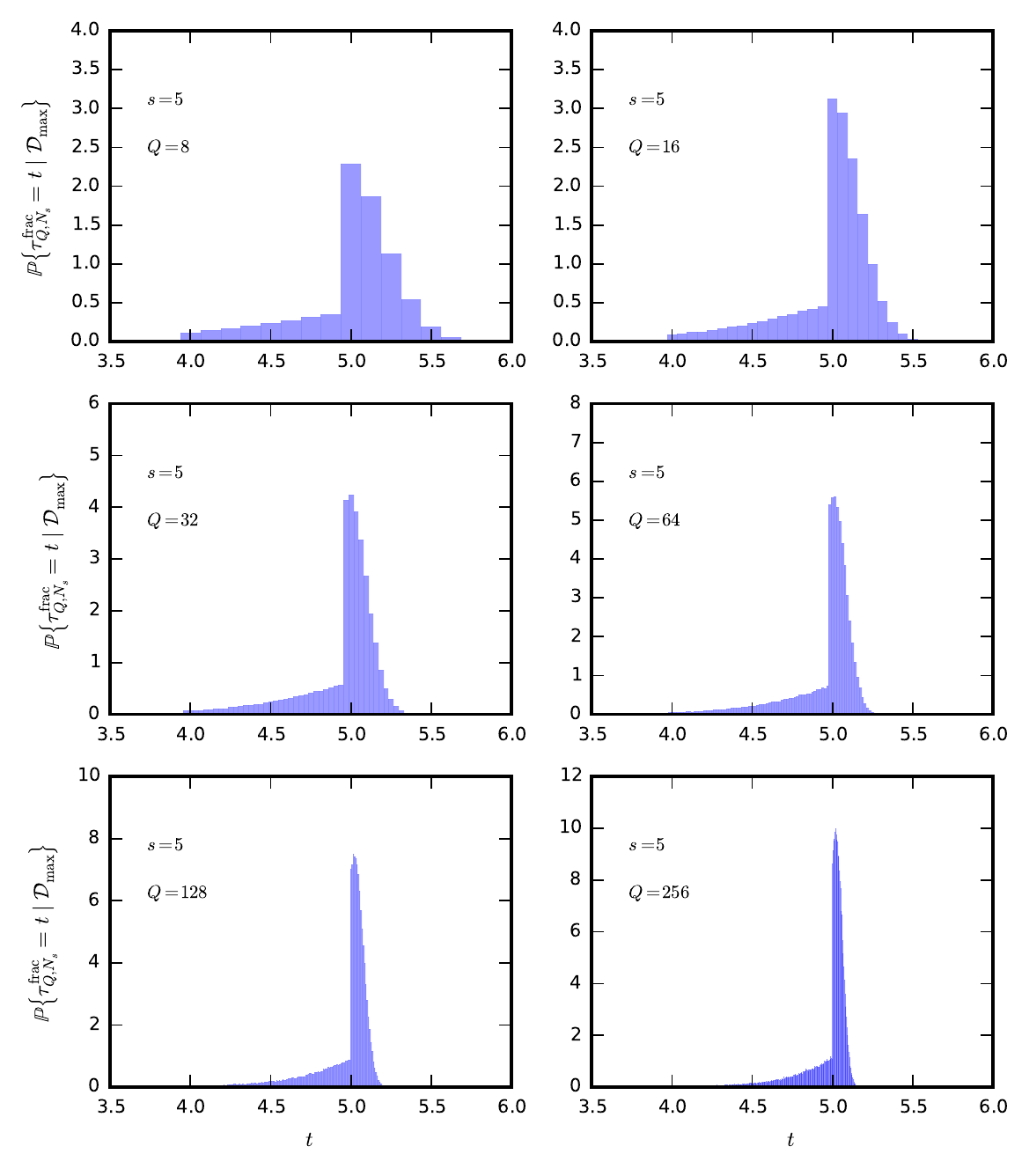}
 \vskip -0.2cm
 \caption{\footnotesize Distribution of $\tau^{\rm\scriptscriptstyle frac}_{Q,N_s}$ for $s=5$, $(\alpha,\nmin)=(2.45,3)$ and $Q=8,\ldots,256$.}
 \label{fig:figsix}
 \vskip -0.2cm
\end{figure}

Upon setting $N\to N_s=h_0Q\mQ^s$, we obtain our final estimate
\begin{empheq}[box=\mybluebox]{align}
  \FQ(x|Q,N_s;\cD_{\max}) \ =\, \hskip -0.1cm \sum_{t=s,s\pm 1, s\pm 2,\ldots}\hskip -0.0cm \dfrac{h_0\mQ^{s-t}\chi(h_0x\mQ^{s-t})}{\int_0^{h_0Q\mQ^{s-t}}\rd y\, \chi(y)}\ \P_{\rm int}\{\tau=t\,|\,\cD_{\max}\}\,,\quad \text{ as } Q,s\to\infty\,,
  \label{eq:fseDmax}  
\end{empheq}
and its asymptotic limit
\begin{align}
  & \FQ(x|Q,N_s;\cD_{\max}) \simeq \frac{\cC(Q)}{\zeta(\alpha-1,\nmin)}\frac{1}{x^\alpha}\,,\qquad \text{ as } x,Q,s\to\infty\,,\qquad \text{with}\nonumber\\[2.0ex]
  & \cC(Q) =\, \frac{\P_{\rm int}\{\tau=s\,|\,\cD_{\max}\}}{\int_0^{h_0Q}\rd y\,\chi(y)} + \frac{1}{\mQ^{\alpha-1}}\frac{\P_{\rm int}\left\{\tau=s-1\,|\,\cD_{\max}\right\}}{\int_0^{h_0Q\mQ}\rd y\,\chi(y)} + \mQ^{\alpha-1}\frac{\P_{\rm int}\left\{\tau=s+1\,|\,\cD_{\max}\right\}}{\int_0^{h_0Q/\mQ}\rd y\,\chi(y)}\,.
  \label{eq:FQmaxthreeterms}
\end{align}

In Fig.~\ref{fig:figsix}, we report the distribution of $\tau^{\rm\scriptscriptstyle frac}_{Q,N_s}$ for $s=5$, $(\alpha,\nmin)=(2.45,3)$ and $Q=8,\ldots,256$. We observe that it is centered around $t=s$. We also notice that the right tail collapses quickly to $t=s$ as $Q$ increases, while the left tail accumulates towards $t=s$ at a slower pace. As a consequence, backward terms in eq.~(\ref{eq:fseDmax}) are more quickly suppressed than forward ones. Simulations of $\tau^{\rm\scriptscriptstyle frac}_{Q,Q\mQ^s}$ at $s=6$ display a rigid shift of the distribution around $t=6$ with no evident change in the structure of the tails. In Table~\ref{tab:tabfour} we report our estimate of $\P_{\rm int}\{\tau=t\,|\,\cD_{\max}\}$, a corresponding estimate of the expectation $\cE[\,x\,|\,Q,N_5;\cD_{\max}]$ and the correction coefficient $\cC(Q)$ for the same choice of parameters. A comparison with Table~\ref{tab:tabthree} shows that $\P_{\rm int}\{\tau=s\,|\,\cD_{\max}\}<\P\{\tau=s\,|\,\cD_{\min}\}$ whereas $\P_{\rm int}\{\tau=t\,|\,\cD_{\max}\}>\P\{\tau=t\,|\,\cD_{\min}\}$ for $t\ne s$. This is reasonable since the system is much less constrained in $\cD_{\max}$ than in $\cD_{\min}$. Notably, $\cE[\,x\,|\,Q,N_s;\cD_{\max}]\to 1$ from below as $Q\to\infty$, whereas $\cE[\,x\,|\,Q,N_s;\cD_{\min}]\to 1$ from above.

\begin{figure}[t!]
  \centering
  \includegraphics[width=0.83\textwidth]{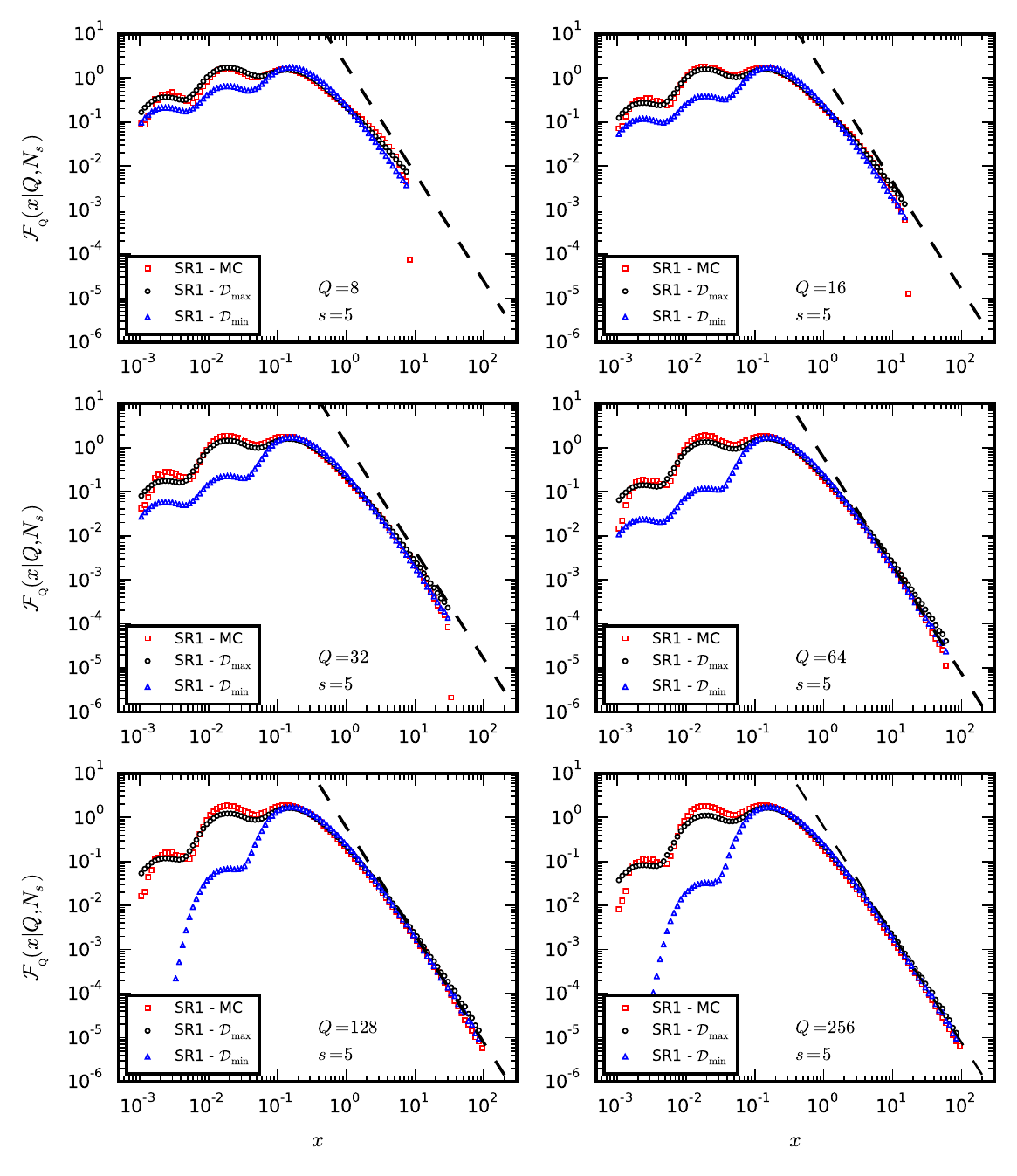}
  \vskip -0.5cm
  \caption{\footnotesize Monte Carlo simulation (MC) of $\FQ(x|Q,N_s)$ with stopping rule SR1 vs. our theoretical estimates of the distribution in $\cD_{\min}$ and $\cD_{\max}$ for $s=5$, $(\alpha,\nmin)=(2.45,3)$, and $Q=8,\ldots,256$. The dashed lines represent the asymptotic limit, eq.~(\ref{eq:FQmaxthreeterms}).}
  \label{fig:figseven}
  \vskip -0.4cm
\end{figure}

In Fig.~\ref{fig:figseven} we compare Monte Carlo simulations of $\FQ(x|Q,N_s)$ under SR1 with our theoretical estimates. For all values of $Q$ the numerical distributions are characterized by three bumps along the left tail, slowly decaying as $Q\to\infty$. These bumps result clearly from the convolution of $\chi(x)$, $\chi(\mQ x)$ and $\chi(\mQ^2 x)$. As can be seen, the bumps are separated by factors of $\mQ$. This confirms qualitatively the correctness of our analysis. The theoretical distribution in $\cD_{\max}$ is in much better agreement with its numerical counterpart than the theoretical distribution in $\cD_{\min}$, specially as $Q\to\infty$. The dashed lines represent the asymptotic limit of the distribution in $\cD_{\max}$ as predicted by eq.~(\ref{eq:FQmaxthreeterms}), with the correction coefficient $\cC(Q)$ as reported in Table~\ref{tab:tabfour}. 

\subsection{$F_{\rm\scriptscriptstyle Q}(x|Q,N)$ with stopping rule SR2}

A second case of practical interest is represented by the algorithm with stopping rule SR2. Under this prescription we run eq.~(\ref{eq:BP2}) up to the earliest level $\texit$ at which the overall number of vertices on all trees exceeds $N$ and we arrest the procedure once {\it all} vertices on the $(\texit-1)$th level  have generated their offspring on the $\texit$th level. In general, we end up with an overall number of vertices $M\ge N$. The \emph{minimal ensemble} $\cD_{\min}$ keeps playing an essential r\^ole in the analytic description of the distribution of the excess--of--votes variable under SR2, yet the adoption of this stopping rule makes the algorithm much less selective than discussed in sects.~3.1 and~3.2. Indeed, the space of all possible vote configurations is now given by
\begin{equation}
  \Sigma_{+}(Q,N) = \left\{(v_1,\ldots,v_Q)\in \dN^Q:\quad \sum_{k=1}^Q v_k \ge N\right\} = \bigcup_{M=N}^\infty \Sigma(Q,M)\,.
\end{equation}
For convenience we also introduce the complement of $\Sigma_{+}(Q,N)$ in $\dN^Q$, namely
\begin{equation}
  \Sigma_{-}(Q,N) = \left\{(v_1,\ldots,v_Q)\in \dN^Q:\quad \sum_{k=1}^Q v_k < N\right\} = \bigcup_{M=1}^{N-1} \Sigma(Q,M)\,.
\end{equation}
For all $T\in\fT$ there exists $t_+\ge 1$ such that $\left(V_{t_+}(T^{(1)}),\ldots,V_{t_+}(T^{(Q)})\right)\in\Sigma_{+}(Q,N)$. In other words, we have $T\in \bigcup_{M=N}^{\infty} \cD_{t_+}(Q,M)$ for some $t_+$. Similarly, for all $T\in\fT$ there exists $t_-\ge 0$ such that $\left(V_{t_-}(T^{(1)}),\ldots,V_{t_-}(T^{(Q)})\right)\in \Sigma_{-}(Q,N)$, provided $N\ge Q$. In other words, we have  $T\in \bigcup_{M=1}^{N-1} \cD_{t_-}(Q,M)$ for some $t_-$. It follows that
\begin{align}
  \hskip -0.1cm\bigcup_{t=0}^\infty\bigcup_{M=N}^\infty \cD_t(Q,M) = \bigcup_{M=N}^\infty \cD_{\min}(Q,M) = \fT\quad \text{and}\quad 
  \bigcup_{t=0}^\infty\bigcup_{M=1}^{N-1} \cD_t(Q,M) = \bigcup_{M=1}^{N-1} \cD_{\min}(Q,M) = \fT\,,
  \label{eq:fTdecompSR2}
\end{align}
independently of $N$, provided $N\ge Q$. Since we request that the algorithm should stop when the overall number of vertices exceeds $N$, the stopping time is correctly described by the variable $\tau_{Q,N}$ introduced in eq.~(\ref{eq:stoptime}). Nevertheless, different values of ${M\ge N}$ are realized with different probabilities, hence we must take into account the way $\sum_{i=1}^Q V^{(i)}_{\texit}=M$ distributes under the condition that $\tau_{Q,N}=\texit$. Specifically, given $(Q,N)$ and $x=\ell Q/N$ with ${\ell=1,2,\ldots,N}$, the discrete probability that the excess--of--votes variable yields $x$ is given by
\begin{equation}
\left\{\begin{array}{l} F_{\rm\scriptscriptstyle Q}(x|Q,N) = \sum_{t=0}^\infty\, F_{\rm\scriptscriptstyle Q}(x|Q,N,t)\, \P\{\tau_{Q,N}=t\}\,,\\[2.0ex]
F_{\rm\scriptscriptstyle Q}(x|Q,N,t) = \displaystyle{\sum_{M=N}^\infty\, \sum_{v\in\Sigma(Q,M)} \delta_{Mx/Q,v_i}\,\Pi_t(v|Q,M)\,\P\biggl\{\, \sum_{j=1}^Q V^{(j)}_t=M\,\biggl|\,\tau_{Q,N}=t \,\biggr\}}\,.
\end{array}\right.
  \label{eq:FproblawSR2}
\end{equation}
Since in general $Mx/Q$ is not an integer, the above expression has to be interpreted as a linear interpolation between the two distributions obtained by replacing $\delta_{Mx/Q,v_i} \to \delta_{\lfloor Mx/Q \rfloor,v_i}$ and $\delta_{Mx/Q,v_i} \to \delta_{\lceil Mx/Q\rceil,v_i}$. The structure of eq.~(\ref{eq:FproblawSR2}) suggests that we discuss first $\P\{\tau_{Q,N}=t\}$ and then $F_{\rm\scriptscriptstyle Q}(x|Q,N,t)$. From our analysis in sects.~3.1 and~3.2 we already know that both quantities have to be eventually evaluated for $N\to N_s = h_0Q\mQ^s$. 

First of all, the probability law of the stopping time $\P\{\tau_{Q,N}=t\}$ is not restricted here to any proper subset of~$\fT$, as it was instead in eqs.~(\ref{eq:PtauDmin}) and~(\ref{eq:PtauDmax}). From eq.~(\ref{eq:fTdecompSR2}) we see that it amounts to
\begin{align}
  \P\{\tau_{Q,N}=t\} \, & = \, \P\left\{\sum_{i=1}^Q V^{(i)}_{t-1}<N\,,\,\sum_{i=1}^Q V^{(i)}_{t}\ge N\right\}\nonumber\\[1.0ex]
  & =  \P\left\{T\,\in\ \left(\bigcup_{M=1}^{N-1}\cD_{t-1}(Q,M)\right) \cap \left(\bigcup_{M=N}^{\infty}\cD_{t}(Q,M)\right)\right\}\,.
\end{align}
At first sight, the limit of $\P\{\tau_{Q,N_s}=t\}$ as $s\to\infty$ is not obvious. To shed light on this, we notice that
\begin{align}
  & \P\{\tau_{Q,N_s}=t\}\, =\, \P\left\{\sum_{i=1}^Q V^{(i)}_{t-1}< h_0Q\mQ^s\,,\,\sum_{i=1}^Q V^{(i)}_t\ge h_0Q\mQ^s\right\} \nonumber\\[1.0ex]
  & = \P\left\{\bar H^{[Q]}_{t-1}<h_0\mQ^{s-t+1}\,,\,\bar H^{[Q]}_t\ge h_0\mQ^{s-t}\right\}\to\P\left\{ h_0\mQ^{s-t} \le \bar H^{[Q]}<h_0\mQ^{s-t+1}\right\}\,, \quad \text{as } s\to\infty\,.
\end{align}

\begin{figure}[t!]
  \centering
  \includegraphics[width=0.85\textwidth]{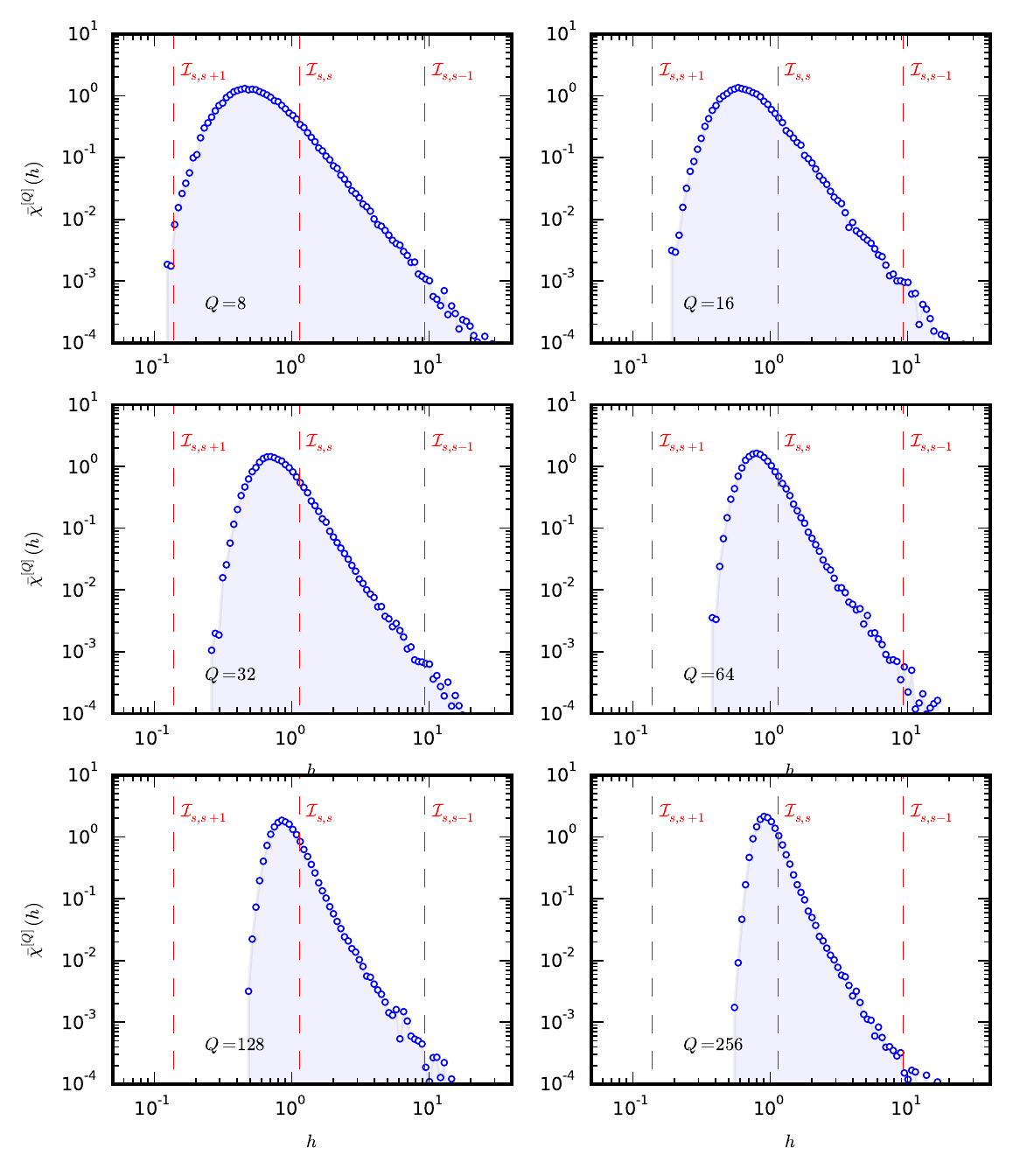}
  \vskip -0.4cm
  \caption{\footnotesize Distribution of $\bar H^{[Q]}$ for $(\alpha,\nmin)=(2.45,3)$ and $Q=8,\ldots,256$. The dashed lines correspond to $h=h_0\mQ^{-1},h_0\mQ^0,h_0\mQ^1$. They separate sectors $\cI_{s,t}$, $t=s,s\pm 1$.}
  \label{fig:figeight}
  \vskip -0.3cm
\end{figure}
\noindent We thus see that the probability law of the stopping time in the thermodynamic limit is related to the continuous distribution of $\bar H^{[Q]}$. More precisely, the domain of the latter splits into contiguous sectors $\cI_{s,t} = (h_0\mQ^{s-t},h_0\mQ^{s-t+1}]$ with ${t=s,s\pm 1,s\pm 2,\ldots}$. The mass of $\bar H^{[Q]}$ in $\cI_{s,t}$ measures the probability of $\tau_{Q,N_s} = t$ as $s\to\infty$. In Fig.~\ref{fig:figeight} we show the distribution of $\bar H^{[Q]}$ obtained from numerical simulations of $\bar H^{[Q]}_7$ (which is very close to $\bar H^{[Q]}$) for $(\alpha,\nmin)=(2.45,3)$ and $Q=8,\ldots,256$. The probability mass lies essentially in the sectors $\cI_{s,s-2},\cI_{s,s-1},\,\cI_{s,s},\,\cI_{s,s+1}$. This is analogous to what we observed in sects.~3.1 and~3.2 and shows that the sum over $t$ in eq.~(\ref{eq:FproblawSR2}) can be restricted with very good approximation to $t=s-2,s-1,s,s+1$. Secondly, the distribution of $\bar H^{[Q]}$ narrows as $Q$ increases. There is a simple explanation for this behaviour. Recall that $\E[\bar H^{[Q]}] = h_0$ independently of $Q$. Consider also that $H$ has a power--law tail $1/h^\alpha$, as eq.~(\ref{eq:chi}) shows, hence it has infinite variance as long as $\alpha<3$. Therefore, $H$ is similar to a Paretian variable. The generalized central limit theorem for such variables (see refs.~\cite[p.~50]{stableone} and \cite[p.~62]{stabletwo}) guarantees the existence of a (cumulative) L\'evy stable distribution $G(h;\alpha,\nmin)$ such that 
\vskip -0.7cm
\begin{equation}
  \lim_{Q\to\infty} \P\left\{\frac{1}{Q^{1/(\alpha-1)}}\left(\sum_{i=1}^Q H^{(i)} - h_0Q\right)<h\right\} = G(h;\alpha,\nmin)\,.
  \label{eq:levystable}
\end{equation}
\vskip -0.4cm
\noindent We have checked numerically the convergence in distribution of $(\sum_{i=1}^Q H^{(i)}-h_0Q)/Q^{1/(\alpha-1)}$ as $Q\to\infty$. Since $\alpha-1>1$, an immediate consequence of eq.~(\ref{eq:levystable}) is that
\begin{equation}
  \lim_{Q\to\infty} \bar\chi^{[Q]}(h)\, =\, \delta(h-h_0)\,.
  \label{eq:barchidelta}
\end{equation}
In particular, the whole probability mass of $\bar H^{[Q]}$ shifts eventually to the sectors $\cI_{s,s}$ and $\cI_{s,s+1}$ as $Q\to\infty$. Accordingly, we have
\begin{equation}
  \lim_{Q\to\infty}\lim_{s\to\infty}\P[\tau_{Q,N_s}=t] = \delta_{t,s}\pi_{s,s} + \delta_{t,s+1}\pi_{s,s+1}\,,\qquad \text{with}\quad \pi_{s,s} + \pi_{s,s+1} = 1\,.
  \label{eq:taudelta}
\end{equation}
\vskip -0.3cm
\noindent It is not easy to determine $\pi_{s,s}$ and $\pi_{s,s+1}$, since the singularity of $\bar\chi^{[Q]}(h)$ as $Q\to\infty$ lies precisely at the common boundary of $\cI_{s,s+1}$ and $\cI_{s,s}$ (this does not represent a problem for the calculation of $F_{\rm\scriptscriptstyle Q}(x|Q,N)$, as we shall see in the sequel). In Table~\ref{tab:tabfive} we quantify the probability of $\tau_{Q,N_s}$ in the large--list limit, for $(\alpha,\nmin)=(2.45,3)$ and $Q=8,\ldots,2048$, via numerical integration of the distributions shown in Fig.~\ref{fig:figeight}. At very small $Q$ a large fraction of the probability mass falls in the sector $\cI_{s,s+1}$; the rest lies essentially within $\cI_{s,s}$, with a residual fraction belonging to $\cI_{s,s-1}$ and $\cI_{s,s-2}$. Things change smoothly as $Q$ increases in accordance with the above discussion: the probability mass shifts progressively from $\cI_{s,s+1}$ to $\cI_{s,s}$, with $\cI_{s,s-1}$ and $\cI_{s,s-2}$ becoming increasingly marginal.  
\begin{table}[t!]
  \small
  \centering
  \begin{tabular}{cr|cccc|}
    & & \multicolumn{4}{c|}{$t$} \\[0.2ex]
     \multicolumn{2}{c|}{$\P\{\bar H^{[Q]}\in\cI_{s,t}\}$}  & $s-2$ & $s-1$ & $s$ & $s+1$ \\
    \hline\\[-3.2ex]
    \multirow{9}{*}{\rotatebox{0}{$\phantom{QQ}Q$}} &          8 & 0.00024(6) & 0.0064(3) & 0.227(2) & 0.766(2) \\
                                                          &   16 & 0.00007(5) & 0.0042(4) & 0.244(3) & 0.752(3) \\[0.0ex]
                                                          &   32 & 0.00003(2) & 0.0033(2) & 0.257(2) & 0.739(2) \\[0.0ex]
                                                          &   64 & \text{n/a} & 0.0020(4) & 0.271(2) & 0.727(2) \\[0.0ex]
                                                          &  128 & \text{n/a} & 0.0016(1) & 0.276(2) & 0.722(2) \\[0.0ex]
                                                          &  256 & \text{n/a} & 0.0012(1) & 0.283(2) & 0.716(2) \\[0.0ex]
                                                          &  512 & \text{n/a} & 0.0008(2) & 0.285(3) & 0.717(4) \\[0.0ex]
                                                          & 1024 & \text{n/a} & 0.0006(2) & 0.293(4) & 0.707(4) \\[0.0ex]
                                                          & 2048 & \text{n/a} & 0.0003(1) & 0.298(4) & 0.702(4) \\[0.0ex]
    \hline
  \end{tabular}
  \caption{\footnotesize Probability of $\bar H^{[Q]}$ for $(\alpha,\nmin)=(2.45,3)$ and $Q=8,\ldots,2048$. Bootstrap uncertainty is quoted in parentheses. N/a entries correspond to measurements for which noise/signal~${\ge 1}$.}
  \label{tab:tabfive}
  \vskip -0.4cm
\end{table}

We now go back to the conditional vote distribution $F_{\rm\scriptscriptstyle Q}(x|Q,N,t)$ given $t$. We have already noticed that it receives contributions from all $Q$--forests $T\in \bigcup_{M=N}^\infty\cD_t(Q,M)$. By performing the same algebra as we did in sect.~3.1, we turn eq.~(\ref{eq:FproblawSR2}) into
\begin{align}
 F_{\rm\scriptscriptstyle Q}(x|Q,N,t) & = \sum_{M=N}^\infty\,  p_t\left( \frac{Mx}{Q}\right)\ \left[{\dfrac{\int_{-\pi}^{\pi}{\rm d}\lambda\, {\rm e}^{-\ri\lambda\frac{M}{Q} \left(Q - x\rfloor\right) }\prod_{j\ne i}^{1\ldots Q}\sum_{v_j=0}^M  {\rm e}^{i\lambda v_j}p_t(v_j)}{\int_{-\pi}^{\pi}{\rm d}\lambda\, {\rm e}^{-i\lambda M} \prod_{j=1}^{ Q}\sum_{v_j=0}^M  {\rm e}^{i\lambda v_j}p_t(v_j) }}\right] \nonumber\\[0.0ex]
   &  \times\, \P\biggl\{\,\sum_{j=1}^{Q} V^{(j)}_t = M \biggr| \tau_{Q,N} = t\,\biggr\}\,.
\end{align}
Provided $x<Q$, we get again highly oscillatory integrals at numerator and denominator of the ratio in square brackets. Therefore, in the large--list limit the above expression converges quickly to
\begin{equation}
  F_{\rm\scriptscriptstyle Q}(x|Q,N,t) = \sum_{t=0}^{\infty} \sum_{M=N}^\infty  \frac{p_t\left( \frac{Mx}{Q} \right)}{\sum_{v_i=0}^M p_t(v_i)}\,\P\left\{\sum_{j=1}^{Q} V^{(j)}_t = M | \tau_{Q,N} = t\right\}\,,\qquad \text{ as } Q,N\to\infty\,.
\end{equation}
\noindent Now we set $N\to N_s=h_0Q\mQ^s$ just as we did in sect.~3.1. We also find it convenient to set $M = N_s + \ell$, with $\ell$ ranging in principle from zero to infinity. Actually, the probability that $\sum_{j=1}^Q V^{(j)}_t=M$ given that $\tau_{Q,N_s}=t$ is wholly confined in the region $M\le \mQ N_s$. Indeed, the equation $\sum_{j=1}^Q V^{(j)}_t=M$ is equivalent to $\bar H^{[Q]}_t = M/Q\mQ^t$, while the condition $\tau_{Q,N_s} = t$ is equivalent to $h_0\mQ^{s-t}\le \bar H_t^{[Q]}\le h_0\mQ^{s-t+1}$ if $t$ is sufficiently large. It follows that $M\le h_0Q\mQ^{s+1} = \mQ N_s$. Therefore, we can restrict the sum over $\ell$ to the range $\{0,1,\ldots, (\mQ-1)N_s\}$. From the scaling equation
\begin{align}
  p_t\left( {Mx}/{Q}\right) & = p_t\left( h_0x\mQ^s +  {x\ell}/{Q}\right) 
   = p_t\left(m^t \left[h_0x\mQ^{s-t} +  {x\ell}/{Q\mQ^t}\right]\right) \nonumber\\[1.0ex]
   &  \to \left(h_0\mQ^{s-t} + {\ell}/{Q\mQ^t}\right) \chi \left(h_0x\mQ^{s-t} +  {x\ell}/{Q\mQ^t}\right)\rd x\,\qquad \text{as }\ s\to\infty\,,
\end{align}
we obtain
\begin{align}
  \FQ(x|Q,N_s,t) & = \sum_{\ell = 0}^{(\mQ-1)N_s} \left(h_0\mQ^{s-t} + \frac{\ell}{Q\mQ^t}\right) \frac{\chi\left(h_0x\mQ^{s-t}+\frac{x\ell}{Q\mQ^t}\right)}{\int_0^{h_0Q\mQ^{s-t} + \ell/\mQ^t}\rd y\, \chi(y)}\nonumber\\[0.0ex]
  & \ \ \times\, \P\biggl\{\,\sum_{j=1}^{Q} V^{(j)}_t = N_s + \ell\, \biggl|\,\tau_{Q,N_s} = t\,\biggr\}\,,\qquad \text{as }\ Q,s\to\infty\,.
\end{align}
The above expression is not in its final form yet, as we are not considering that the discrete sum over $\ell$ converges to a Riemann integral in the thermodynamic limit. Specifically, from the scaling law 
\begin{align}
& \P\biggl\{\sum_{j=1}^Q V^{(j)}_t = N_s + \ell\,\biggr|\, \tau_{Q,N_s} =t\biggr\} = \P\left\{\bar H_t^{[Q]} = h_0\mQ^{s-t} + \frac{\ell}{Q\mQ^t}\,\biggr|\, \tau_{Q,N_s} = t\right\}\nonumber\\[0.0ex]
& \hskip 2.5cm \to\ \bar\chi^{[Q]}\left(h_0\mQ^{s-t} + h\,\bigr|\, \bar H^{[Q]}\in \cI_{s,t}\right)\rd h \qquad  \text{as } \ s\to\infty\,,
\end{align}
it follows that
\begin{equation}
 \FQ(x|Q,N_s,t) = \int_{\cI_{s,t}} \rd h\ \frac{h\,\chi(hx)}{\int_0^{hQ}\rd y\,\chi(y)}\,\bar\chi^{[Q]}\left(h\,\bigr|\, \bar H^{[Q]}\in \cI_{s,t}\right)\,,\qquad \text{ as } Q,s\to\infty\,.
  \label{eq:FtSR2}
\end{equation}
\noindent The distribution on the r.h.s. of eq.~(\ref{eq:FtSR2}) is correctly normalized, as can be seen upon integrating both sides over ${x\in [0,Q]}$. We see from the above equation that the main effect of the stopping rule SR2 is to smooth the distributions $h_0\mQ^{s-t}\chi(h_0\mQ^{s-t}x)/\int_0^{h_0Q\mQ^{s-t}}\rd y\,\chi(y)$ contributing to eqs.~(\ref{eq:fse}) and~(\ref{eq:fseDmax}) via a convolution with the conditional distribution ${\bar\chi^{[Q]}(h\,|\,\bar H^{[Q]}\in \cI_{s,t})}$.

We finally insert eq.~(\ref{eq:FtSR2}) into eq.~(\ref{eq:FproblawSR2}). We notice that since $\bar\chi^{[Q]}(h\,|\,\bar H^{[Q]}\in\cI_{s,t}) = 0$ if $h\notin\cI_{s,t}$, the integral in eq.~(\ref{eq:FtSR2}) can be extended to $h\in(0,\infty)$ without affecting its value. Accordingly, we obtain our final estimate
\begin{empheq}[box=\mybluebox]{align}
  \FQ(x|Q,N_s) & = \sum_{t=0}^\infty\, \FQ(x|Q,N_s,t)\,\P\{\tau_{Q,N_s}=t\}\nonumber\\[1.0ex]
  & = \int_0^\infty \rd h\  \frac{h\,\chi(hx)}{\int_0^{hQ}\rd y\,\chi(y)}\,\sum_{t=0}^\infty \bar\chi^{[Q]}\left(h\,\bigr|\, \bar H^{[Q]}\in \cI_{s,t}\right)\,\P[\tau_{Q,N_s}=t] \nonumber\\[2.0ex]
  & = \int_0^\infty \rd h\  \frac{h\,\chi(hx)}{\int_0^{hQ}\rd y\,\chi(y)}\,\bar\chi^{[Q]}(h)\,,\qquad \text{ as } Q,s\to\infty\,.
\label{eq:fseSR2}
\end{empheq}
From eq.~(\ref{eq:barchidelta}) it follows that
\begin{equation}
  \lim_{Q\to\infty}\lim_{s\to\infty}\FQ(x|Q,N_s) =  h_0\,\chi(h_0x)\,.
  \label{eq:fseSR2asymp}
\end{equation}
\begin{figure}[t!]
  \centering
  \includegraphics[width=0.80\textwidth]{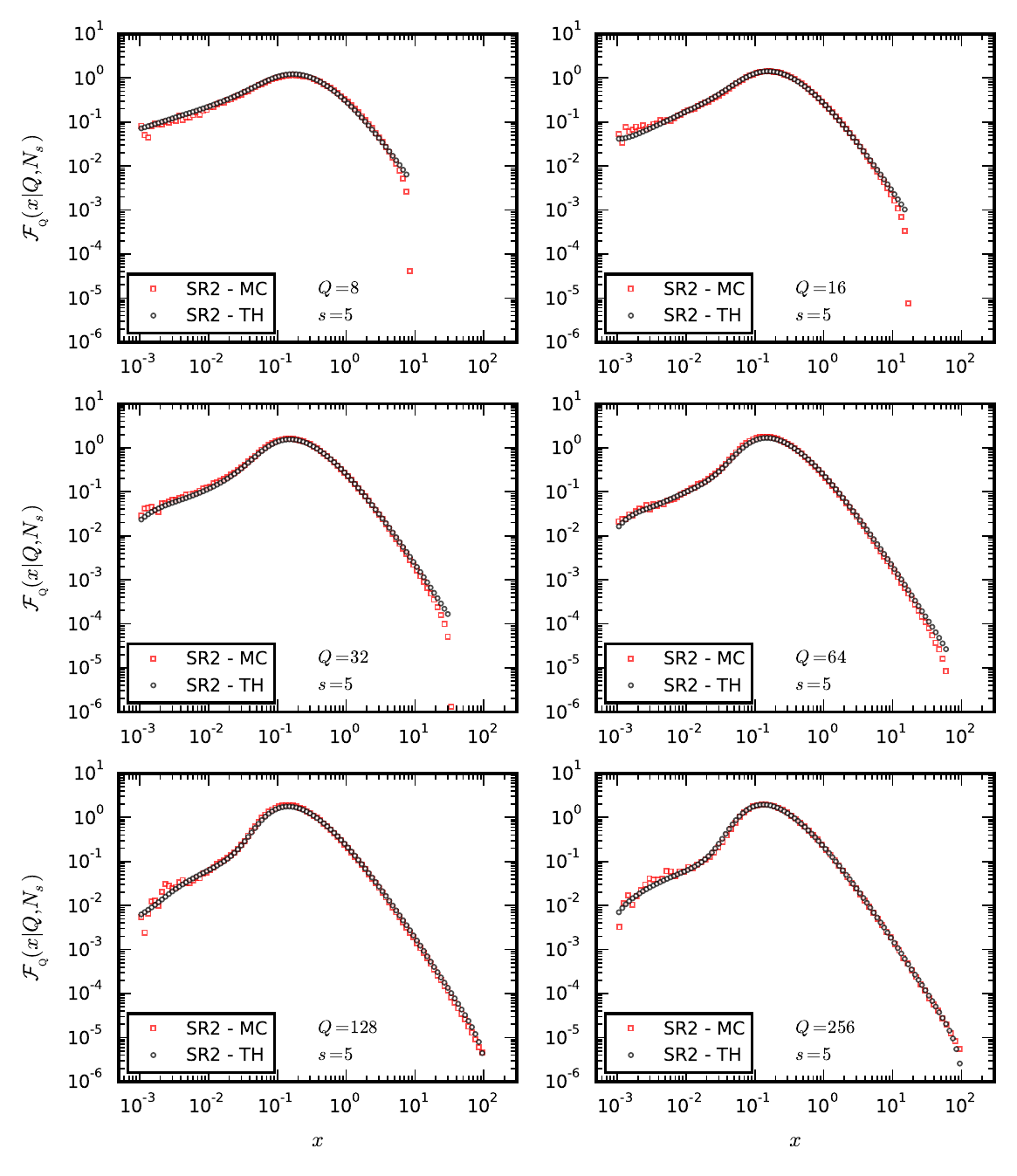}
  \vskip -0.5cm
  \caption{\footnotesize Monte Carlo simulation (MC) of $\FQ(x|Q,N_s)$ with stopping rule SR2 vs. theoretical estimate (TH) for $(\alpha,\nmin)=(2.45,3)$, $s=5$ and $Q=8,\ldots,256$.}
  \label{fig:fignine}
  \vskip -0.4cm
\end{figure}
\vskip -0.5cm
\noindent The power--law regime of the limit distribution is reached once more as soon as $x\gtrsim h_0^{-1}\simeq 1$. Eq.~(\ref{eq:fseSR2asymp}) tells us that the vote distribution in the large--list limit is universal under a change of stopping rule. 
\vskip -0.0cm
Notice that both functions $\chi(x)$ and $\bar\chi^{[Q]}(h)$ are known only for a finite number of points, depending on the choice of the bins in the histograms which represent them, see Figs.~\ref{fig:figfive} and~\ref{fig:figeight}. Eq.~(\ref{eq:fseSR2}) can be integrated numerically provided we interpolate $\chi(x)$ and $\bar\chi^{[Q]}(h)$ between subsequent observations. We show in Fig.~\ref{fig:fignine} a comparison between the Monte Carlo simulation of $\FQ(x|Q,N_s)$ and eq.~(\ref{eq:fseSR2}). The three bumps characterizing the distribution with stopping rule SR1 are now absent due to the convolution with $\bar\chi^{[Q]}(h)$. We conclude that only the left tail of the distribution is sensitive to the stopping rule adopted\footnote{This is analogous to the sensitivity of the left tail to the choice of the model parameters $r$ and $\kmin$, firstly observed by FC in ref.~\cite{fcscaling}.}. The agreement between numerical and theoretical results is very good at all scales.

\subsection{$F_{\rm\scriptscriptstyle Q}(x|Q,N)$ with stopping rule SR3}

We finally consider the algorithm with stopping rule SR3. According to this prescription, given $(Q,s)$ we stop the generation of new vertices as soon as all variables $(Z^{(i)}_t)_{i=1}^Q$ have been realized for $t\le s$. Under SR3 the stopping time fulfills $\tau_{Q,M}=s$ with certainty, whereas the final number of vertices $M$ on all trees fluctuates freely. In particular, we know that $M\sim\text{O}(h_0Q\mQ^s)$. Therefore, in order to calculate $F_{\rm\scriptscriptstyle Q}(x|Q,s)$ it is sufficient that we properly take the fluctuations of $\sum_{j=1}^Q V^{(j)}_s=M$ into account. Specifically, we let ${x_+=\ell Q/\lfloor h_0Q\mQ^s\rfloor}$ with $\ell=1,2,\ldots,\lfloor h_0Qm^s\rfloor$ and ${x_-=\ell Q/\lceil h_0Q\mQ^s\rceil}$ with $\ell=1,2,\ldots,\lceil h_0Qm^s\rceil$ denote all possible values of the excess--of--votes variable. The discrete probability that the latter yields $x$ is obtained by interpolating the distribution
\begin{equation}
  F_{\rm\scriptscriptstyle Q}(x|Q,s) = \sum_{M=0}^\infty\, \sum_{v\in\Sigma(Q,M)}\delta_{Mx/Q,v_i}\,\Pi_s(v|Q,M)\,\P\biggl\{\,\sum_{j=1}^QV^{(j)}_s = M\, \biggr|\,\tau_{Q,M} = s\biggr\}\,
  \label{eq:ProblawSR3}
\end{equation}
between its values at the nearest points $x_\pm$ of $x$. Since the event $\tau_{Q,M}=s$ implies that ${\sum_{j=1}^QV^{(j)}_s \ge M}$, the conditional probability $\P\{\,\sum_{j=1}^QV^{(j)}_s = M\,|\,\tau_{Q,M} = s \}$ can be safely replaced by its full counterpart, namely $\P\{\,\sum_{j=1}^QV^{(j)}_s = M\}$. Just as in sect.~3.3, the thermodynamic limit is taken by letting $s\to\infty$ after setting $M=h_0 Q\mQ^s + \ell$. Notice that this time $\ell$ ranges over the set $\{-h_0Q\mQ^s, -h_0Q\mQ^s+1,\ldots,\infty\}$. Repeating the calculation should be at this point trivial and tedious. The reader can easily get convinced that
\begin{empheq}[box=\mybluebox]{align}
  \cF_{\rm\scriptscriptstyle Q}(x|Q,s) = \int_0^\infty \rd h\ \frac{h\,\chi(hx)}{\int_0^{hQ}\rd y\,\chi(y)}\,\bar\chi^{[Q]}(h)\,,\qquad \text{ as } Q,s\to\infty\,.
\end{empheq}
We conclude that SR2 and SR3 are equivalent in the large--list limit. Numerical simulations confirm the equivalence with very good accuracy. 

\section{Exponential scaling in the FC model}

The approach we followed to work out the large--list limit of the vote distribution in the quenched model relies ultimately on the exponential scaling of Galton--Watson trees. This led us to perform the thermodynamic limit along the sequence $N_s=h_0Q\mQ^s$, $s=1,2,3,\ldots,\infty$, which corresponds to having the $s$th level of the $Q$--forest generated in the algorithm fully realized on average. Given that FC trees are not Galton--Watson, the reader may wonder whether our analysis can be extended to the original FC model. Specifically, since the levels of a FC tree are progressively populated at subsequent times according to eq.~(\ref{eq:FC}), one could suspect that the overall number of preferences generated at a given time $t$ on a single tree follows a super--exponential behaviour as a function of $t$. To get confident that this is not the case, it suffices to observe that $Y^{(i)}_n(t)\le X^{(i)}_n$ for all $t$ and that $X^{(i)}_n$ scales exponentially with $n$ (see Remark 2.1).  The exponential growth rate of $V_{\rm\scriptscriptstyle FC}(t)$ can be read off from its expectation value: all we have to do to identify it is to work out the averages $\E[Y^{(i)}_n(t)]$ for $n=1,2,\ldots,t$ and add them all. Before we embark on this calculation, we recall that in sect.~2 we let $\mB$ denote the average offspring of trees in $\cP_{\rm\scriptscriptstyle B}(\alpha,r,\kmin)$. If, in addition, we let $\hatmQ$ denote the average offspring of trees in $\cP_{\rm\scriptscriptstyle Q}(\alpha,\kmin)$, then we have approximately $\mB = r\hatmQ$. This relation is not exact since it leaves out the effects of fluctuations, yet it is fulfilled with more than acceptable accuracy. We shall use it several times in the next few lines. Incidentally we notice that $\hatmQ$ is just the growth rate of $X^{(i)}_n$, \ie $\E[X^{(i)}_n] = \hatmQ^n$.

\begin{figure}[t!]
  \centering
  \includegraphics[width=0.9\textwidth]{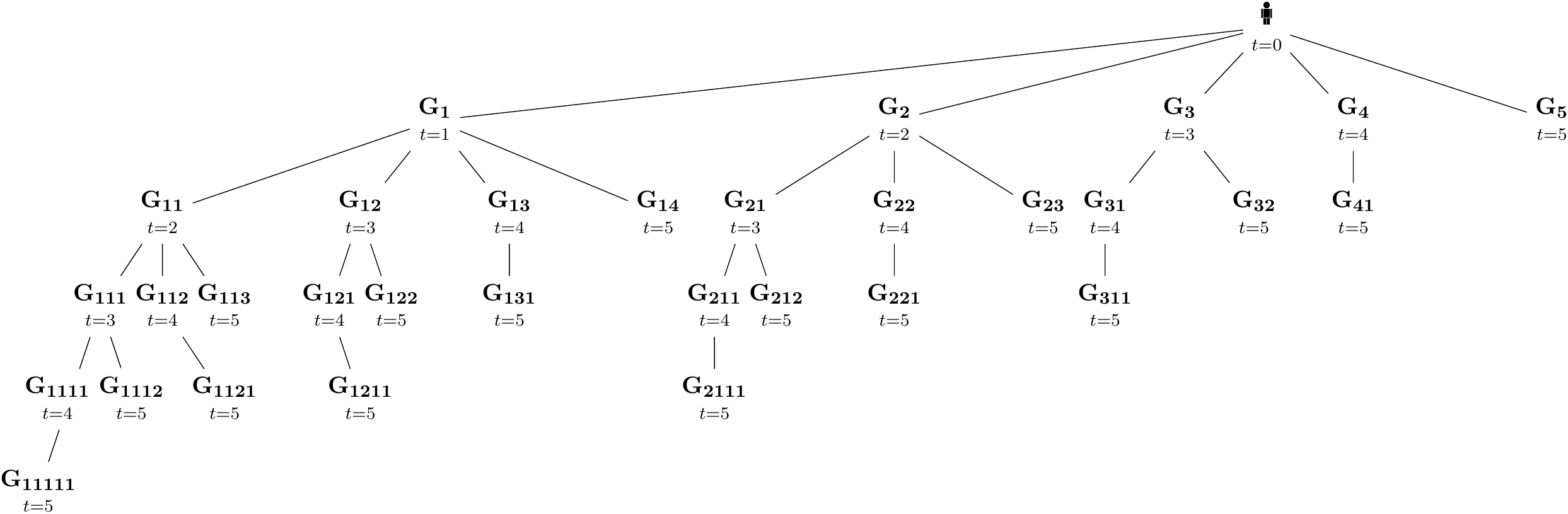}
  \vskip -0.3cm
  \caption{\footnotesize Structure of a generic FC tree at time $t=5$.}
  \label{fig:figten}
  \vskip -0.3cm
\end{figure}

In Fig.~\ref{fig:figten} we show the structure of a generic FC tree at time $t=5$. The candidate persuades a first group $G_1$ of agents at time $t=1$, a second group $G_2$ at $t=2$ and so forth. All groups $\{G_k\}_{k=1,2,\ldots}$ lie on the 1st level of the tree. Agents belonging to $G_1$ persuade in turn a first group of agents $G_{11}$ at time $t=2$, a second group $G_{12}$ at $t=3$, \emph{etc}. All groups $\{G_{ij}\}_{i,j=1,2,\ldots}$ lie on the 2nd level of the tree. In full generality, agents belonging to one of $\{G_{i_1i_2\ldots i_n}\}_{i_1,\ldots,i_n=1,2,\ldots}$ lie on the $n$th level of the tree; they have been persuaded at time $t=i_1+\ldots+i_n$; given $k\ge 1$, they persuade at time $t+k$ a new group of agents $G_{i_1i_2\ldots i_n k}$ lying on the $(n+1)$th level of the tree.

The average size of $G_1$ can be calculated at a glance: since the candidate has $\hatmQ$ neighbours on average and each of these is persuaded with probability $r$, it follows that $\E[|G_1|] = r\hatmQ = \mB$. The average size of $G_2$ can be easily calculated too: at time $t=2$ there are on average $(1-r)\hatmQ$ undecided agents on the 2nd level and each of these is equally persuaded with probability $r$, hence $\E[|G_2|]=\mB(1-r)$. By the same argument we can prove that $\E[|G_3|] = \mB (1-r)^2$, \emph{etc}. In full generality we have $\E[|G_k|] = \mB(1-r)^{k-1}$ for all groups lying on the 1st level. We can similarly calculate the average size of $G_{k1}$ with $k=1,2,\ldots$ To this aim we observe that each agent in $G_k$ has $\hatmQ$ neighbours on average and, once more, each of these is persuaded with probability~$r$. Therefore $\E[|G_{k1}|]=\mB\E[|G_{k}|] = \mB^2(1-r)^{k-1}$. As a consequence, we have $\E[|G_{k2}|] = r(\hatmQ|G_k| - |G_{k1}|) = \mB^2(1-r)^k$. We infer that $\E[|G_{ij}|] = \mB^2(1-r)^{i+j-2}$ holds in general for all groups lying on the 2nd level. It should be clear by now that a simple diagrammatic rule determines $\E[|G_{i_1\ldots i_n}|]$, namely
\vskip 0.3cm
\begin{center}
\framebox[0.8\textwidth]{Each subindex $k=1,2,\ldots$ contributes to $\E[|G_{i_1\ldots i_n}|]$ by a power of $\mB(1-r)^{k-1}$.}
\end{center}
\vskip 0.1cm
It follows that
\begin{equation}
  \E[|G_{i_1\ldots i_n}|] = \mB^n(1-r)^{i_1+\ldots+i_n - n}\,.
\end{equation}
For later convenience, we list below the average size of all groups shown in Fig.~\ref{fig:figten}:
\vskip 0.2cm
\noindent\underline{\bf 1st level}
\begin{align}
  & \E[|G_1|] = \mB\,,\quad \E[|G_2|] = \mB(1-r)\,,\quad \E[|G_3|] = \mB(1-r)^2\,,\nonumber\\[2.0ex]
  & \E[|G_4|] = \mB(1-r)^3\,,\quad \E[|G_5|] = \mB(1-r)^4\,,
\end{align}
\vskip 0.1cm\noindent\underline{\bf 2nd level}
\begin{align}
  & \E[|G_{11}|] = \mB^2\,,\quad \E[|G_{12}|] = \mB^2(1-r)\,,\quad \E[|G_{13}|] = \mB^2(1-r)^2\,,\quad \E[|G_{14}|] = \mB^2(1-r)^3\,,\nonumber\\[2.0ex] 
  & \E[|G_{21}|] = \mB^2(1-r)\,,\quad \E[|G_{22}|] = \mB^2(1-r)^2\,,\quad \E[|G_{23}|]=\mB^2(1-r)^3\nonumber\\[1.0ex]
  & \E[|G_{31}|] = \mB^2(1-r)^2\,,\quad \E[|G_{32}|] = \mB^2(1-r)^3\,,\quad \E[|G_{41}|] = \mB^2(1-r)^3\,,
\end{align}
\vskip 0.1cm\noindent\underline{\bf 3rd level}
\begin{align}
  & \E[|G_{111}|] = \mB^3\,,\quad\E[|G_{112}|] = \mB^3(1-r)\,,\quad\E[|G_{113}|]=\mB^3(1-r)^2\,,\nonumber\\[1.0ex]
  & \E[|G_{121}|] = \mB^3(1-r)\,,\quad\E[|G_{122}|] = \mB^3(1-r)^2\,,\quad\E[|G_{131}|]=\mB^3(1-r)^2\,,\nonumber\\[1.0ex]
  & \E[|G_{211}|] = \mB^3(1-r)\,,\quad\E[|G_{212}|] = \mB^3(1-r)^2\,,\quad\E[|G_{221}|]=\mB^3(1-r)^2\,,\nonumber\\[1.0ex]
  & \E[|G_{311}|] = \mB^3(1-r)^2\,,
\end{align}
\vskip 0.1cm\noindent\underline{\bf 4th level}
\begin{align}
  & \E[|G_{1111}|] = \mB^4\,,\quad\E[|G_{1112}|] = \mB^4(1-r)\,,\quad\E[|G_{1121}|]=\mB^4(1-r)\,,\nonumber\\[1.0ex]
  & \E[|G_{1211}|] = \mB^4(1-r)\,,\quad\E[|G_{2111}|] = \mB^4(1-r)\,,
\end{align}
\vskip 0.1cm\noindent\underline{\bf 5th level}
\begin{align}
  & \E[|G_{11111}|] = \mB^5\,,
\end{align}
\noindent We can use the above expectations to calculate $\E[Y^{(i)}_n(t)]$ for $t=1,\ldots,5$ and $n\le t$. Then we infer $\E[Y^{(i)}_n(t)]$ for generic $t$ and $n$ (we leave to the reader the exercise of proving by induction the general formulae given below). We first examine the $t$th level. We have 
\begin{align}
  & \E[Y^{(i)}_1(1)] = \E[|G_1|] = \mB\,,\nonumber\\[0.8ex]
  & \E[Y^{(i)}_2(2)] = \E[|G_{11}|] = \mB^2\,,\nonumber\\[0.8ex]
  & \E[Y^{(i)}_3(3)] = \E[|G_{111}|] = \mB^3\,,\nonumber\\[0.8ex]
  & \E[Y^{(i)}_4(4)] = \E[|G_{1111}|] = \mB^4\,,\nonumber\\[0.8ex]
  & \E[Y^{(i)}_5(5)] = \E[|G_{11111}|] = \mB^5\,,\nonumber\\[-1.0ex]
  &\hskip 1.7cm \vdots\nonumber\\
  & \E[Y^{(i)}_t(t)] = \mB^t\,.
  \label{eq:lastlev}
\end{align}
We thus see that the $t$th level of trees in $\cP_{{\rm\scriptscriptstyle FC},t}(\alpha,r,\kmin)$ scales just like trees in $\cP_{\rm\scriptscriptstyle B}(\alpha,r,\kmin)$. Things get more interesting on the $(t-1)$th level. In this case we have
\begin{align}
  & \E[Y^{(i)}_1(2)] = \E[|G_1|+|G_2|] = \mB[1+(1-r)]\,,\nonumber\\[0.8ex]
  & \E[Y^{(i)}_2(3)] = \E[|G_{11}|+|G_{12}|+|G_{21}|] = \mB^2[1+2(1-r)]\,,\nonumber\\[0.8ex]
  & \E[Y^{(i)}_3(4)] = \E[|G_{111}|+|G_{112}|+|G_{121}|+|G_{211}|] = \mB^3[1+3(1-r)]\,,\nonumber\\[0.8ex]
  & \E[Y^{(i)}_4(5)] = \E[|G_{1111}|] + \E[|G_{1112}|] + \E[|G_{1121}|] + \E[|G_{1211}|] + \E[|G_{2111}|]  = \mB^4[1+4(1-r)]\,,\nonumber\\[-1.0ex]
  & \hskip 1.75cm \vdots \nonumber\\
  & \E[Y^{(i)}_{t-1}(t)] = [1+(1-r)(t-1)]\,\to\,(1-r)t\ \mB^{t-1}\,,\quad \text{ as } t\to\infty\,.
  \label{eq:scndlastlev}
\end{align}
\begin{figure}[t!]
  \centering
  \includegraphics[width=0.9\textwidth]{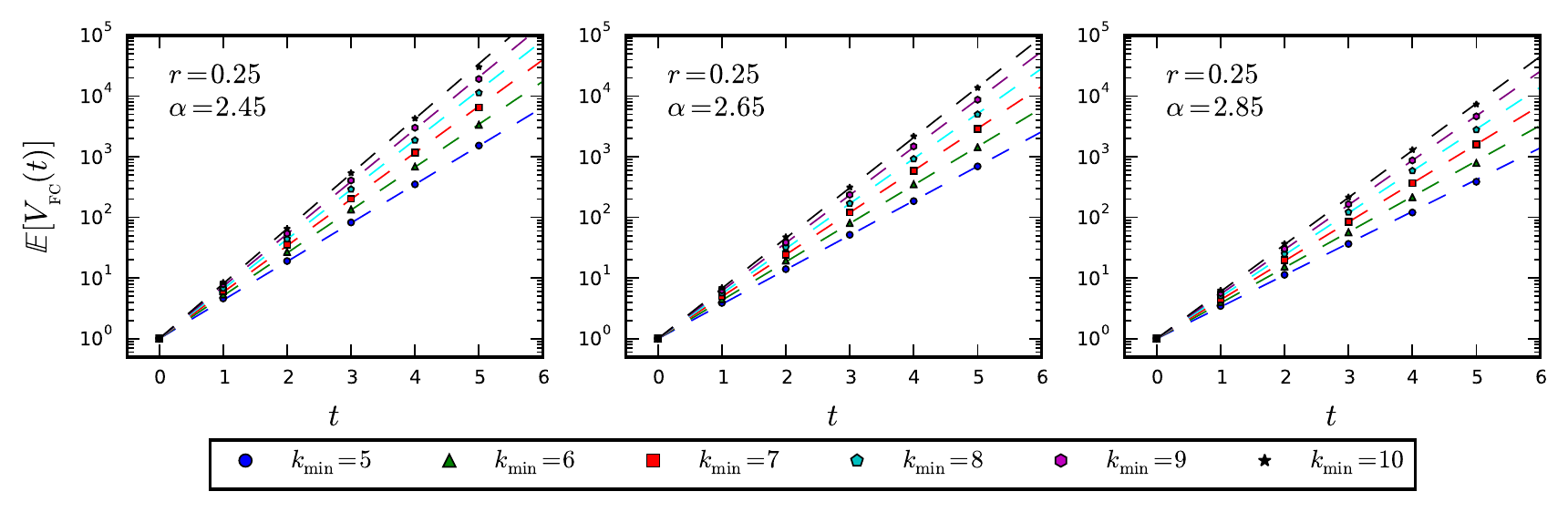}
  \vskip 0.0cm
  \caption{\footnotesize Scaling of $V_{\rm\scriptscriptstyle FC}(t)$ from numerical simulations. Dashed lines represent eq.~(\ref{eq:VFCscaling}).}
  \label{fig:figeleven}
  \vskip 0.0cm
\end{figure}
\hskip -0.12cm We now see that the exponential growth is broken by power corrections. In particular, for the $(t-1)$th level the correction is linear in $t$. The additional factor $[1+(1-r)(t-1)]$ quantifies the effect of the second interaction the undecided agents have with their parents. More generally, we can show that the correction factor to the exponential scaling of persuaded agents on the $(t-k)$th level amounts to a polynomial in $t$ of degree $k$. This quantifies the effect of the $k$ additional interactions the undecided agents have with their parents up to time $t$. Since we are mainly interested in the scaling of trees in~$\cP_{{\rm\scriptscriptstyle FC},t}(\alpha,r,\kmin)$ as $t\to\infty$, we keep only the leading term of the polynomial and leave out the subleading ones. Actually, power corrections yield the Taylor expansion of an unknown function of $t$, which emerges only when we add persuaded agents on all levels. To confirm this, we need to examine at least the $(t-2)$th level. In this case, we have
\begin{align}
  & \E[Y^{(i)}_1(3)] = \E[|G_1|+|G_2|+|G_3|] = \mB[1+(1-r)+(1-r)^2]\,,\nonumber\\[1.0ex]
  & \E[Y^{(i)}_2(4)] = \E[|G_{11}|+|G_{12}|+|G_{13}|+|G_{21}|+|G_{22}|+|G_{31}|] = \mB^2[1+2(1-r)+3(1-r)^2]\,,\nonumber\\[1.0ex]
  & \E[Y^{(i)}_3(5)] = \E[|G_{111}|+|G_{112}|+|G_{113}|+|G_{121}||G_{122}|+|G_{131}|+|G_{211}|+|G_{212}|+|G_{221}|+|G_{311}|]\nonumber\\[1.0ex]
  & \hskip 1.52cm = \mB^3[1+3(1-r)+6(1-r)^2]\,,\nonumber\\[-1.0ex]
  & \hskip 1.74cm \vdots \nonumber\\
  & \E[Y^{(i)}_{t-2}(t)] = \mB^{t-2}\left[1+(1-r)(t-2) + \frac{1}{2}(1-r)^2(t-1)(t-2)\right]\nonumber\\[1.0ex]
  & \hskip 1.5cm \to\,\frac{1}{2}(1-r)^2 t^2\,\mB^{t-2}\,,\ \text{ as } t\to\infty\,.
  \label{eq:thrdlastlev}
\end{align}
At this point we can calculate $\E[V_{\rm\scriptscriptstyle FC}(t)]$ as $t\to\infty$ by just summing eqs.~(\ref{eq:lastlev}),~(\ref{eq:scndlastlev}),~(\ref{eq:thrdlastlev}), \emph{etc}. In first approximation this yields
\begin{align}
  \E[V_{\rm\scriptscriptstyle FC}(t)] & = \sum_{s=0}^t \E[Y^{(i)}_s(t)] \simeq \mB^t\left\{1 + \frac{(1-r)t}{\mB} + \frac{1}{2}\frac{(1-r)^2t^2}{\mB^2} + \ldots\right\}  \nonumber\\[0.0ex]
  & = \left\{\mB \exp\left[\frac{(1-r)}{\mB}\right]\right\}^t \equiv \mFC^t\,,\qquad \text{as } t\to\infty\,.
\label{eq:VFCscaling}
\end{align}
Hence, we see that $V_{\rm\scriptscriptstyle FC}(t)$ scales exponentially with growth rate close to $\mFC = \mB\exp\{(1-r)/\mB\}$. In Fig.~\ref{fig:figeleven} we show numerical simulations of $\E[V_{\rm\scriptscriptstyle FC}(t)]$ for $r=0.25$, $\alpha=2.45,2.65,2.85$, $\kmin=5,6,\ldots,10$ and $t\le5$. The dashed lines represent our theoretical estimates. As can be seen, the agreement with simulation data is rather good. In spite of this we must bear in mind that $\mFC$ is just an approximation to the true growth rate of $V_{\rm\scriptscriptstyle FC}(t)$. Indeed, as we noticed above, to resum in closed form the power corrections we had to drop all subleading terms on each level. By doing this, we did not consider that each subleading term mixes with the leading term of some upper level, thus producing a small shift of its coefficient proportional to some inverse power of $\mB$. We conclude that $\mFC$ represents the correct growth rate up to $\text{O}(1/\mB)$. We shall come back to this in a while. 

\begin{figure}[t!]
  \centering
  \includegraphics[width=0.45\textwidth]{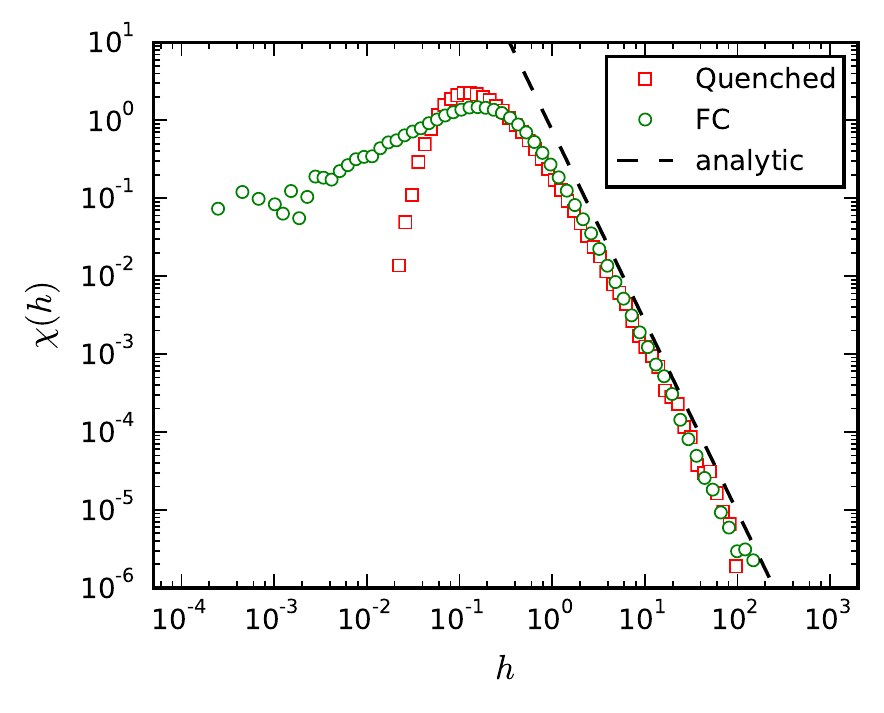}
  \vskip -0.3cm
  \caption{\footnotesize $\chi(h)$ for $(\alpha,\nmin)=(2.45,3)$ vs. $\chi_{_{\rm\scriptscriptstyle FC}}(h)$ for $(\alpha,r,\kmin)=(2.45,0.25,10)$.}
  \label{fig:figtwelve}
\end{figure}

In consideration of eq.~(\ref{eq:VFCscaling}), we find it convenient to introduce the rescaled variable
\begin{equation}
  H_{\rm\scriptscriptstyle FC}(t) = \frac{V_{\rm\scriptscriptstyle FC}(t)}{\mFC^t}\,.
\end{equation}
From the above discussion it follows that $\E[H_{\rm\scriptscriptstyle FC}(t)] = 1 + \text{O}(1/\mB)$. As $t\to\infty$ the sequence $(H_{\rm\scriptscriptstyle FC}(t))_{t\ge0}$ converges to a finite limit $H_{\rm\scriptscriptstyle FC}$ with continuous \emph{p.d.f.} $\chi_{\rm\scriptscriptstyle FC}(h)$. To prove this, it suffices to show that $H_{\rm\scriptscriptstyle FC}(t)$ is a martingale. To this aim we need to  evaluate $\E[V_{\rm\scriptscriptstyle FC}(t)\,|\,\{G_{i_1\ldots i_n}\}_{i_1+\ldots+i_n\le t-1}]$, \ie the conditional expectation of $V_{\rm\scriptscriptstyle FC}(t)$ given all groups which have been generated up to time $t-1$. As previously, we perform the calculation for $t=1,2,\ldots$ and then we extrapolate to generic $t$. With reference to Fig.~\ref{fig:figten}, we have
\begin{align}
  V_{\rm\scriptscriptstyle FC}(1) & = 1 + |G_1|\,,\nonumber\\[0.8ex]
  V_{\rm\scriptscriptstyle FC}(2) & = 1 + |G_1| + |G_2| + |G_{11}|\,,\nonumber\\[0.8ex]
  V_{\rm\scriptscriptstyle FC}(3) & = 1 + |G_1| + |G_2| + |G_3| + |G_{11}| + |G_{12}| + |G_{21}| + |G_{111}|\,,\nonumber\\[-1.0ex]
  & \hskip 0.18cm \vdots
\end{align}
At time $t=2$ the conditional expectation yields
\begin{align}
  \E[V_{\rm\scriptscriptstyle FC}(2)\,|\,G_1]\, & =\, 1 + |G_1| +\, \E\bigl[|G_2|\,\bigl|\,G_1\bigr]\, +\,  \E\bigl[|G_{11}|\,\bigl|\,G_1\bigr] = 1 + |G_1| + r(\hatmQ - |G_1|) + \mB|G_1|\nonumber\\[1.0ex]
  & \hskip -1.18cm = [\mB + (1-r)]V_{\rm\scriptscriptstyle FC}(1) + r\,.
\end{align}
At time $t=3$ it yields
\begin{align}
  \E[V_{\rm\scriptscriptstyle FC}(3)\,|\,G_1,G_2,G_{11}]\, & =\, 1 + |G_1| + |G_2| + |G_{11}| \nonumber\\[1.0ex]
  & \hskip -2.53cm  +\, \E\bigl[|G_3|\,\bigl|\,G_1,G_2\bigr]\, +\, \E\bigl[|G_{12}|\,\bigr|\,G_1,G_{11}\bigr]\, +\, \E\bigl[|G_{21}|\,\bigl|\,G_2\bigr]\, +\, \E\bigl[|G_{111}|\,\bigl|\,G_{11}\bigr]\nonumber\\[1.0ex]
  & \hskip -2.53cm =\, 1 + |G_1| + |G_2| + |G_{11}| + r(\hatmQ - |G_1| - |G_2|) +  r(\hatmQ|G_1| - |G_{11}|) + \mB |G_2| + \mB|G_{11}| \nonumber\\[1.0ex]
  & \hskip -2.53cm = [\mB + (1-r)]V_{\rm\scriptscriptstyle FC}(2) + r\,.
\end{align}
At time $t=4$ we can equally show that $\E[V_{\rm\scriptscriptstyle FC}(4)\,|\,\{G_{i_1\ldots i_n}\}_{i_1+\ldots+i_n\le 3}] = [\mB + (1-r)]V_{\rm\scriptscriptstyle FC}(3) + r$. Hence, we conclude that the equation
\begin{empheq}[box=\mybluebox]{align}
  \E[V_{\rm\scriptscriptstyle FC}(t)\,|\,\{G_{i_1\ldots i_n}\}_{i_1+\ldots+i_n\le t-1}] = [\mB + (1-r)]V_{\rm\scriptscriptstyle FC}(t-1) + r\,
  \label{eq:HFCmartingale}
\end{empheq}
holds for generic $t$. Notice that $\mB + (1-r) = \mFC+\text{O}(1/\mB)$. Since the additive term $r$ on the r.h.s. of eq.~(\ref{eq:HFCmartingale}) represents an infinitesimal correction as $t\to\infty$, we conclude that $H_{\rm\scriptscriptstyle FC}(t)$ is asymptotically a martingale. In Fig.~\ref{fig:figtwelve} we compare $\chi_{\rm\scriptscriptstyle FC}(h)$ with $\chi(h)$. The two \emph{p.d.f.}'s look very similar for $h\ge 1$. Their tails have the same power--law exponent and essentially the same scale factor. As we know, the latter is sufficiently well approximated by $h_0^{\alpha-1}/\zeta(\alpha-1,\nmin)$. In practice, they differ only for $h\le 1$, where the effects of the activism of the agents and the noise of the agent--agent interactions become important. To conclude, we observe that the analysis presented in sect.~3 can now be easily adapted to the FC model, provided we replace $\mQ\to m_{\rm\scriptscriptstyle FC}$, $\chi(h)\to\chi_{\rm\scriptscriptstyle FC}(h)$ and $h_0\to 1$.  

\section{One--step transition probabilities}

\begin{figure}[t!]
    \centering
    \includegraphics[width=0.9\textwidth]{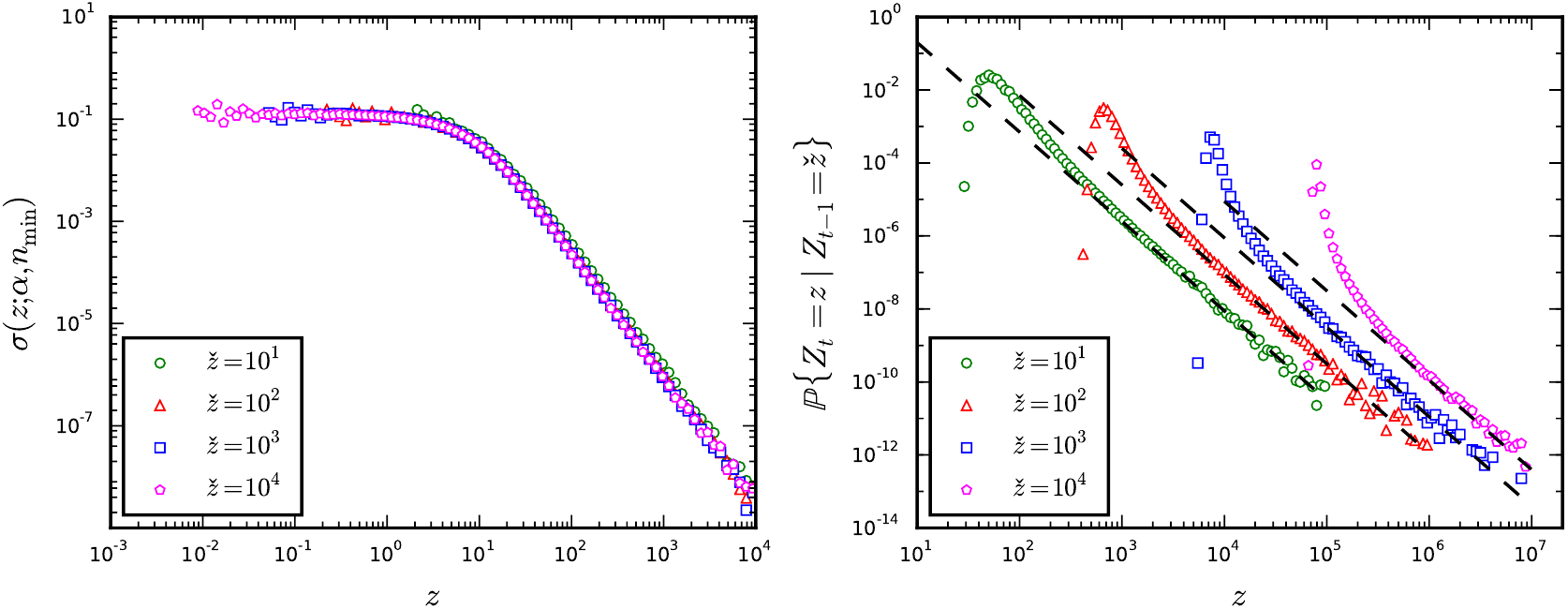}
    \caption{\footnotesize (Left) \emph{p.d.f.} $\sigma(z;\alpha,\nmin)$ for $\alpha=2.45$ and $\nmin=3$ as obtained from numerical simulations of $[\sum_{j=1}^{\check{z}} \xi_{j} - \mQ\check{z}]/\check{z}^{1/(\alpha-1)}$ with $\check{z}=10^1,10^2,10^3,10^4$; (Right) OSTP for the same choice of parameters. The dashed lines represent eq.~(\ref{eq:OSTPasymp}).}
    \label{fig:figthirteen}
    \vskip -0.3cm
  \end{figure}

We now go back to the distribution of $W$. In sect.~2 we wrote eq.~(\ref{eq:phi}) without explaining where it comes from. In principle $\phi(w)$ can be derived from $\P\{Z_t=z\}$ by setting $z=w \mQ^t$ and by using the scaling law $\P\{Z_t=w\mQ^t\} = \P\{W_t=w\}\to \phi(w)\rd w$ as $t\to\infty$. Since $\E[Z_t]=\mQ^t$, the tail the distribution is observed for $z\gg \mQ^t$, that is for $w\gg 1$. In order to work out  $\phi(w)$ in this limit, we adopt a strategy based on the one--step transition probabilities (OSTP) $\P\{Z_{t}\, |\, Z_{t-1}\}$. Although these objects are seldom used for the analysis of branching trees\footnote{Most known results are indeed based on the use of probability generating functions.}, they can be regarded to all purposes as building blocks for theoretical calculations, in that they propagate complete information from one level to the next one. Specifically, we have
\begin{align}
\label{eq:onestep}
  &  \P\{Z_t=z\} = \sum_{z_{t-1},\ldots,z_1}\P\{Z_{t}=z\,|\, Z_{t-1}=z_{t-1}\}\cdot\P\{Z_{t-1}=z_{t-1}\,|\, Z_{t-2}=z_{t-2}\}\cdot\ldots\cdot\P\{Z_{1}=z_1\}\,,\\[2.0ex]
&  \P\{V_t=v\} = \sum_{\substack{z_{t},\ldots,z_1 \\[0.5ex] |z|=v-1}}\,\P\{Z_{t}=z_t\,|\, Z_{t-1}=z_{t-1}\}\cdot\P\{Z_{t-1}=z_{t-1}\,|\, Z_{t-2}=z_{t-2}\}\cdot\ldots\cdot\P\{Z_{1}=z_1\}\,.
\end{align}
Here we are interested in particular in eq.~(\ref{eq:onestep}), which is also equivalent to 
\begin{equation}
   \P\{Z_t=z\} = \sum_{\check{z}}\P\{Z_{t}=z\,|\, Z_{t-1}=\check{z}\}\cdot \P\{Z_{t-1}=\check{z}\}\, =\, \E\bigl[\P\{Z_t=z\,|\,Z_{t-1}\}\bigr]\,.
   \label{eq:PZzE}
\end{equation}
The law of iterated expectations is useful only if we are able to both calculate $\P\{Z_{t}=z\,|\, Z_{t-1}=\check{z}\}$ and average it over $\check{z}$. Before we embark on this, we recall that $Z_t = \sum_{j=1}^{Z_{t-1}}\xi_{j}$ with $\xi_{j}\sim \pQ(n)$ and $Z_{t-1}\ge (\nmin)^{t-1}$. Therefore, if $\nmin>1$ and $t$ is not too small, $Z_t$ is the sum of a large number of i.i.d. Mandelbrot variables. Since $\xi_{j}$ has infinite variance for $\alpha<3$, from the generalized central limit theorem (see refs.~\cite[p.~50]{stableone} and \cite[p.~62]{stabletwo}) it follows that $\P\{Z_{t}=z\,|\, Z_{t-1}=\check{z}\}$ falls in  the domain of attraction of a stable law of index $(\alpha-1)$ as $\check{z}\to\infty$, \ie there exists a L\'evy stable distribution $\Sigma(z;\alpha,\nmin)$ with \emph{p.d.f.} $\sigma(z;\alpha,\nmin)$ such that
\begin{equation}
  \lim_{\check{z}\to\infty}\P\biggl\{\frac{1}{{\check{z}}^{1/(\alpha-1)}}\biggl(\sum_{j=1}^{\check{z}}\xi_{j} - \mQ\check{z}\biggr)<z \biggr\} = \Sigma(z;\alpha,\nmin)\,.
  \label{eq:Levysigma}
\end{equation}
In Fig.~\ref{fig:figthirteen} (left) we plot the distribution of $(\,\sum_{j=1}^{\check{z}}\xi_{j} - \mQ\check{z}\,)/\check{z}^{1/(\alpha-1)}$ for $\alpha=2.45$, $\nmin=3$ and \linebreak ${\check{z}=10^1,\ldots,10^4}$. The plot shows that the convergence to the limit distribution $\sigma(z;\alpha,\nmin)$ is very fast. In view of this, we conclude that $\P\{Z_{t}=z\,|\, Z_{t-1}=\check{z}\}$ has a power--law tail as a function of $z$ even for moderate values of $\check{z}$. Although we can calculate the OSTP, we are not able to work out its expectation in closed form. In first approximation, we can get the tail of the full probability $\P\{Z_t=z\}$ by expanding $\P\{Z_t=z\,|\,Z_{t-1}=\check{z}\}$ asymptotically in inverse powers of $z$ as $z\to\infty$ and by then averaging over $\check{z}$ just the leading term of the expansion. We anticipate that the resulting estimate is not uniformly good in $\alpha$. In particular, our estimate is mathematically consistent with the scaling law of $\P\{Z_t=z\}$ only for $\alpha\to 2$. In this limit eq.~(\ref{eq:phi}) holds true. We shall see, however, that the error we make for $\alpha> 2$ is not exceedingly large.  

In order to perform the calculation of $\P\{Z_{t}=z\,|\, Z_{t-1}=\check{z}\}$, we follow ref.~\cite{Roehner}, where a similar study is carried out for (continuous) Paretian variables. In fact, the only difference with that paper is that our variables are discrete. We introduce preliminarily the Lerch transcendent
\begin{equation}
  \Phi(z,\alpha,n) = \sum_{j=0}^\infty \frac{z^j}{(j+n)^\alpha}\,.
\end{equation}
This special function generalizes the Hurwitz $\zeta$--function, which is indeed obtained for $z=1$. We notice that $\Phi(z,\alpha,n)$ is analytic for $|z|<1$ and, if $\text{Re}\{\alpha\}>1$, also for $|z|=1$, provided $n\ne 0,-1,-2,\ldots$ For the other values of $z$, $\Phi(z,\alpha,n)$ is defined by analytic continuation. In particular, $\Phi(z,\alpha,n)$ can be differentiated infinitely many times at $z=0$. It is also useful to recall that $\Phi(z,\alpha,n)$ can be represented as a Taylor series, namely (see ref.~\cite[p.~29]{Erdelyi})
\begin{equation}
  \Phi(z,\alpha,n) = z^{-n}\left[\Gamma(1-\alpha)[-\log(z)]^{\alpha-1} + \sum_{a=0}^{\infty}\zeta(\alpha-a,n)\frac{\log(z)^a}{a!}\right]\,.
\label{eq:lerchexp}
\end{equation}
This series converges for $|\log(z)|<2\pi$, $\alpha \ne 1,2,3,\ldots$ and $n\ne 0,-1,-2,\ldots$ The special character of $\Phi(z,\alpha,n)$ is clearly expressed by the presence of a fractional power in the Taylor series. For $\alpha<3$ the term $[-\log(z)]^{\alpha-1}$ is the ultimate source of the power--law behaviour of $\P\{Z_{t}=z\,|\, Z_{t-1}=\check{z}\}$, as we shall see in the next few lines.

Since we have ${Z_t = \sum_{j=1}^{Z_{t-1}}\xi_{j}}$, the OSTP can be written as a convolution of $Z_{t-1}=\check{z}$ copies of the Mandelbrot distribution, namely
\begin{align}
  \P\{Z_t = z\,|\, Z_{t-1}=\check{z}\} & \, = \, \sum_{n_1,\ldots,n_{\check{z}}=0}^\infty \delta_{z,n_1+\ldots+n_{\check{z}}}\ \pQ(n_1)\cdot \ldots\cdot \pQ(n_{\check{z}}) \nonumber\\[1.0ex]
  &  =\,[\zeta(\alpha,\nmin)]^{-\check{z}}\sum_{n_1,\ldots,n_{\check{z}}=0}^\infty\frac{1}{2\pi}\int_{-\pi}^{\pi}\rd\phi\ \re^{\ri\phi(n_1+\ldots+n_{\check{z}}-z)}\ \frac{\theta_{n_1,\nmin}}{n_1^\alpha}\cdot\ldots\cdot\frac{\theta_{n_{\check{z}},\nmin}}{n_{\check{z}}^\alpha}\nonumber\\[1.0ex]
  &  =\, \frac{1}{2\pi}\int_{-\pi}^{\pi}\rd\phi\ \re^{-i\phi (z-\check{z}\nmin)}\,\left[\frac{\Phi(\re^{\ri\phi},\alpha,\nmin)}{\zeta(\alpha,\nmin)}\right]^{\bar z}\,,
  \label{eq:Pconvol}
\end{align}
where we used the Fourier representation of the Kronecker delta and we let $\theta$ denote the discrete Heaviside step function, namely $\theta_{a,b}=1$ if $a\ge b$ and $\theta_{a,b}=0$ otherwise. We insert the Taylor expansion of the Lerch transcendent into eq.~(\ref{eq:Pconvol}). This yields the expression
\begin{equation}
  \P\{Z_t = z\,|\, Z_{t-1}=\check{z}\} = \frac{1}{2\pi}\int_{-\pi}^{\pi}\rd\phi\ \re^{-\ri\phi z}\,\biggl[\,\frac{\Gamma(1-\alpha)}{\zeta(\alpha,\nmin)}\left(\frac{\phi}{\ri}\right)^{\alpha-1} + \sum_{a=0}^\infty\frac{(-1)^{a}}{a!}\frac{\zeta(\alpha-a,\nmin)}{\zeta(\alpha,\nmin)}\left(\frac{\phi}{\ri}\right)^a\,\biggr]^{\check{z}}\,.
\end{equation}
Now we observe that if $2<\alpha<3$ the fractional power $(\phi/\ri)^{\alpha-1}$ lies between the integer powers $(\phi/\ri)$ and $(\phi/\ri)^2$, while if $\alpha>3$ it lies between $(\phi/\ri)^2$ and $(\phi/\ri)^3$. The first integer power $(\phi/\ri)$ is related to the conditional expectation $\E[Z_t|Z_{t-1}]$, as can be seen from its coefficient which is precisely $-\mQ$. To take advantage of this, we insert $1 = \exp\{\ri\phi \mQ\check{z}\} \exp\{-\ri\phi \mQ\check{z}\}$ under the integral sign and then expand the exponential $\exp\{-\ri\phi \mQ\check{z}\} = (\exp\{-\ri\phi \mQ\})^{\check{z}}$ in Taylor series. Accordingly, we obtain
\begin{equation}
  \P\{Z_t = z\,|\, Z_{t-1}=\check{z}\} = \frac{1}{2\pi}\int_{-\pi}^{\pi}\rd\phi\ \re^{-\ri\phi (z-\mQ\check{z})}\,\biggl[\, 1 + \frac{\Gamma(1-\alpha)}{\zeta(\alpha,\nmin)}\left(\frac{\phi}{\ri}\right)^{\alpha-1} +\ \text{O}(\phi^2),\biggr]^{\check{z}}\,,
\end{equation}
for $2<\alpha<3$, and
\begin{equation}
  \P\{Z_t = z\,|\, Z_{t-1}=\check{z}\} = \frac{1}{2\pi}\int_{-\pi}^{\pi}\rd\phi\ \re^{-\ri\phi (z-\mQ\check{z})}\,\biggl[\, 1 + \frac{1}{2}\frac{\zeta(\alpha-2,\nmin)}{\zeta(\alpha,\nmin)}\left(\frac{\phi}{\ri}\right)^{2} +\ \text{o}(\phi^2),\biggr]^{\check{z}}\,,
\end{equation}
for $\alpha>3$. We are interested in how the OSTP looks as $\check{z}\to\infty$. For this, we rescale the integration variable according to $\phi\to \check{z}^{1/(\alpha-1)}\phi$ if $2<\alpha<3$ and according to $\phi\to\check{z}^{1/2}\phi$ if $\alpha>3$, then we use the notable limit $\lim_{x\to\infty}(1+\lambda/x)^{x}=\exp\{\lambda\}$ and finally we scale back the integration variable. This yields
\begin{empheq}[box=\mybluebox]{align}
  \P\{Z_t = z\,|\, Z_{t-1}=\check{z}\} & = \frac{1}{2\pi}\int_{-\infty}^{\infty}\rd\phi\ \exp\biggl\{-\ri\phi (z-\mQ\check{z}) + \frac{\Gamma(1-\alpha)}{\zeta(\alpha-1,\nmin)}\mQ\check{z}\left(\frac{\phi}{\ri}\right)^{\alpha-1}\biggr\}\nonumber\\[2.0ex]
  & \hskip -3.2cm = \frac{1}{2\pi}\int_{-\infty}^{\infty}\rd\phi\ \re^{-\ri\phi z}\nonumber\\[1.0ex]
  & \hskip -2.7cm \cdot \exp\biggl\{\ri\phi \mQ\check{z} + \frac{\Gamma(1-\alpha)}{\zeta(\alpha-1,\nmin)}\mQ\check{z}\cos\biggl(\frac{\pi}{2}(\alpha-1)\biggr)|\phi|^{\alpha-1}\biggl[1-\ri\,\text{sign}(\phi)\tan\biggl(\frac{\pi}{2}(\alpha-1)\biggr)\biggr]\biggr\}\,,
  \label{eq:levy}
\end{empheq}
for $2<\alpha<3$ and
\begin{equation}
 \P\{Z_t = z\,|\, Z_{t-1}=\check{z}\} = \frac{1}{2\pi}\int_{-\infty}^{\infty}\rd\phi\ \re^{-\ri\phi z}\,\exp\biggl\{\ri\,\phi \mQ\check{z} - \frac{1}{2}\frac{\zeta(\alpha-2,\nmin)}{\zeta(\alpha-1,\nmin)}\mQ\check{z}\phi^2\biggr\}\,,
 \label{eq:gauss}
\end{equation}
for $\alpha>3$. A comment is in order concerning the integration limits. From eq.~(\ref{eq:Pconvol}) we know that ${\P\{Z_t=z\,|\,Z_{t-1}=\check{z}\}=0}$ unless $z\ge \nmin\check{z}$, therefore the probability mass of the OSTP shifts progressively towards $z\to\infty$ as $\check{z}\to\infty$. Moreover, since $Z_t\in\dN$ its minimum variation amounts to $\Delta Z_t=1$ independently of $Z_t$. It follows that $\Delta Z_t/Z_t = 1/z\to 0$ as $\check{z}\to\infty$, hence $\P\{Z_t=z\,|Z_{t-1}=\check{z}\}$ becomes to all practical purposes a continuous distribution in $z$. This explains why the integral transform in eqs.~(\ref{eq:levy})--(\ref{eq:gauss}) is over ${\phi\in(-\infty,\infty)}$. 

If we compare the second exponential in eq.~(\ref{eq:levy}) with the general expression of the characteristic function of a stable distribution of index $(\alpha-1)$ given in ref.~\cite[p.~164]{Gnedenko}, namely
\begin{equation}
  \Psi(\phi) = \exp\biggl\{\ri\gamma\phi - c|\phi|^{\alpha-1}\biggl[1+\ri\,\beta\,\text{sign}(\phi)\tan\biggl(\frac{\pi}{2}(\alpha-1)\biggr)\biggr]\biggr\}\,,
\end{equation}
we see that
\begin{equation}
\gamma = \mQ\check{z}\,,\qquad c = -\frac{\Gamma(1-\alpha)}{\zeta(\alpha-1,\nmin)}\mQ\check{z}\cos\left(\frac{\pi}{2}(\alpha-1)\right)\,,\qquad \beta = -1.\qquad 
\end{equation}
In particular, $\gamma$ is  the location parameter of the distribution (its value here is in agreement with \linebreak ${\E[Z_t\,|\,Z_{t-1}] = \mQ Z_{t-1}}$), $c$ is the scale parameter (it is proportional to $\gamma$ in our case) and $\beta$ represents the skewness of the distribution (in general $\beta$ fulfills $|\beta|\le 1$, hence it takes here the maximum negative value). 
The second exponential in eq.~(\ref{eq:gauss}) is the characteristic function of a normal variable with mean $\mu = \mQ\check{z}$ and variance $\sigma^2 = \mQ\check{z}\, \zeta(\alpha-2,\nmin)/\zeta(\alpha-1,\nmin)$. Since in the quenched model we are always interested in $2<\alpha<3$, we focus on eq.~(\ref{eq:levy}) and forget about eq.~(\ref{eq:gauss}) in the following.

\subsection{Tail of the distribution}

A general formula describing the asymptotic behaviour of the \emph{p.d.f.} of a  L\'evy stable distribution is given in {ref.~\cite[ch.~1]{Nolan}}. For the sake of completeness (and to be sure that the reader interprets correctly the formula), we reproduce it here. We start from eq.~(\ref{eq:levy}), that we recast in the form
\begin{equation}
\P\{Z_t=z\,|\,Z_{t-1}=\check{z}\} = \frac{1}{\pi}\text{Re}\,{\left\{\int_0^{\pi}\rd\phi\,\re^{\ri\phi(z-\mQ\check{z})}\,\re^{-\rho\phi^{\alpha-1}}\right\}}\,, 
\end{equation}
with $\rho = -\frac{\Gamma(1-\alpha)}{\zeta(\alpha-1,\nmin)}\exp\left\{-\ri\frac{\pi}{2}(\alpha-1)\right\}\mQ\check{z}$. We expand the exponential $\exp\{-\rho\phi^{\alpha-1}\}$ in Taylor series. Then, we integrate the series term by term. In this way, we obtain
\begin{align}
\P\{Z_{t}=z\,|\,Z_{t-1}=\check{z}\} & = \frac{1}{\pi}\text{Re}\,\left\{\sum_{\ell=0}^\infty \frac{(-1)^\ell\, \rho^\ell}{\ell!}\int_{0}^{\infty}\rd\phi\,\phi^{\ell(\alpha-1)}\re^{-\ri(z-\mQ\check{z})\phi}\right\} \nonumber\\[1.0ex]
& = \frac{1}{\pi}\text{Re}\,\left\{\sum_{\ell=0}^\infty\frac{(-1)^\ell\, \rho^\ell}{\ell!}\frac{\Gamma\bigl(1+(\alpha-1)\ell\bigr)}{[\ri(z-\mQ\check{z})]^{1+\ell(\alpha-1)}}\right\}\,.
\label{eq:OSTPtaylexp}
\end{align}
Apart from the first term of the series, which is a Dirac delta $\delta(z-\mQ\check{z})$, all the subsequent terms are inverse powers of $(z-\mQ\check{z})$. Therefore, at leading order the expansion reads
\begin{equation}
   \P\{Z_t=z\,|\,Z_{t-1}=\check{z}\}\ =\ \delta(z-\mQ\check{z}) - \frac{1}{\pi}\frac{\Gamma(\alpha)\,\text{Re}\left\{\rho\re^{-\ri\alpha\pi/2}\right\}}{(z-\mQ\check{z})^\alpha} + \text{O}\left(\frac{1}{(z-\mQ\check{z})^{2\alpha-1}}\right)\,.
\label{eq:asymprepr}
\end{equation}
For $z\gg \mQ\check{z}$ we can drop both the Dirac delta and the subleading terms of the expansion. Moreover, from the Euler reflection formula $\Gamma(\alpha)\Gamma(1-\alpha) = \pi/\sin(\pi\alpha)$, it follows that 
\begin{align}
  \frac{\Gamma(\alpha)\text{Re}\left\{\rho\re^{-\ri\alpha\pi/2}\right\}}{\pi} = -\frac{1}{\pi}\frac{\Gamma(\alpha)\Gamma(1-\alpha)}{\zeta(\alpha-1,\nmin)}\mQ\check{z}\,\text{Re}\left\{\re^{-\ri\pi(\alpha-1/2)}\right\} = -\frac{\mQ\check{z}}{\zeta(\alpha-1,\nmin)}\,.
\end{align}
Therefore, we end up with the asymptotic estimate
\begin{empheq}[box=\mybluebox]{align}
   \P\{Z_t=z\,|\,Z_{t-1}=\check{z}\} =  \frac{1}{\zeta(\alpha-1,\nmin)}\frac{\mQ \check{z}}{z^\alpha}\,,\qquad \text{as } z\to\infty\,.
\label{eq:OSTPasymp}
\end{empheq}
In Fig.~\ref{fig:figthirteen} (right) we show the OSTP for $\alpha=2.45$, $\nmin = 3$ and $\check{z}=10^1,\ldots,10^4$. The dashed lines on the plot represent the power--law tail as predicted by eq.~(\ref{eq:OSTPasymp}). We see that the theoretical estimates are in perfect agreement with the simulations.

\begin{figure}[t!]
    \centering
    \includegraphics[width=0.45\textwidth]{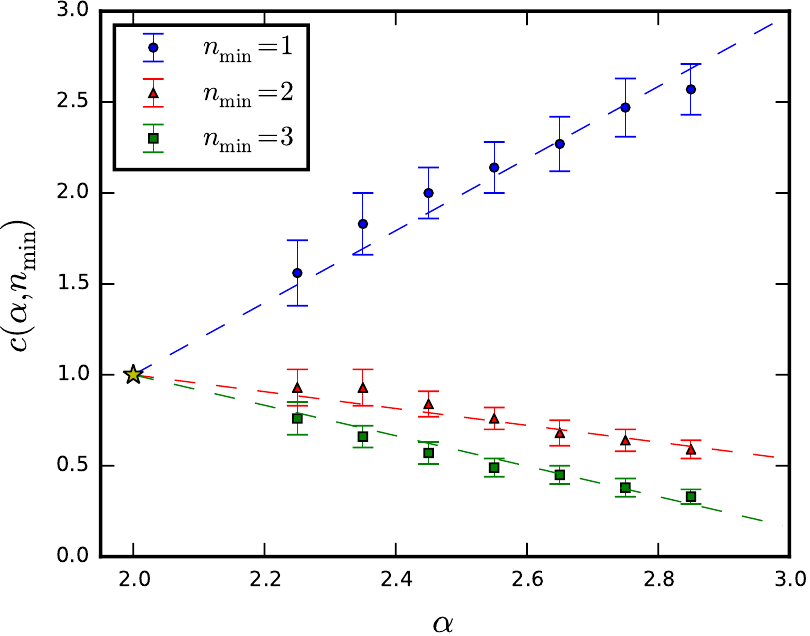}
    \caption{\footnotesize Coefficient $c(\alpha,\nmin)$ (see eq.~(\ref{eq:phiansatz})) as obtained from numerical simulations of the distribution of $W_6$ for ${\alpha = 2.25,\ldots,2.85}$ and $\nmin = 1,2,3$.}
    \label{fig:figfourteen}
  \end{figure}

Now, averaging eq.~(\ref{eq:OSTPasymp}) over $Z_{t-1}=\check{z}$ yields
\begin{equation}
  \P\{Z_t=z\} = \frac{1}{\zeta(\alpha-1,\nmin)}\frac{\mQ^t}{z^\alpha}\,,\qquad \text{as } z\to\infty\,.
\end{equation}
As anticipated, this formula is inconsistent with the scaling law of $\P\{Z_t=z\}$ unless $\alpha\to 2$. Indeed, by setting $z = w\mQ^t$, we get
\begin{equation}
  \P\{W_t=w\} = \frac{1}{\mQ^{(\alpha-1)t}}\frac{1}{\zeta(\alpha-1,\nmin)}\frac{1}{w^\alpha}\,,\qquad \text{as } w\to\infty\,.
  \label{eq:anomaly}
\end{equation}
Since the minimum variation of $W_t$ amounts to ${\Delta W_t = 1/\mQ^t\to\rd w}$ as $t\to \infty$, we conclude that eq.~(\ref{eq:anomaly}) scales anomalously with $\rd w^{\alpha-1}$. This is not totally surprising in consideration that \emph{i}) we obtained our estimate from a representation of the OSTP which formally holds only in the limit $\check{z}\to\infty$ and \emph{ii}) we dropped additional terms in the OSTP in the limit $z\to\infty$. Notice, however, that the scaling anomaly disappears as $\alpha\to 2$. In this limit eq.~(\ref{eq:anomaly}) scales correctly with $\rd w$.  It is only for $\alpha\gtrsim 2$ that we expect $\phi(w)$ to be accurately described by eq.~(\ref{eq:phi}). If we make the ansatz
\begin{equation}
  \phi(w) = \frac{c(\alpha,\nmin)}{\zeta(\alpha-1,\nmin)}\frac{1}{w^\alpha}\,,\qquad \text{as } w\to\infty\,,
  \label{eq:phiansatz}
\end{equation}
we can measure the constant $c(\alpha,\nmin)$ by means of numerical simulations. In Fig.~\ref{fig:figfourteen} we report our determinations of $c(\alpha,\nmin)$ for $\alpha = 2.25,\ldots,2.85$ and $\nmin=1,2,3$ from simulations of $W_6$. The plot confirms that $c(\alpha,\nmin)\to 1$ as $\alpha\to2$. Morover, it shows that $1/2 \lesssim c(\alpha,\nmin)\lesssim 2$ for $\alpha\lesssim 2.5$ and $\nmin\le 3$. For these values of the model parameters eq.~(\ref{eq:phi}) provides a reasonably good approximation. 

To conclude, we stress once more that the above derivation relies on eq.~(\ref{eq:levy}), which holds formally in the limit $\check{z}\to\infty$. The reader may wonder how difficult it would be to calculate the OSTP for generic $\check{z}$. To answer this question, we derive in App.~B an alternative representation of ${\P\{Z_t=z\,|\,Z_{t-1}=\check{z}\}}$, based on multiple polylogarithms and shuffle products, as a series of inverse powers $z^{-\alpha_1}$ with $\alpha_1=\alpha,\alpha+1,\ldots$ This representation is valid for integer $\alpha$ and makes no assumptions on $\check{z}$.

\section{Discussion and outlook}

The word--of--mouth model for proportional elections, proposed by Fortunato and Castellano in ref.~\cite{fcscaling}, reproduces with great accuracy the scaling distribution $\Femp(x)$, universally observed in elections held in different countries and years. As such, the model represents a significant breakthrough in the field of opinion dynamics, where a qualitative agreement between empirical data and theoretical models is often the best result one can achieve. In spite of this, the analytic structure of the model had never been studied so far, neither by its authors nor by other scholars, and the only available information was based on computer simulations. In the present paper we made a first step to fill this gap.

It was known, in particular, that the conditional distribution $F_{\rm\scriptscriptstyle FC}(x|Q,N)$ predicted by the model develops a power--law right tail as $Q\to\infty$. The amount of empirical observations available with given $(Q,N)$ is at present largely insufficient for confirming or disproving this prediction with crystal clear evidence. Yet, we found that a signature of the presence of power--law structures in the empirical data can be observed (at least for some countries) in the conditional distribution $\cF_{\rm\scriptscriptstyle EMP}(x|Q>Q_0)$, provided $Q_0$ is sufficiently large. This yields indirect evidence that $F_{\rm\scriptscriptstyle EMP}(x|Q,N)$ has a power--law tail. It turns out indeed that $\cF_{\rm\scriptscriptstyle EMP}(x|Q>Q_0)$ is close to $F_{\rm\scriptscriptstyle EMP}(x|Q_0,\bar N)$, with $\bar N = \E[N|Q>Q_0]\gg 1$, for $x\ll Q_0$ and $Q_0\gg 1$. Moreover, it is possible to find a range of values for $Q_0$ where the pool of data contributing to $\cF_{\rm\scriptscriptstyle EMP}(x|Q>Q_0)$ is sufficiently large to yield an acceptable signal--to--noise ratio and the domain of $F_{\rm\scriptscriptstyle EMP}(x|Q_0,\bar N)$ is sufficiently extended to reveal the asymptotic shape of the right tail of the distribution.

In consideration of this, we studied the equations of the model in the large--list limit, \ie the double limit ${Q,N\to\infty}$. For pedagogical reasons, we first presented a derivation of the vote distribution in a quenched model, where the original branching--like process is replaced by a supercritical branching process having similar features. The main results of our analysis are that the vote distribution converges quickly to a convolution of single--tree distributions, with weights given by the probabilities of the stopping time of the model, and that this convolution can be discrete or continuous depending on the stopping rule of the model. As a second step we showed that the original branching--like process scales just like a supercritical branching process, hence the solution we found for our quenched model applies \emph{mutatis mutandis} also to the original one. Finally, we presented a derivation of the power--law tail of the vote distribution in the quenched model. The resulting estimate holds, within a reasonable approximation, also for the vote distribution of the original model. 

While developing the ideas presented along this exploratory paper, we ran into questions that still look for an answer and represent directions of future research. We list them below in the same order they arise in the text:
\begin{itemize}
  \setlength\itemsep{0.03em}
\item{Finland is the only country in the U group of ref.~\cite{fempanal} for which $\Femp(x|Q>Q_0)$ displays a lognormal behavior independently of $Q_0$. It is not clear at present what makes it different from the other countries.}
\item{A relevant difference between eq.~(\ref{eq:BP1}) and eq.~(\ref{eq:BP2}) is that that the extinction probability is positive for the former while it vanishes for the latter. In other words, the subspace of finite trees has a positive measure under eq.~(\ref{eq:BP1}). It is not clear  whether/how this feature affects the vote distribution.}
\item{We derived the analytic structure of the vote distribution under SR1 in the \emph{minimal} and \emph{maximal ensembles}. It is not clear  how to extend the derivation so as to take into account the whole space of trees.}
\item{We computed the distribution of the stopping time by means of numerical simulations. Although we know from general arguments that this is related to a L\'evy stable distribution under SR2 and SR3, it is not clear how to go beyond numerical simulations under SR1 in the \emph{minimal} and \emph{maximal ensembles}.}
\item{We calculated the exponential growth rate and proved the martingale property of the branching--like trees introduced by Fortunato and Castellano in a sort of mean--field approximation. Our results hold true up to corrections. It is not clear at present how to improve the estimates.}
\item{A more orthodox derivation of eq.~(\ref{eq:phi}) than provided in sect.~5 is necessary to ensure a full control of the power--law tail of the vote distribution.}
\item{The representation of the OSTP given in App.~B follows from a totally different approach than pursued in sect.~5. It is not clear whether/how it could be used effectively to calculate the coefficient $c(\alpha,\nmin)$ introduced in eq.~(\ref{eq:phiansatz}).}
\end{itemize}

To conclude, we observe that a complete understanding of the analytic structure of $F_{\rm\scriptscriptstyle FC}(x|Q,N)$ seems to be a necessary condition to pin down the exact formula relating the microscopic parameters of the word--of--mouth model to the macroscopic parameters of the universal scaling distribution. This will be maybe a premise to shed light on the ultimate mechanisms lying behind the scaling and universality properties of vote distributions in proportional elections.

\section*{Acknowledgments}

The computing resources used for our numerical study and the related technical support have been provided by the CRESCO/ENEA\-GRID High Performance Computing infrastructure and its staff \cite{HPCS2014}. CRESCO ({\color{red}C}omputational  {\color{red} RES}earch centre on {\color{red} CO}mplex systems) is funded by ENEA and by Italian and European research programmes.

\begin{appendices}

  \section{Proof of eq.~(\ref{eq:appone})}
  
The distribution $\Femp(x)$ can be regarded as a convolution of distributions $F_{\rm\scriptscriptstyle EMP}(x|Q,N)$ with weights $p(Q,N)$ and $Q,N=1,2,\ldots$ We can split the sum over $Q$ into sums over $Q\le Q_0$ and $Q>Q_0$, namely
  \begin{align}
    \Femp(x) & = \sum_{Q\le Q_0}\sum_{N}\,p(Q,N)\,F_{\rm\scriptscriptstyle EMP}(x|Q,N) + \sum_{Q> Q_0}\sum_{N}\,p(Q,N)\,F_{\rm\scriptscriptstyle EMP}(x|Q,N) \nonumber\\[1.0ex]
    & = \Femp^{(Q\le Q_0)}(x) + \Femp^{(Q>Q_0)}(x)\,.
  \end{align}
  Notice, however that neither $\Femp^{(Q\le Q_0)}$ nor $\Femp^{(Q>Q_0)}$ are probability densities, in that they fulfill
  \begin{equation}
    \int_0^\infty\rd x\, \Femp^{(Q\le Q_0)}(x) = \sum_{Q\le Q_0}\sum_N\,p(Q,N) < 1\, \quad \text{and}\quad \int_0^\infty\rd x\, \Femp^{(Q> Q_0)}(x) = \sum_{Q> Q_0}\sum_N\,p(Q,N) < 1\,.
  \end{equation}
Since $F_{\rm\scriptscriptstyle EMP}(x|Q,N)\ne 0$ only for $Q/N\le x \le Q$, it follows that $\Femp^{(Q\le Q_0)}(x)=0$ for $x>Q_0$. Hence,
  \begin{equation}
    \Femp(x) =  \Femp^{(Q>Q_0)}(x)\,,\qquad \text{for } x>Q_0\,.
  \end{equation}
Analogoulsy, the distribution $\Femp(x|Q>Q_0)$ can be regarded as a convolution of distributions $F_{\rm\scriptscriptstyle EMP}(x|Q,N)$ with weights $p(Q,N|Q>Q_0)$ and $Q>Q_0$, $N=1,2,\ldots$, see eqs.~(\ref{eq:FQQ0})--(\ref{eq:pqncond}).  Since $p(Q,N|Q>Q_0)$ is related to $p(Q,N)$ by
  \begin{equation}
    p(Q,N) = p(Q,N|Q>Q_0)p(Q>Q_0)\,, \quad \text{ for } Q>Q_0\,,
  \end{equation}
  it follows that 
  \begin{equation}
    \Femp(x)\, =\, \Femp^{(Q>Q_0)}(x)\, =\,   p(Q>Q_0)\,\Femp(x|Q>Q_0)\,<\,\Femp(x|Q>Q_0)\,,
  \end{equation}
  for $x>Q_0$. 
  
  \section{OSTP and generalized multiple $\zeta$--values}

  \newcommand{\Li}{\text{Li}}
  \newcommand*\mystrut[1]{\vrule width0pt height0pt depth#1\relax}
  \newcommand{\Czero}{\mathbf{\omega_0}}
  \newcommand{\Cone}{{\color{red}\mathbf{\omega_1}}}

  We can represent $\P\{Z_t=z\,|\,Z_{t-1}=\check{z}\}$ beyond the asymptotic regime $\check{z}\to\infty$ as an inverse power series in $z$ with coefficient functions amounting to linear combinations of generalized multiple ${\zeta\text{--values}}$. Our derivation assumes $\alpha\in\dN$. It is not clear at present whether/how it could be extended to non--integer $\alpha$.

  We start by recalling a simple property of the generating functions of discrete probability distributions. Let $(a_n)_{n=0}^\infty$ and $(b_n)_{n=0}^\infty$ be two sequences of real numbers and let $A(y) = \sum_{n=0}^\infty a_ny^n$ and $B(y) = \sum_{n=0}^\infty b_ny^n$ denote their respective generating functions. Consider also the convolution of $(a_n)_{n=0}^\infty$ and $(b_n)_{n=0}^\infty$, \ie the sequence $(c_n)_{n=0}^\infty$ defined by
  \begin{equation}
    c_n = \sum_{n_1,n_2=0}^{\infty} \delta_{n,n_1+n_2}\,a_{n_1}b_{n_2}\,,
  \end{equation}
  and let $C(y) =\sum_{n=0}^\infty c_n y^n$ denote in turn the generating function of $(c_n)_{n=0}^\infty$. It can be easily checked that
  \begin{equation}
    C(y) = A(y)B(y)\,.
    \label{eq:genprod}
  \end{equation}
  Eq.~(\ref{eq:genprod}) can be generalized recursively to the convolution of an arbitrary number of sequences. Since it holds without any restriction on $(a_n)_{n=0}^\infty$ and $(b_n)_{n=0}^\infty$, it applies in particular when $a_n\ge 0$, $b_n\ge 0$ and $A(1)=B(1)=1$, \ie when $a_n=a(n)$ and $b_n=b(n)$ are discrete probabilities. In this case we have $C(1)=1$ and thus $c_n=c(n)\ge 0$ is a discrete probability too.

  We have already observed that $\P\{Z_t=z\,|\,Z_{t-1}=\check{z}\}$ is the $\check{z}$--fold convolution of $\pQ(n)$, see eq.~(\ref{eq:Pconvol}) (first line). Since $G_{\rm\scriptscriptstyle Q}(y) = y^\nmin \Phi(y,\alpha,\nmin)/\zeta(\alpha,\nmin)$ is the generating function of $\pQ(n)$, it follows by differentiation that
  \begin{equation}
    \P\{Z_t=z\,|\,Z_{t-1}=\check{z}\} = \frac{\left[\zeta(\alpha,\nmin)\right]^{-\check{z}}}{z!}\frac{\rd^z}{\rd y^{z}}\Psi(y,\alpha,\nmin)^{\check{z}}\biggr|_{y=0}\,.
    \label{eq:OSTPLerch}
  \end{equation}
  with $\Psi(y,\alpha,\nmin) \equiv y^{\nmin}\Phi(y,\alpha,\nmin) = \sum_{n=\nmin}^\infty y^n/n^\alpha$. This function is a generalization of the polylogarithm $\Li_\alpha(y)$. Indeed, we have $\Li_\alpha(y) = \Psi(y,\alpha,1)$. Since the product of polylogarithms can be expressed in terms of multiple polylogarithms, we expect a similar representation to hold also for $\Psi(y,\alpha,\nmin)^{\check{z}}$. In sect.~B.1 we recall some basic elements of the theory of multiple polylogarithms and shuffle products (for this we follow ref.~\cite{Waldschmidt}). In sect.~B.2 we discuss how the theory can be adapted to our case and be used to work out eq.~(\ref{eq:OSTPLerch}). We notice incidentally that over the past few years multiple polylogarithms and shuffle products have been gaining popularity in the context of quantum field theory, see ref.~\cite{Brown} for a review. It is funny to see that they can be used to describe complex systems too.

  \subsection{Review of multiple polylogarithms and shuffle products}

  The polylogarithm function generalizes the usual logarithm and the Riemann $\zeta$--function. For $\alpha\in\dN$ and $y\in\dC$ with  $|y|\le 1$ and $(\alpha,y)\ne(1,1)$, the polylogarithm is defined by the Taylor series
  \begin{equation}
    \Li_\alpha(y) = \sum_{n=1}^\infty \frac{y^n}{n^\alpha}\,.
    \label{eq:polydef}
  \end{equation}
  For $\alpha=1$ and $y\ne 1$ it fulfills $\Li_1(y) = -\log(1-y)$. For $\alpha\ge 2$ and $y=1$ it fulfills $\Li_\alpha(1) = \zeta(\alpha)$. Polylogarithms can be also defined recursively via differential equations, namely
  \begin{equation}
    \left\{\begin{array}{ll}
    \dfrac{\rd}{\rd y}\Li_1(y) = \dfrac{1}{1-y}\,, & \\[3.0ex]
    \dfrac{\rd}{\rd y}\Li_\alpha(y) = \dfrac{1}{y}\Li_{\alpha-1}(y)\,,\qquad  \text{for }\ \alpha\ge 2\,,
    \end{array}\right.
    \label{eq:polydiff}
  \end{equation}
  with initial condition $\Li_\alpha(0)=0$, as can be seen by differentiating eq.~(\ref{eq:polydef}) term by term. Recursive integration of eq.~(\ref{eq:polydiff}) yields the nested integral representation
  \begin{align}
    \Li_\alpha(y) & = \int_0^y \frac{\rd t_1}{t_1}\,\Li_{\alpha-1} = \int_{0}^y \frac{\rd t_1}{t_1}\int_{0}^{t_1} \frac{\rd t_2}{t_2}\ldots\int_{0}^{t_{\alpha-2}}\frac{\rd t_{\alpha-1}}{t_{\alpha-1}}\int_0^{t_{\alpha-1}}\frac{\rd t_\alpha}{1-t_\alpha}\nonumber\\[2.0ex]
    & = \int_{\cJ_\alpha(y)} \frac{\rd t_1}{t_1} \frac{\rd t_2}{t_2}\ldots \frac{\rd t_{\alpha-1}}{t_{\alpha-1}} \frac{\rd t_\alpha}{1-t_\alpha}\,,
    \label{eq:intrepr}
  \end{align}
  where $\cJ_{\alpha}(y)$ is  the $\alpha$--dimensional domain
  \begin{equation}
    \cJ_{\alpha}(y) = \bigl\{\, (t_1,\ldots,t_\alpha):\quad y>t_1>t_2>\ldots>t_\alpha>0 \, \bigl\}\,.
  \end{equation}
  The rightmost integral in eq.~(\ref{eq:intrepr}) belongs to the class of Chen iterated integrals \cite{Chen}. Specifically, given the holomorphic 1--forms $\omega_0 = \rd t/t$ and $\omega_1 = \rd t/(1-t)$, we can recast eq.~(\ref{eq:intrepr}) in the form
  \begin{equation}
    \Li_\alpha(y) = \int_0^y \underbrace{\mystrut{1.0ex}\omega_0 \circ\ldots\circ\omega_0}_{(\alpha-1)\text{ times}}\circ\,\,\omega_1 = \int_0^z\omega_0^{\circ(\alpha-1)}\circ\,\omega_1\,,
    \label{eq:Lichen}
  \end{equation}
  where the nesting operator $\circ$ is defined recursively by $\int_{0}^y \phi_1\circ\ldots\circ\phi_k = \int_0^y\phi_1(t)\int_0^t\phi_2\circ\ldots\circ\phi_k$. In the following we let $\tilde \omega_\alpha = \omega_0^{\circ(\alpha-1)}\circ\omega_1$ and accordingly we write $\Li_\alpha(y)$ in the more compact notation $\Li_\alpha(y) = \int_0^y \tilde \omega_\alpha$.

\begin{figure}[t!]
    \centering 
    \begin{minipage}{0.20\textwidth}
      \centering
      \includegraphics[width=1.0\textwidth]{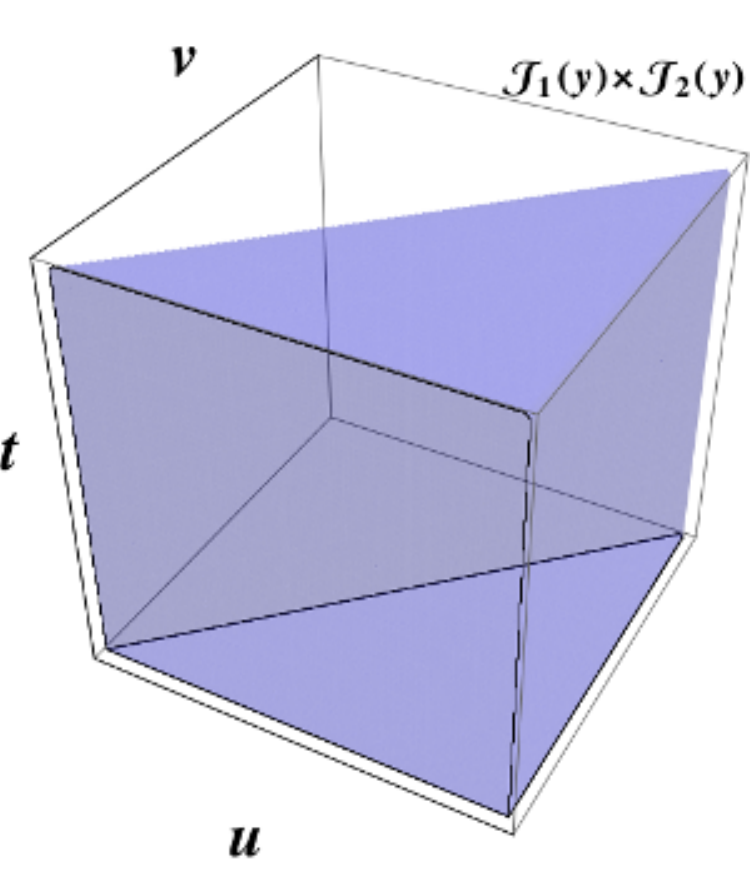}
    \end{minipage}
    \hskip 0.4cm\begin{minipage}{0.20\textwidth}
    \centering
    \includegraphics[width=1.0\textwidth]{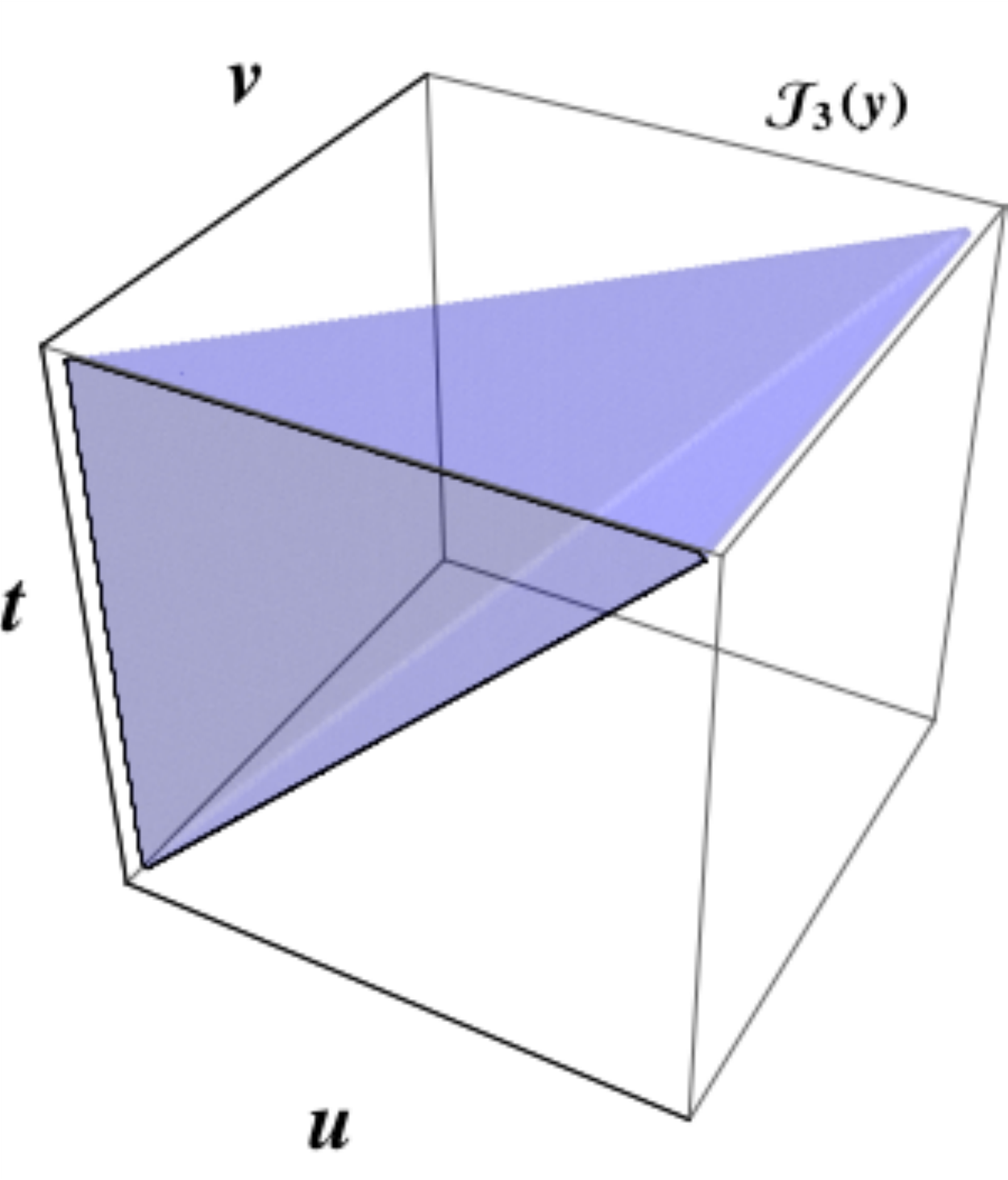}
    \end{minipage}
    \hskip 0.4cm\begin{minipage}{0.20\textwidth}
    \centering
    \includegraphics[width=1.0\textwidth]{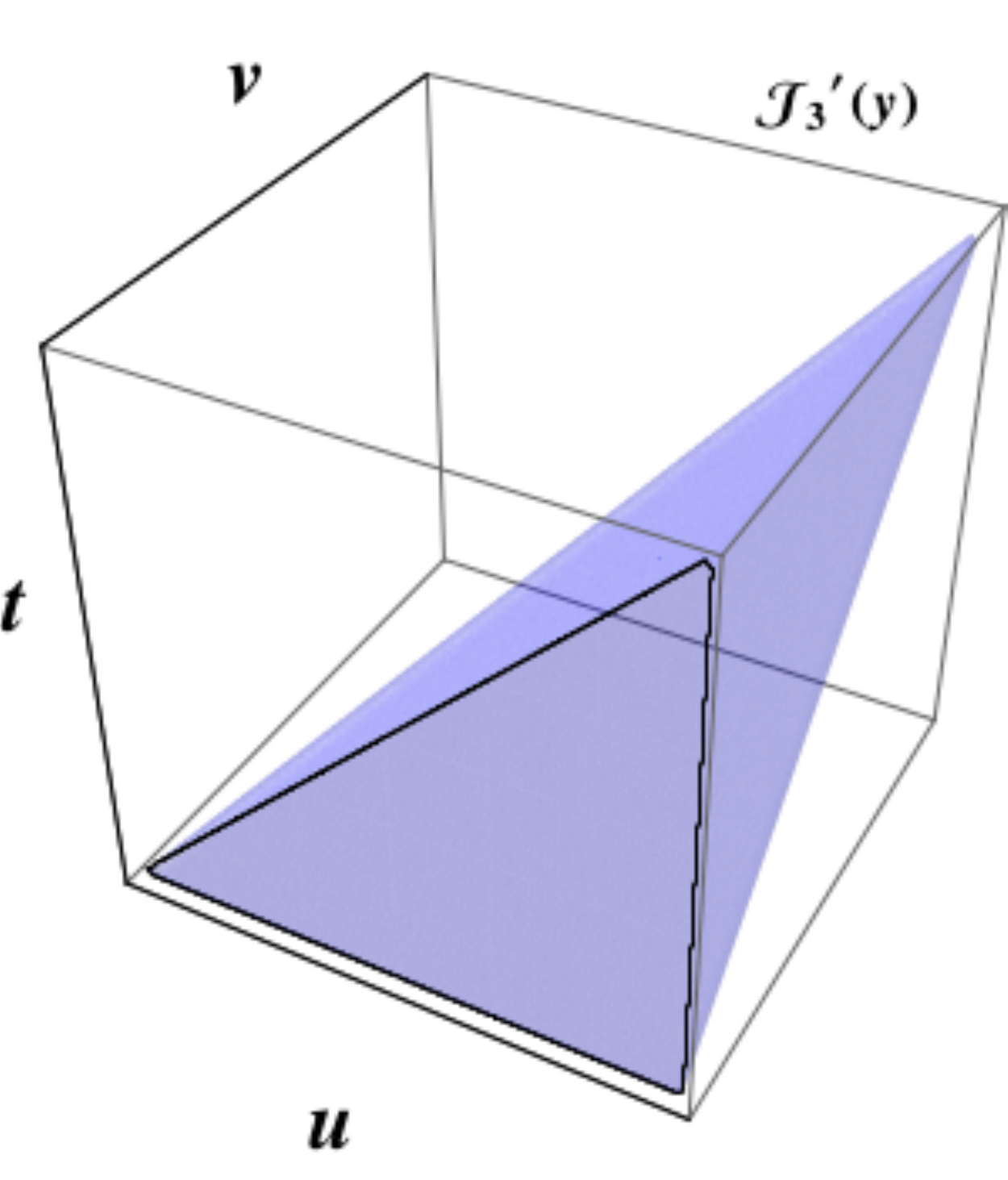}
    \end{minipage}
    \hskip 0.4cm\begin{minipage}{0.20\textwidth}
    \centering
    \includegraphics[width=1.0\textwidth]{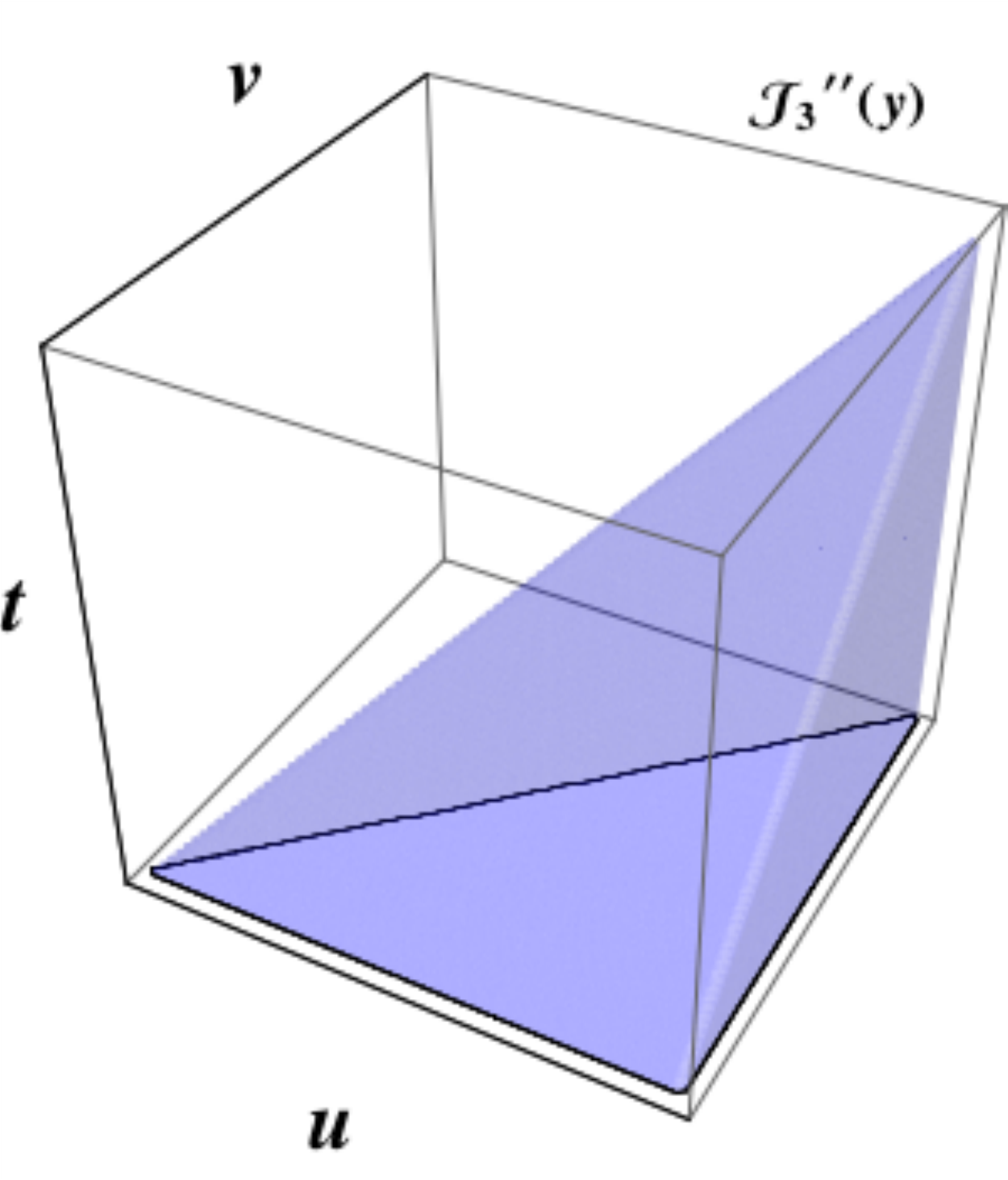}
    \end{minipage}
    \caption{\footnotesize Decomposition of $\cJ_1(y)\times\cJ_2(y)$ into the union of $\cJ_3(y)$, $\cJ'_3(y)$ and $\cJ''_3(y)$.}
    \label{fig:figfifteen}
    \vskip -0.3cm
  \end{figure}

  The product of polylogarithms can be expressed in terms of Chen iterated integrals. As an example, we consider the product $\Li_1(y)\Li_2(y)$ for $y\in\dR$ and $0<y<1$. This is given by
  \begin{equation}
    \Li_1(y)\Li_2(y) = \int_{\cJ_1(y)}\frac{\rd t_1}{1-t_1}\int_{\cJ_2(y)}\frac{\rd t_2}{t_2}\frac{\rd t_3}{1-t_3} = \int_{\cJ_1(y)\times\cJ_2(y)}\frac{\rd t_1}{1-t_1}\frac{\rd t_2}{t_2}\frac{\rd t_3}{1-t_3}\,.
  \end{equation}
The expression on the r.h.s. is not a Chen iterated integral. Nevertheless, the 3--dimensional domain $\cJ_1(y)\times\cJ_2(y)=\{(t_1,t_2,t_3):\ y>t_1>0 \text{ and } y>t_2>t_3>0\}$ can be decomposed into nested domains. Indeed, we have
  \begin{align}
    \cJ_1(y)\times\cJ_2(y)=\cJ_3(y)\cup\cJ'_3(y)\cup\cJ''_3(y)\,,
    \label{eq:decomp12}
  \end{align}
up to zero--measure sets, with
  \begin{align}
    \cJ_3'(y) & = \bigl\{\, (t_1,t_2,t_3):\quad y>t_2>t_1>t_3>0\,\bigr\}\,,\\[2.0ex]
    \cJ_3''(y) & = \bigl\{\, (t_1,t_2,t_3):\quad y>t_2>t_3>t_1>0\,\bigr\}\,.
  \end{align}
  The decomposition of $\cJ_1(y)\times\cJ_2(y)$ into the disjoint union of $\cJ_3(y)$, $\cJ'_3(y)$ and $\cJ_3''(y)$ is shown in Fig.~\ref{fig:figfifteen}. Thanks to it, the integral representing $\Li_1(y)\Li_2(y)$ splits into three contributions, each taking the form of a Chen iterated integral, namely
  \begin{equation}
    \Li_1(y)\Li_2(y) = \int_0^y \omega_1\circ\omega_0\circ\omega_1 + 2\int_{0}^y\omega_0\circ\omega_1^2 = \int_0^y\left(\omega_1\circ\omega_0\circ\omega_1 + 2\,\omega_0\circ\omega_1^2\right)\,.
  \end{equation}
  The rightmost differential form is obtained by interlacing the 1--form ${\color{blue}\mathbf{\omega_1}}$ of $\Li_1(z)$ with the differential form $\omega_0\circ\omega_1$ of $\Li_2(z)$ in all possible ways, \ie
  \vskip -0.5cm
  \begin{equation}
    \begin{tikzpicture}[>=stealth,baseline,anchor=base,inner sep=0pt]
      \matrix (foil) [matrix of math nodes,nodes={minimum height=0.5em}] {
        {{\color{blue}\mathbf{\omega_1}}} &\ ,\ & \omega_0 &\, \circ \, & \omega_1 & \quad \longrightarrow \quad & {\color{blue}\mathbf{\omega_1}} &\,\circ\, & \omega_0 &\,\circ\, &\omega_1 & \,,\quad &  \omega_0 &\,\circ\,\, & {\color{blue}\mathbf{\omega_1}} &\,\circ\, & \omega_1 &\,,\quad & \omega_0 &\,\circ\,\, & \omega_1 &\,\circ\, & {\color{blue}\mathbf{\omega_1}}\,. \\
      };
      \path[->] ($(foil-1-2.north)+(-2ex,1.2ex)$)   edge[black,bend left=10]    ($(foil-1-7.north)+(-1ex,1.2ex)$);
      \path[->] ($(foil-1-2.north)+(-2ex,1.2ex)$)   edge[black,bend left=10]    ($(foil-1-15.north)+(-1ex,1.2ex)$);
      \path[->] ($(foil-1-2.north)+(-2ex,1.2ex)$)   edge[black,bend left=10]    ($(foil-1-23.north)+(-1ex,1.2ex)$);
    \end{tikzpicture}
    \label{eq:shuffletwo}
  \end{equation}
  If we think about $\omega_0$ and $\omega_1$ as playing cards in a deck, the above operation consists in shuffling the sets of cards $\color{blue}\mathbf{\omega_1}$ and $\omega_0\circ\omega_1$ in all possible ways. 

  The above example can be generalized to the product of an arbitrary number of polylogarithms. Given the 1--forms $\phi_1,\ldots,\phi_k,\psi_1\,\ldots,\psi_n$ with $k,n\ge 1$, we define the shuffle product of $\phi_1\circ\ldots \circ\phi_k$ and $\psi_1\circ\ldots\circ\psi_n$ recursively via
  \begin{align}
    (\phi_1\circ\ldots\circ\phi_k) \shuffle (\psi_1\circ\ldots\circ\psi_n) & = \phi_1\circ[(\phi_2\circ\ldots\circ\phi_k) \shuffle(\psi_1\circ\ldots\circ\psi_n)] \nonumber\\[1.0ex]
    & + \psi_1\circ[(\phi_2\circ\ldots\circ\phi_k) \shuffle(\psi_2\circ\ldots\circ\psi_n)] \,,
    \label{eq:shuffleprod}
  \end{align}
  with $\phi_1\shuffle\psi_1 = \phi_1\circ\psi_1 + \psi_1\circ\phi_1$. For instance, for $k=1$ and $n=2$ we have
  \begin{equation}
    \phi_1 \shuffle (\psi_1\circ\psi_2) = \phi_1\circ\psi_1\circ\psi_2 + \psi_1\circ\phi_1\circ\psi_2 + \psi_1\circ\psi_2\circ\phi_1\,,
  \end{equation}
 for $k=n=2$ we have
  \begin{align}
    (\phi_1\circ\phi_2) \shuffle (\psi_1\circ\psi_2) & = \phi_1\circ\phi_2\circ\psi_1\circ\psi_2 + \phi_1\circ\psi_1\circ\phi_2\circ\psi_2 \nonumber\\[1.0ex]
    & + \phi_1\circ\psi_1\circ\psi_2\circ\phi_2 + \psi_1\circ\phi_1\circ\phi_2\circ\psi_2\nonumber\\[1.0ex]
    & + \psi_1\circ\phi_1\circ\psi_2\circ\phi_2 + \psi_1\circ\psi_2\circ\phi_1\circ\phi_2\,,
  \end{align}
and so on and so forth. These examples confirm that the shuffle product interlaces the sets $\{\phi_1,\ldots,\phi_k\}$ and $\{\psi_1,\ldots,\psi_n\}$ in all possible ways without changing the order of the 1--forms within each set. By arguments similar to those presented above it can be proved in full generality that
  \begin{equation}
    \int_0^y \phi_1\circ\ldots\circ\phi_k \int_0^y \psi_1\circ\ldots\circ\psi_n = \int_0^y (\phi_1\circ\ldots\circ\phi_k) \shuffle (\psi_1\circ\ldots\circ\psi_n)\,.
    \label{eq:intprod}
  \end{equation}
  Eq.~(\ref{eq:intprod}) applies immediately to the product of polylogarithms. For instance, we have
  \begin{equation}
    \Li_\alpha(y)\Li_{\alpha'}(y) = \int_0^y \tilde\omega_\alpha\shuffle\tilde\omega_{\alpha'}\qquad \text { and }\qquad [\Li_\alpha(y)]^{\check{z}} = \int_0^y \underbrace{\tilde\omega_\alpha\shuffle\tilde\omega_{\alpha}\shuffle\ldots\shuffle\tilde\omega_{\alpha}}_{\check{z}\text{ times}} = \int_0^y\tilde\omega_\alpha^{\,\shuffle\check{z}}\,.
  \end{equation}
  In particular, each term in the expansion of $[\Li_\alpha(y)]^{\check{z}}$ is a Chen iterated integral of the form
    \begin{equation}
      \Li_{\alpha_1,\ldots,\alpha_{\check{z}}}(y) = \int_0^y \tilde\omega_{\alpha_1}\circ\ldots\circ\tilde\omega_{\alpha_{\check{z}}} = \int_0^y \omega_0^{\circ(\alpha_1-1)}\circ\omega_1\circ\ldots\circ\omega_0^{\circ(\alpha_{\check{z}}-1)}\circ\omega_1\,,
      \label{eq:multichen}
    \end{equation}
    with the index vector $\{\alpha_1,\ldots,\alpha_{\check{z}}\}$ fulfilling $\alpha_1 + \ldots + \alpha_{\check{z}} = \check{z}\alpha$ and representing one of many possible ways of interlacing $\check{z}$ differential forms $\tilde\omega_\alpha$. Reviewing the combinatorics of the indices goes beyond our aims here. We simply define
\begin{equation}
  \{\alpha\}_{\check{z}} = \underbrace{\{\alpha,\ldots,\alpha\}}_{\check{z}\text{ times }}\,,
\end{equation}
 and we let $\shuffle\{\alpha\}_{\check{z}}$ denote the set of all possible indices $\{\alpha_1,\ldots,\alpha_{\check{z}}\}$ contributing to the shuffle product. Formally, we have
    \begin{equation}
      [\Li_\alpha(y)]^{\check{z}} = \sum_{\{\alpha_1,\ldots,\alpha_{\check{z}}\}\,\in\,\shuffle\{\alpha\}_{\check{z}}}\Li_{\alpha_1,\ldots,\alpha_{\check{z}}}(y)\,.
    \end{equation}

    The function $\Li_{\alpha_1,\ldots,\alpha_{\check{z}}}(y)$ is the multiple polylogarithm of the $k$th order in one variable. It generalizes the polylogarithm function in the index space. Although eq.~(\ref{eq:multichen}) provides an integral representation of it, $\Li_{\alpha_1,\ldots,\alpha_{\check{z}}}(y)$ can be equivalently defined as a Taylor series, namely 
    \begin{equation}
      \Li_{\alpha_1,\ldots,\alpha_{\check{z}}}(y)\, = \sum_{n_1\ge n_2\ge\ldots\ge n_{\check{z}}\ge 1}\frac{y^{n_1}}{n_1^{\alpha_1}\ldots n_k^{\alpha_{\check{z}}}}\,,\qquad \text{for }\ y\in\dC \text{ and } |y|\le 1\,,
      \label{eq:multipolytaylor}
    \end{equation}
    with $\alpha_1,\ldots,\alpha_{\check{z}}\ge 1$ and $(y,\alpha_1)\ne (1,1)$. By differentiating this series term by term, we get an alternative representation of $\Li_{\alpha_1,\ldots,\alpha_{\check{z}}}(y)$ as the solution of the recursive differential equations
  \begin{equation}
    \left\{\begin{array}{ll}
    \dfrac{\rd}{\rd y}\Li_{1,\alpha_2,\ldots,\alpha_{\check{z}}}(y) = \dfrac{1}{1-y}\Li_{\alpha_2,\ldots,\alpha_{\check{z}}}\,, & \\[3.0ex]
    \dfrac{\rd}{\rd y}\Li_{\alpha_1,\alpha_2,\ldots,\alpha_{\check{z}}}(y) = \dfrac{1}{y}\Li_{\alpha_1-1,\alpha_2,\ldots,\alpha_{\check{z}}}(y)\,,\qquad  \text{for }\ \alpha_1\ge 2\,,
    \end{array}\right.
    \label{eq:multipolydiff}
  \end{equation}
  with initial condition $\Li_{\alpha_1,\ldots,\alpha_{\check{z}}}(0) = 0$. Eq.~(\ref{eq:multichen}) is then obtained by recursively integrating eq.~(\ref{eq:multipolydiff}).

  We conclude this short review by introducing the multiple $\zeta$--function
  \begin{equation}
    \zeta(\alpha_1,\ldots,\alpha_{\check{z}})\, = \sum_{n_1\ge n_2\ge\ldots\ge n_{\check{z}}\ge 1}\frac{1}{n_1^{\alpha_1}\ldots n_k^{\alpha_{\check{z}}}}\,,\qquad \text{ for }\ \alpha_1\ge 2\ \text{ and }\ \alpha_2,\ldots,\alpha_{\check{z}}\ge 1\,.
  \end{equation}
  This function represents a multi--dimensional generalization of the ordinary Riemann $\zeta$--function and fulfills $\zeta(\alpha_1,\ldots,\alpha_{\check{z}}) = \Li_{\alpha_1,\ldots,\alpha_{\check{z}}}(1)$ for $\alpha_1\ge 2$.

  \subsection{Adaptation to the OSTP}
  
  The function $\Psi(y,\alpha,\nmin)$ can be regarded as an upper incomplete (or lower--truncated) polylogarithm, \ie it satisfies the relation
  \begin{equation}
    \Psi(y,\alpha,\nmin) = \Li_{\alpha}(y) - \sum_{j=0}^{\nmin-1}\frac{y^j}{j^\alpha}\,.
  \end{equation}
  From eq.~(\ref{eq:polydiff}) and recalling that $\sum_{k=0}^s y^k = (1-y^{s-1})/(1-y)$ for $0<y<1$, we see that $\Psi(y,\alpha,n)$ fulfills the recursive differential equations
  \begin{equation}
    \left\{\begin{array}{ll}
    \dfrac{\rd}{\rd y}\Psi(y,1,\nmin) = \dfrac{y^{\nmin-1}}{1-y}\,, & \\[3.0ex]
    \dfrac{\rd}{\rd y}\Psi(y,\alpha,\nmin) = \dfrac{1}{y}\Psi(y,\alpha-1,\nmin)\,,\qquad  \text{for }\ \alpha\ge 2\,,
    \end{array}\right.
    \label{eq:lerchdiff}
  \end{equation}
  with initial condition $\Psi(0,\alpha,\nmin) = 0$. As a consequence, it can be represented as a Chen iterated integral,
  \begin{equation}
    \Psi(y,\alpha,\nmin) = \int_0^y \underbrace{\omega_0\circ\ldots\circ\omega_0}_{\alpha-1 \text{ times}}\circ\,\omega_{1,\nmin} = \int_0^y \omega_0^{\circ(\alpha-1)}\circ\omega_{1,\nmin} = \int_0^y \tilde\omega_{\alpha,\nmin}\,,
    \label{eq:lerchchen}
  \end{equation}
  where we let $\omega_{1,\nmin} = t^{\nmin-1}\rd t/(1-t)$ and $\tilde\omega_{\alpha,\nmin} = \omega_0^{\alpha-1}\circ\,\omega_{1,\nmin}$. Since the only difference between eq.~(\ref{eq:lerchchen}) and eq.~(\ref{eq:Lichen}) is that $\omega_1$ is replaced by $\omega_{1,\nmin}$, we conclude that  $\Psi(y,\alpha,\nmin)^{\check{z}}$ can be analogously represented as the integral of the shuffle product of the corresponding nested differential forms, \ie
  \begin{equation}
    [\Psi(y,\alpha,\nmin)]^{\check{z}} = \int_0^y \underbrace{\tilde\omega_{\alpha,\nmin}\shuffle\tilde\omega_{\alpha,\nmin}\shuffle\ldots\shuffle\tilde\omega_{\alpha,\nmin}}_{\check{z}\text{ times}} = \int_0^y\,\tilde\omega_{\alpha,\nmin}^{\,\shuffle\check{z}}\,.
  \end{equation}
  As previously, each term in the expansion of $\omega_{\alpha,\nmin}^{\,\shuffle\check{z}}$ gives rise to a Chen iterated integral
  \begin{align}
    \Psi(y,\{\alpha_1,\ldots,\alpha_{\check{z}}\},\nmin) & = \int_0^y \tilde\omega_{\alpha_1,\nmin}\circ\ldots\circ\tilde\omega_{\alpha_{\check{z}},\nmin} \nonumber\\[1.0ex]
    & = \int_0^y \omega_0^{\circ(\alpha_1-1)}\circ\,\omega_{1,\nmin}\circ\ldots\circ\omega_0^{\circ(\alpha_{\check{z}}-1)}\circ\,\omega_{1,\nmin}\,,
    \label{eq:truncmulerchint}
  \end{align}
  with the index vector $\{\alpha_1,\ldots,\alpha_{\check{z}}\}$ fulfilling $\alpha_1+\ldots+\alpha_{\check{z}}=\check{z}\alpha$ and representing one of many possible ways of interlacing $\check{z}$ differential forms $\tilde\omega_{\alpha,\nmin}$. Formally, we have
  \begin{equation}
    [\Psi(y,\alpha,\nmin)]^{\check{z}} = \sum_{\{\alpha_1,\ldots,\alpha_{\check{z}}\}\,\in\,\shuffle\{\alpha\}_{\check{z}}}\Psi(y,\{\alpha_1,\ldots,\alpha_{\check{z}}\},\nmin)\,.
  \end{equation}
  The function $\Psi(y,\{\alpha_1,\ldots,\alpha_{\check{z}}\},\nmin)$ is a lower--truncated multiple polylogarithm. It generalizes the function $\Psi(y,\alpha,\nmin)$ in the index space. Upon differentiating eq.~(\ref{eq:truncmulerchint}), we see that $\Psi(y,\{\alpha_1,\ldots,\alpha_{\check{z}}\},\nmin)$ fulfills the recursive differential equations
  \begin{equation}
    \left\{\begin{array}{ll}
    \dfrac{\rd}{\rd y}\Psi(y,\{1,\alpha_2,\ldots,\alpha_{\check{z}}\},\nmin) = \dfrac{y^{\nmin-1}}{1-y}\Psi(y,\{\alpha_2,\ldots,\alpha_{\check{z}}\},\nmin)\,, & \\[3.0ex]
    \dfrac{\rd}{\rd y}\Psi(y,\{\alpha_1,\ldots,\alpha_{\check{z}}\},\nmin) = \dfrac{1}{y}\Psi(y,\{\alpha_1-1,\alpha_2,\ldots,\alpha_{\check{z}}\},\nmin)\,,\qquad  \text{for }\ \alpha_1\ge 2\,,
    \end{array}\right.
    \label{eq:truncmulerchdiff}
  \end{equation}
  with initial condition $\Psi(0,\{\alpha_1,\ldots,\alpha_{\check{z}}\},\nmin)=0$. Similar to the multiple polylogarithm, also the function $\Psi(y,\{\alpha_1,\ldots,\alpha_k\},\nmin)$ can be represented as a power series, \ie
  \begin{equation}
    \Psi(y,\{\alpha_1\ldots,\alpha_{\check{z}}\},\nmin) = \sum_{n_1=\check{z}\nmin}^\infty\ \sum_{n_2=(\check{z}-1)\nmin}^{n_1-\nmin}\ \ldots\ \sum_{n_{\check{z}}=\nmin}^{n_{\check{z}-1}-\nmin} \frac{y^{n_1}}{n_1^{\alpha_1}\ldots n_{\check{z}}^{\alpha_{\check{z}}}}\,.
    \label{eq:Psiseries}
  \end{equation}
  The reader can easily check that the r.h.s of eq.~(\ref{eq:Psiseries}) fulfills eq.~(\ref{eq:truncmulerchdiff}) for $\alpha_1\ge 2$. We want to prove that it fulfills eq.~(\ref{eq:truncmulerchdiff}) also for $\alpha_1=1$. To this aim, we let $\tilde\Psi(y,\{\alpha_1,\ldots,\alpha_{\check{z}}\},\nmin)$ denote the r.h.s. of eq.~(\ref{eq:Psiseries}). Moreover, to simplify the notation, we let
  \begin{equation}
    \tilde\Psi(y,\{1,\alpha_2,\ldots,\alpha_{\check{z}}\},\nmin) = \sum_{n_1=\check{z}\nmin}^\infty c_{n_1}\frac{y^{n_1}}{n_1}\,,
  \end{equation}
  with the coefficient $c_{n_1}$ being given by
  \begin{equation}
    c_{n_1} = \sum_{n_2=(\check{z}-1)\nmin}^{n_1-\nmin}\ \ldots\ \sum_{n_{\check{z}}=\nmin}^{n_{\check{z}-1}-\nmin} \frac{1}{n_2^{\alpha_2}\ldots n_{\check{z}}^{\alpha_{\check{z}}}}\,.
    \label{eq:cn1}
  \end{equation}
Now, we have
  \begin{align}
    (1-y)\frac{\rd}{\rd y}\tilde\Psi(y,\{1,\alpha_2,\ldots,\alpha_{\check{z}}\},\nmin) & = \sum_{n_1=\check{z}\nmin}^\infty c_{n_1}y^{n_1-1} - \sum_{n_1=\check{z}\nmin}^\infty c_{n_1}y^{n_1}\nonumber\\[1.0ex]
    & \hskip 0.0cm = c_{\check{z}\nmin}y^{\check{z}\nmin-1} + \sum_{n_1=\check{z}\nmin}^\infty(c_{n_1+1}-c_{n_1})z^{n_1}\,.
  \end{align}
The coefficient $c_{n_1+1}-c_{n_1}$ is obtained from eq.~(\ref{eq:cn1}) by removing the sum over $n_2$ and by calculating the rest at $n_2 = n_1+1-\nmin$, \ie
  \begin{equation}
  c_{n_1+1}-c_{n_1} = \sum_{n_3=(\check{z}-2)\nmin}^{n_1+1-2\nmin}\sum_{n_4=(\check{z}-3)\nmin}^{n_3-\nmin}\ldots\sum_{n_{\check{z}}=\nmin}^{n_{\check{z}-1}-\nmin}\frac{1}{(n_1+1-\nmin)_{\vphantom{0.4ex}}^{\alpha_2}n_3^{\alpha_3}\ldots n_{\check{z}}^{\alpha_{\check{z}}}}\,.
  \end{equation}
Inserting this expression into the previous equation yields
  \begin{align}
    & (1-y)\frac{\rd}{\rd y}\tilde\Psi(y,\{\alpha_1,\ldots,\alpha_{\check{z}}\},\nmin)  \nonumber\\[1.0ex]
    & \hskip 1.0cm  = c_{\check{z}\nmin}y^{\check{z}\nmin-1} + \sum_{n_1=\check{z}\nmin}^\infty \sum_{n_3=(\check{z}-2)\nmin}^{n_1+1-2\nmin}\ldots\sum_{n_{\check{z}}=\nmin}^{n_{\check{z}-1}-\nmin}\frac{y^{n_1}}{(n_1+1-\nmin)_{\vphantom{0.4ex}}^{\alpha_2}n_3^{\alpha_3}\ldots n_{\check{z}}^{\alpha_{\check{z}}}}\nonumber
  \end{align}
  \begin{align}
    & \hskip 1.0cm = y^{\nmin-1}\,\biggl\{\sum_{n_2=(\check{z}-1)\nmin}^{\infty}\sum_{n_3=(\check{z}-2)\nmin}^{n_2-\nmin}\ldots \sum_{n_{\check{z}}=\nmin}^{n_{\check{z}-1}-\nmin}\frac{y^{n_2}}{n_2^{\alpha_2}n_3^{\alpha_3}\ldots n_{\check{z}}^{\alpha_{\check{z}}}}\biggr\} \nonumber\\[1.5ex]
    & \hskip 1.0cm = y^{\nmin-1}\tilde\Psi(y,\{\alpha_2,\ldots,\alpha_{\check{z}}\},\nmin) \,.
  \end{align}
  It follows that $\tilde\Psi(y,\{1,\alpha_2,\ldots,\alpha_{\check{z}}\},\nmin) = \Psi(y,\{1,\alpha_2,\ldots,\alpha_{\check{z}}\},\nmin)$.

  Just as the multiple $\zeta$--function $\zeta(\alpha_1,\ldots,\alpha_{\check{z}})$ is obtained from the multiple polylogarithm $\Li_{\alpha_1,\ldots,\alpha_{\check{z}}}(y)$ by taking the latter at $y=1$, we define a truncated multiple $\zeta$--function $\zeta(\{\alpha_1,\ldots,\alpha_{\check{z}}\},\nmin)$ from the truncated multiple polylogarithm $\Psi(y,\{\alpha_1,\ldots,\alpha_{\check{z}}\},\nmin)$ by taking the latter at $y=1$, namely
  \begin{equation}
    \zeta(\{\alpha_1\ldots,\alpha_{\check{z}}\},\nmin) = \sum_{n_1=\check{z}\nmin}^\infty\ \sum_{n_2=(\check{z}-1)\nmin}^{n_1-\nmin}\ \ldots\ \sum_{n_{\check{z}}=\nmin}^{n_{\check{z}-1}-\nmin} \frac{1}{n_1^{\alpha_1}\ldots n_{\check{z}}^{\alpha_{\check{z}}}}\,.
  \end{equation}
  This function is a multi--dimensional generalization of the Hurwitz $\zeta$--function $\zeta(\alpha,\nmin)$. 

  By means of the above formalism we can work out eq.~(\ref{eq:OSTPLerch}). Since  $y^{\check{z}\nmin}$ is the monomial of lowest degree contributing to $\Psi(y,\{\alpha_1,\ldots,\alpha_{\check{z}}\},\nmin)$, we have
  \begin{align}
    \frac{1}{z!}\frac{\rd^z}{\rd y^z}\Psi(y,\{\alpha_1,\ldots,\alpha_{\check{z}}\},\nmin)\biggr|_{y=0} & \, = \, \frac{\theta_{z,\check{z}\nmin}}{z!}\frac{\rd^z}{\rd y^z}\biggl\{\frac{y^{z}}{z^{\alpha_1}}\sum_{n_2=(\check{z}-1)\nmin)}^{z-\nmin}\ldots\sum_{n_{\check{z}}=\nmin)}^{n_{k-1}-\nmin}\frac{1}{n_2^{\alpha_2}\ldots n_{\check{z}}^{\alpha_{\check{z}}}}\biggr\}\nonumber\\[2.0ex]
    &\, =\,  \theta_{z,\check{z}\nmin}\,\frac{\zeta(z,\{\alpha_2,\ldots,\alpha_{\check{z}}\},\nmin)}{z^{\alpha_1}}\,,
  \end{align}
  with 
  \begin{equation}
    \zeta(z,\{\alpha_2,\ldots,\alpha_{\check{z}}\},\nmin) = \sum_{n_2=(\check{z}-1)\nmin}^{z-\nmin}\ldots\sum_{n_{\check{z}}=\nmin}^{n_{k-1}-\nmin}\frac{1}{n_2^{\alpha_2}\ldots n_{\check{z}}^{\alpha_{\check{z}}}}\,,
  \end{equation}
  representing a doubly truncated multiple $\zeta$-function. It follows that
  \begin{empheq}[box=\mybluebox]{align}
    \P\{Z_t=z\,|\,Z_{t-1}=\check{z}\}\, =\, \theta_{z,\check{z}\nmin}\sum_{\{\alpha_1,\ldots,\alpha_{\check{z}}\}\,\in\,\shuffle\{\alpha\}_{\check{z}}} \frac{1}{z^{\alpha_1}}\frac{\zeta(z,\{\alpha_2,\ldots,\alpha_{\check{z}}\},\nmin)}{\zeta(\alpha,\nmin)^{\check{z}}}\,.
    \label{eq:zetarepr}
  \end{empheq}
  The presence of $\theta_{z,\check{z}\nmin}$ is perfectly natural, since each of the $\check{z}$ vertices on the $(t-1)$th level of a quenched tree generates at least $\nmin$ vertices. It can be easily shown that $\alpha_1\ge\alpha$ for each $\{\alpha_1,\ldots,\alpha_{\check{z}}\}\,\in\,\shuffle\{\alpha\}_{\check{z}}$. Suppose indeed that you have a deck of playing cards with only two different types of cards, namely $\mathbf{\omega_0}$ and $\color{red}\mathbf{\omega_1}$. The deck is made of $(\alpha-1)\check{z}$ cards of type $\mathbf{\omega_0}$ and $\check{z}$ cards of type $\color{red}\mathbf{\omega_1}$. These are initially stacked in iterated groups $\underbrace{\Czero,\ldots,\Czero}_{\alpha-1\text{ times}},\Cone$, namely
  \begin{equation} 
    \rlap{$\overbracket{\phantom{\hskip-2pt\Czero^2,\ldots,\Czero,\,\Cone}}^{\mystrut{1.0ex}1\text{st group}}$}\underbrace{\mystrut{1.0ex}\Czero,\ldots,\Czero}_{\alpha-1\text{ times}},\,\Cone\,,\rlap{$\overbracket{\phantom{\hskip -3pt\Czero^2,\ldots,\Czero,\,\Cone}}^{\mystrut{1.0ex}2\text{nd group}}$}\underbrace{\mystrut{1.0ex}\Czero\,\ldots,\Czero}_{\alpha-1\text{ times}},\,\Cone\,,\ldots,\rlap{$\overbracket{\phantom{\hskip -7pt\Czero^2,\ldots,\Czero,\,\Cone}}^{\mystrut{1.0ex}\check{z}\text{th group}}$}\underbrace{\mystrut{1.0ex}\Czero,\ldots\Czero}_{\alpha-1\text{ times}},\,\Cone\,.
 \end{equation}
The basic rule of the shuffle product states that upon shuffling the deck, each card of type $\Cone$ must always stay to the right of all cards $\Czero$ belonging to its group. It follows that each allowed reshuffling of cards has at least $\alpha-1$ cards of type $\Czero$ on the left, \ie $\alpha_1-1\ge \alpha-1$. Hence, eq.~(\ref{eq:zetarepr}) represents the OSTP as a series of inverse powers $z^{-\alpha_1}$ with degree $\alpha_1\ge\alpha$ and non--negative modulating coefficient functions $\zeta(z,\{\alpha_2,\ldots,\alpha_{\check{z}}\},\nmin)/\zeta(\alpha,\nmin)^{\check{z}}$. A comparison between eq.~(\ref{eq:zetarepr}) and eq.~(\ref{eq:asymprepr}) inspires the conjecture
\begin{equation}
\lim_{\check{z}\to\infty}\lim_{z\to\infty}\ \frac{1}{\check{z}} \sum_{\substack{\{\alpha_1,\ldots,\alpha_{\check{z}}\}\,\in\,\shuffle\{\alpha\}_{\check{z}} \\[1.0ex] \text{with } \alpha_1=\,\alpha}}\frac{\zeta(z,\{\alpha_2,\ldots,\alpha_{\check{z}}\},\nmin)}{\zeta(\alpha,\nmin)^{\check{z}-1}}\ = \ 1\,.
\end{equation}
It is not clear whether this appealing formula is correct at all, since eq.~(\ref{eq:asymprepr}) has been derived under the hypothesis $\alpha\ne 1,2,3,\ldots$ while eq.~(\ref{eq:zetarepr}) has been obtained for integer $\alpha$. We just notice that each term on the l.h.s. has exactly $\check{z}-1$ nested sums at numerator and $\check{z}-1$ independent sums at denominator. We do not even know how eq.~(\ref{eq:zetarepr}) could be extended to non--integer values of $\alpha$. 

\end{appendices}

\bibliographystyle{hunsrt}
\bibliography{main}

\end{document}